\documentclass[11pt]{article}
\pdfoutput=1 
\usepackage{bbm,bm,graphicx,mathtools,color,slashed}
\usepackage{amssymb,amsmath}
\usepackage[colorlinks=true, pdfstartview=FitV, linkcolor=red, citecolor=blue, urlcolor=blue]{hyperref}
\usepackage[T1]{fontenc}
\usepackage[caption=false]{subfig}
\usepackage[nottoc]{tocbibind}
\usepackage[toc,page]{appendix}
\usepackage{authblk}
\usepackage[numbers,sort&compress]{natbib}
\usepackage{youngtab}
\usepackage[top=25truemm,bottom=25truemm,left=15truemm,right=15truemm]{geometry}
\usepackage{url}
\usepackage[margin=15truemm]{caption}
\usepackage{caption}
\usepackage{slashed}
\usepackage{ulem}
\numberwithin{equation}{section}
\allowdisplaybreaks

\usepackage{qcircuit}
\usepackage{mathtools}
\DeclarePairedDelimiter\bra{\langle}{\rvert}
\DeclarePairedDelimiter\ket{\lvert}{\rangle}
\DeclarePairedDelimiterX\braket[2]{\langle}{\rangle}{#1 \delimsize\vert #2}

\graphicspath{{./figures/}}

\newcommand{\ri}{\mathrm{i}}

\newcommand{\diff}{\mathrm{d}}

\newcommand{\up}{\uparrow}
\newcommand{\down}{\downarrow}
\newcommand{\be}{\begin{equation}}      
\newcommand{\ee}{\end{equation}}      
\newcommand{\bea}{\begin{eqnarray}}      
\newcommand{\eea}{\end{eqnarray}}

\newcommand{\tr}{\mathrm{tr}}
\newcommand{\im}{\mathrm{i}}
\newcommand{\e}{\mathrm{e}}

\newcommand{\calH}{\mathcal{H}}

\newcommand{\calO}{\mathcal{O}}

\newcommand{\calQ}{\mathcal{Q}}

\newcommand{\Tr}{\mathrm{Tr}}

\newcommand{\red}[1]{{\color{red}{#1}}}

\newlength{\fighskip} \fighskip=2pt
\newlength{\figvskip} \figvskip=3pt

\newcommand*{\figbox}[2]{{
 \def\figscale{#1}
 \def\arraystretch{0.8}
 \arraycolsep=0pt
 \begin{array}{c}
\vbox{\vskip\figscale\figvskip
  \hbox{\hskip\figscale\fighskip
    \includegraphics[scale=\figscale]{#2}}}
 \end{array}}}

\begin{document}
\title{Diagnosis of information scrambling\\ from Hamiltonian evolution under decoherence }
\author[1]{Tomoya Hayata\thanks{hayata@keio.jp}}
\author[2,3,4]{Yoshimasa Hidaka\thanks{hidaka@post.kek.jp}}
\author[5]{Yuta Kikuchi\thanks{ykikuchi@bnl.gov}}

\affil[1]{\it Departments of Physics, Keio University, 4-1-1 Hiyoshi, Kanagawa 223-8521, Japan}
\affil[2]{\it KEK Theory Center, Tsukuba 305-0801, Japan}
\affil[3]{\it Graduate University for Advanced Studies (Sokendai), Tsukuba 305-0801, Japan}
\affil[4]{\it RIKEN iTHEMS, RIKEN, Wako 351-0198, Japan}
\affil[5]{\it Department of Physics, Brookhaven National Laboratory, Upton, New York, 11973-5000}

\date{}
\maketitle

\abstract
We apply a quantum teleportation protocol based on the Hayden-Preskill thought experiment to quantify how scrambling a given quantum evolution is.
It has an advantage over the direct measurement of out-of-time ordered correlators when used to diagnose the information scrambling in the presence of decoherence effects stemming from a noisy quantum device. We demonstrate the protocol by applying it to two physical systems: Ising spin chain and SU(2) lattice Yang-Mills theory.
To this end, we numerically simulate the time evolution of the two theories in the Hamiltonian formalism. The lattice Yang-Mills theory is implemented with a suitable truncation of Hilbert space on the basis of the Kogut-Susskind formalism. On a two-leg ladder geometry and with the lowest nontrivial spin representations, it can be mapped to a spin chain, which we call {\it Yang-Mills-Ising model} and is also directly applicable to future digital quantum simulations. We find that the Yang-Mills-Ising model 
shows the signal of information scrambling at late times.

\maketitle

\section{Introduction}

Information scrambling refers to the spread of quantum information due to complex dynamics in a quantum system. The original information tossed into such a scrambling system is not accessible via a local measurement, and hence, the quantum information is hidden from a local observer.
Quantum information theoretic aspects of the scrambling dynamics were initially studied in the context of black hole physics, where the black hole dynamics is modeled by a Haar random unitary evolution to reflect a nature of black hole as a fast scrambler~\cite{Hayden:2007cs,Sekino:2008he,Shenker:2013pqa,Hosur:2015ylk} (see Refs.~\cite{Devetak2004,Abeyesinghe:2006,Hayden:2007} for preceding works in the context of quantum information theory).
Since then, the information scrambling has attracted huge attention in a wide range of disciplines: quantum gravity, quantum many-body physics, quantum information, and many others. Along the progress in understanding complex quantum dynamics, various diagnostic tools of scrambling have been proposed and scrutinized. Among others, out-of-time ordered correlators (OTOCs) provide a way to quantify the {\it scrambling-ness} by capturing  operator growth due to the associated quantum dynamics~\cite{larkin1969quasiclassical,kitaev2015,Shenker:2013pqa,Roberts:2014isa}.

The quantum teleportation protocol allows a sender to transfer a quantum state to a distant receiver with the use of classical communications by exploiting an entangled pair of states  shared between the sender and receiver~\cite{Bennett1993,nielsen2002quantum}.
Inspired by the seminal paper~\cite{Hayden:2007cs}, a distinct type of quantum teleportation protocol, Hayden-Preskill protocol, has been proposed~\cite{Yoshida:2017non,Yoshida:2018vly} (see Refs.~\cite{Maldacena_2017,Gao_2017,Bao_2018,Gao:2019nyj,Brown:2019hmk,Nezami:2021yaq,Schuster:2021uvg} for other closely related protocols). 
While the information thrown into a black hole whose dynamics is modeled by a Haar random unitary evolution cannot be retrieved by a local measurement, Hayden and Preskill pointed out that the local measurement combined with the Hawking radiation emitted before the information is thrown into a black hole enables the retrieval of the input information~\cite{Hayden:2007cs}.
The Hawking radiation collected earlier may be interpreted as a part of entangled states shared between the black hole and local observer from the viewpoint of quantum teleportation protocol.
Later, an explicit protocol to reconstruct the input information from the local measurement was given in Ref.~\cite{Yoshida:2017non}.
This Hayden-Preskill protocol can be interpreted as a teleportation of input information to a distant local observer by leveraging the initially shared entanglement.
Furthermore, it turns out that the Hayden-Preskill protocol works in other class of quantum systems as long as the dynamics is scrambling.
Hence, the success of teleportation in the protocol depends not only on whether entanglement is initially shared between the involved parties but also crucially on how scrambling the quantum evolution is.
Recently, the interpretation of the Hayden-Preskill protocol mentioned above has led to a new diagnostic tool of information scrambling~\cite{Yoshida:2017non,Landsman:2018jpm}.
Namely, the quality of teleportation with the Hayden-Preskill protocol serves as a proxy of information scrambling.
In the absence of noises and errors, the diagnosis of scrambling by the Hayden-Preskill protocol extracts the same information as one by the averaged OTOC~\eqref{eq:avOTOC}. However, the protocol has an advantage over the direct measurement of the averaged OTOC when certain noises and errors exist~\cite{Yoshida:2018vly,Landsman:2018jpm}.
Relatedly, the verification of information scrambling in the system coupled to external systems has been studied in for example~\cite{Bao_2018,Agarwal:2019gjk,Zhang2019,Piroli:2020dlx,Bao:2020zdo,Touil2021}. It is further linked to a new type of phase transition, so called the entanglement phase transition~\cite{Li:2018zeno,Li2019:hybrid,Skinner:2019,Vasseur:2019,Zabalo:2020,Bao:2020phase,Choi:2020qec,Gullans:2020scalable}.

In the context of nuclear and high-energy physics, the scrambling dynamics of nonabelian gauge theories play a key role in understanding the thermalization of quark-gluon plasma produced in the relativistic heavy-ion collision experiments since fast scrambling is closely tied to rapid thermalization~\cite{Murthy:2019fgs}.
In the theoretical studies of the heavy-ion collision experiments, rapid thermalization is predicted, for instance, by gauge-gravity duality and hydrodynamic models~\cite{Horowitz:1999jd,Chesler:2009cy,Heller:2011ju,Heller:2012km,Baier:2000sb,Baier:2002bt} (see Refs.~\cite{Busza:2018rrf,Berges:2020fwq} for reviews), while fast scrambling dynamics is predicted only by gauge-gravity duality~\cite{Sekino:2008he,Shenker:2013pqa,Roberts:2014isa} and no direct analysis of nonabelian gauge theories can handle their scrambling properties  (as well as rapid thermalization) so far. Conventionally, such rapid thermalization is attributed to the strong coupling nature of the underlying nonabelian gauge theory, that is, quantum chromodynamics, in a relatively large system. It is, however, observed that even small systems exhibit a hydrodynamic behavior and has led to a significant controversy~\cite{ALICE:2014dwt,CMS:2017kcs,ATLAS:2017hap} (see also Ref.~\cite{Berges:2020fwq}). Hence it is of great interest to carry out quantum simulation of the real-time dynamics of strongly-coupled gauge theory in a small system and study its scrambling properties, which is potentially performed on a near term quantum hardware while it is still hard task for classical computers.

In the present work, we demonstrate the usefulness of the Hayden-Preskill protocol for the purpose of verifying the information scrambling by applying it to two physical systems: Ising spin chain and SU(2) lattice Yang-Mills theory, in the presence of decoherence effects.
The major difficulty of measuring OTOCs as a proxy of scrambling on a noisy quantum device lies in the indistinguishability of decays caused by information scrambling and noise effects, while the Hayden-Preskill protocol does not suffer from the same issue and thus is resilient against decoherence effects as we will discuss in detail.
Our noisy simulation indeed shows that the protocol is better suited to extract the signal of information scrambling than measuring averaged OTOC. 
We formulate the protocol in a way suitable for digital quantum simulation and simulated on a Qiskit classical simulator provided by IBM~\cite{Qiskit}, foreseeing its implementation on a noisy quantum device in the Noisy Intermediate-Scale Quantum era~\cite{Preskill2018quantumcomputing}.

Another important ingredient in this paper is the implementation of SU(2) lattice Yang-Mills (YM) theory in the Hamiltonian formalism given a limited computational resource to represent its Hilbert space. In particular, gauge theories tend to require a large amount of memories to represent their states primarily because (i) it has unphysical states associated with gauge redundancy, and (ii) each bosonic state is represented by infinite-dimensional Hilbert space.
Thus, finding efficient implementations of gauge theories to circumvent the issues has been a target of intensive studies; see Refs.~\cite{Zohar_2015,Banuls:2019bmf,Davoudi:2020yln} for review of recent progress. 
We employ the Kogut-Susskind Hamiltonian formalism~\cite{Kogut:1974ag} and efficient representation of the Hilbert space by eliminating gauge redundant states and truncating infinite-dimensional Hilbert space following~\cite{Hayata:2020xxm}.
We call the resultant theory {\it Yang-Mills-Ising (YM-Ising) model} throughout the present paper. It should be emphasized that the employed truncation reduces the size of Hilbert space with the exact SU(2) gauge symmetry retained.
Upon application of the Hayden-Preskill protocol to the YM-Ising model, we find the saturation of the teleportation quality at the value close to that of Haar random evolution at late times, which implies the dynamics scrambles the information of input state. Furthermore, the simulation in the presence of decoherence still shows the qualitatively same behavior. Our results of noisy simulations open up a possibility to diagnose the signal of scrambling in non-abelian gauge theories on a near-term quantum device.

The organization of the rest of this work goes as follows. In Sec.~\ref{sec:Hayden-Preskill protocol}, we will review the Hayden-Preskill protocol, and its relation to the information scrambling and the averaged OTOC. The contents are presented in a self-contained manner so as to be accessible for the readers without the relevant quantum information background. In Sec.~\ref{sec:Diagnosis of scrambling in Ising spin chain}, we will simulate the Hayden-Preskill protocol with the Hamiltonian evolution of the Ising spin chain. The noise effects will be examined by employing the depolarizing channel to model the decoherence. In Sec.~\ref{sec:Diagnosis of scrambling in Yang-Mills-Ising model}, we will apply the Hayden-Preskill protocol to the YM-Ising model in the absence/presence of decoherence effects. Finally, we will conclude with a discussion of our results and their implications in Sec.~\ref{sec:Summary and discussion}.
We summarize the technical details in appendices: Some computations involving Haar random unitary evolutions are presented in Appendix~\ref{app:HP_Haar}. A state-teleportation protocol based on the Hayden-Preskill protocol is explained in Appendix~\ref{app:stateTel}. Explicit implementations of the Hamiltonian evolution operators with elementary quantum gates are given in Appendix~\ref{app:ciruits}. The Hamiltonian formulation of SU(2) lattice Yang-Mills theory is described in detail in Appendix~\ref{app:truncatedYM}.
Throughout this paper, the base of logarithmic function is taken to be 2.

\section{Hayden-Preskill protocol}\label{sec:Hayden-Preskill protocol}

Given a quantum evolution $U_{AB}:\calH_A\otimes\calH_B\to \calH_C\otimes\calH_D$,
we consider a state representation of $U_{AB}$ by augmenting reference states on $\calH_R$ and $\calH_{B'}$~\cite{Hayden:2007cs}:
\begin{align}
\label{eq:HPstate}
    \ket{\Psi} = (I_{R}\otimes U_{AB}\otimes I_{B'})(\ket{\text{EPR}}_{RA}\otimes \ket{\text{EPR}}_{BB'})
    =\figbox{0.4}{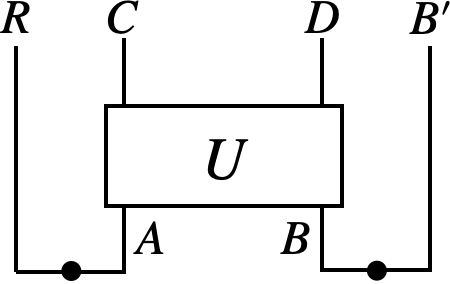},
\end{align}
where the time flows from bottom to top in the quantum circuit diagram. Dimensions of Hilbert spaces satisfy $d_Ad_B=d_Cd_D\eqqcolon d$, $d_B=d_{B'}$ and $d_A=d_{R}$, with $d_A\coloneqq\text{dim}\,\calH_A$ and so on. 
The EPR states $\ket{\text{EPR}}_{RA}$ and $\ket{\text{EPR}}_{BB'}$ have been introduced:
\begin{align}
\begin{split}
    \ket{\text{EPR}}_{RA} &= \frac{1}{\sqrt{d_A}}\sum_{i=1}^{d_A} \ket{i}_R\otimes\ket{i}_A
    = \figbox{0.4}{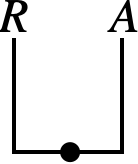},
    \\
    \ket{\text{EPR}}_{BB'} &= \frac{1}{\sqrt{d_B}}\sum_{i=1}^{d_B} \ket{i}_B\otimes\ket{i}_{B'}
    = \figbox{0.4}{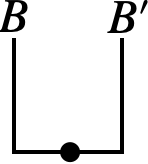},
\end{split}
\end{align}
where the black blobs stand for the normalization factors $1/\sqrt{d_A}$ and $1/\sqrt{d_B}$, respectively.
The Hayden-Preskill thought experiment asks whether an input quantum state on $\calH_A$ can be retrieved from the quantum information available on $\calH_{B'}$ and $\calH_D$.
It was argued that, provided the quantum evolution $U_{AB}$ is a Haar random unitary operator, the retrieval is possible if $N_D\ge N_A +\epsilon$ for $N_A\coloneqq\log d_A$, $N_D\coloneqq\log d_D$, and some small positive number $\epsilon$~\cite{Hayden:2007cs}.
In this section, we review the recovery protocol and its relation to the information scrambling.
In this work, we do not set $U_{AB}$ to be a Haar random unitary operator but let it be a general unitary operator unless otherwise stated.

\subsection{Mutual information as an indicator of information scrambling}

Informally speaking, the information scrambling caused by $U_{AB}$ refers to the spread of input information on $\calH_A$ over the output Hilbert space $\calH_C\otimes\calH_D$.
It is characterized by the information about $A$ nonlocally shared by the outputs $C$ and $D$ which cannot be extracted from either $C$ or $D$ exclusively.
Given the state $\ket{\Psi}$~\eqref{eq:HPstate}, such a quantity may be represented by the negativity of the tripartite information $I(R,C,D)$~\cite{Hosur:2015ylk},
\begin{align}
 -I(R,C,D) \coloneqq I(R,CD) - I(R,C) - I(R,D).
\end{align}
Here, $I(R,C)\coloneqq S(R)+S(C)-S(RC)$ is the mutual information. $S(R)$, $S(C)$, and $S(RC)$ are the von Neumann entropies of associated reduced density operators. For instance, $S(R)\coloneqq-\Tr[\rho_R\log(\rho_R)]$ with the reduced density operator $\rho_R\coloneqq\Tr_{B'CD}[\ket{\Psi}\bra{\Psi}]$.


In the context of the Hayden-Preskill thought experiment, we are interested in the mutual information $I(R,B'D)$ of $\ket{\Psi}$ under the assumptions $d_A\ll d_B$ and $d_D\ll d_C$. To this end, we examine $I(R,B',D)=I(R,C,D)$. We say that $U_{AB}$ is strongly scrambling when the tripartite information is nearly minimum, $I(R,C,D)\approx -2N_A$ with $N_A=\log d_A$. Then, we find
\begin{align}
\label{eq:I(RBD)}
-2N_A \approx I(R,B',D) = I(R,B') + I(R,D) - I(R,B'D) \approx - I(R,B'D),
\end{align}
where we have used $I(R,B')=0$ and $I(R,D)\approx0$.\footnote{
This may be understood from the argument that a local observer on $D$ cannot retrieve the input information on $A$ after the scrambling dynamics $U_{AB}$.
Indeed, provided $U_{AB}$ is a Haar random unitary, one can show that $I(R,D)=N_A+N_D-S(RD) \le N_A+N_D-S^{(2)}(RD)= I^{(2)}(R,D)\approx (d_Ad_D)/(d_Bd_C)$ with $S^{(2)}(RD)=-\log\Tr[\rho_{RD}^2]$. We here used $\Tr[\rho_{RD}^2]\approx (d_Ad_D)^{-1}+(d_Bd_C)^{-1}$. See Appendix~\ref{app:HP_Haar} for details.
} Thus, $I(R,B'D)\approx 2N_A$ is nearly maximal.

The mutual information decreases if there are noise effects stemming from the interactions with the environment~\cite{Esposito2010,Touil2021}. Such interactions create entanglement between the principle system and environment, which in turn reduces the entanglement within the principle system, and hence, the information shared between $A$ and $BD$, $I(R,B'D)$, decreases.

In what follows, we employ the R\'enyi-2 mutual information 
\begin{align}
    I^{(2)}(R,B'D) \coloneqq S^{(2)}(R)+S^{(2)}(B'D)-S^{(2)}(RB'D),
\end{align}
with $S^{(2)}(R)\coloneqq-\log\Tr[\rho_R^2]$.
Although it is different from the mutual information $I(R,B'D)$, by noting $S(\rho) \ge S^{(2)}(\rho)$ for an arbitrary density operator $\rho$,\footnote{More generally, $S^{(n)}(\rho) \ge S^{(n+1)}(\rho)$ holds for $S^{(n)}(\rho)\coloneqq\frac{1}{1-n}\log\Tr[\rho^n]$ and $n\ge1$. Note that $S^{(1)}(\rho)\coloneqq\lim_{n\to1}S^{(n)}(\rho)=S(\rho)$.
} we find that $I^{(2)}(R,B'D)$ gives a lower bound on $I(R,B'D)$:
\begin{align}
    I^{(2)}(R,B'D) = \log d_A +S^{(2)}(B'D)-\log d_C
    \le \log d_A +S(B'D)-\log d_C
    =I(R,B'D).
\end{align}
Therefore, the large $I^{(2)}(R,B'D)$ guarantees the large $I(R,B'D)$, and thus the small $I(R,C,D)$. The Hayden-Peskill protocol measures $I^{(2)}(R,B'D)$ via the quality of teleportation.

\subsection{Ideal protocol}
\label{sec:ideal}

The Hayden-Peskill thought experiment implies that there exists a quantum channel $V_{DB'}$ such that the input state on $A$ can be transferred to $R'$ if the quantum channel $U_{AB}$ is given by a scrambling dynamics and $N_D\ge N_A+\epsilon$ for a small number $\epsilon$~\cite{Hayden:2007cs},
\begin{align}
\label{eq:Vpsi}
    V_{DB'}\ket{\Psi}
    =\figbox{0.4}{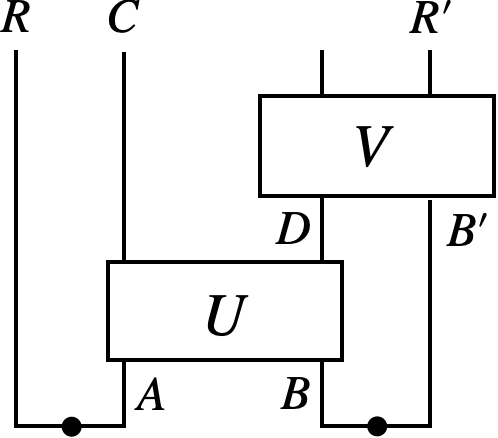}\,.
\end{align}
An analogous protocol to transmit an arbitrary state on $A$ to $R'$ is explained in Appendix~\ref{app:stateTel}.
Yoshida and Kitaev found such a quantum channel $V_{DB'}$~\cite{Yoshida:2017non} that consists of the following three operations (Hayden-Preskill protocol):
\begin{enumerate}
    \item Extend the Hilbert space by attaching $\calH_{A'}\otimes\calH_{R'}$ and prepare an EPR state, $\ket{\text{EPR}}_{A'R'}$.
    \item Apply a  backward evolution $U^\ast_{A'B'}$ on $\calH_{A'}\otimes\calH_{B'}$ to find a state $\ket{\tilde{\Phi}}$.
    \item Project the state $\ket{\tilde{\Phi}}$ onto $\ket{\text{EPR}}_{DD'}$.
\end{enumerate}
Then, the resultant quantum state $\ket{\Phi}$ is given by
\begin{align}
\begin{split}
    \ket{\tilde{\Phi}} 
    &= (U_{AB}\otimes U^\ast_{A'B'})(\ket{\text{EPR}}_{RA}\otimes \ket{\text{EPR}}_{BB'})\otimes \ket{\text{EPR}}_{A'R'}),
    \\
    \ket{\Phi} 
    &= \frac{\Pi_{DD'} \ket{\tilde{\Phi}}}{\sqrt{P_\text{EPR}}}
    =\frac{1}{\sqrt{P_\text{EPR}}}\figbox{0.4}{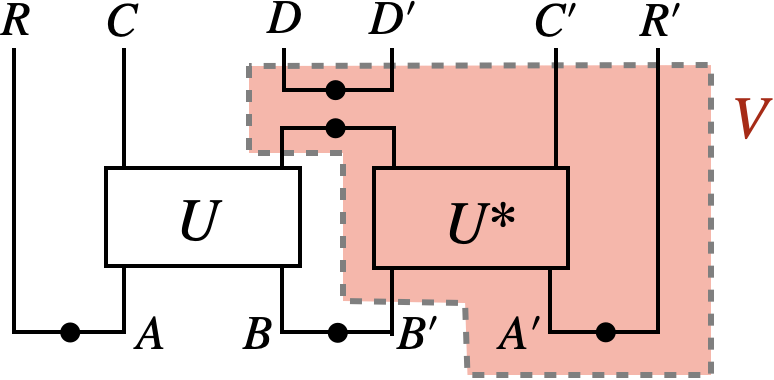}\,,
\end{split}
\end{align}
where $\Pi_{DD'}\coloneqq\ket{\text{EPR}}\bra{\text{EPR}}_{DD'}$ is the projection operator onto $\ket{\text{EPR}}_{DD'}$. The circuit shown in the red region corresponds to $V$ in~Eq.~\eqref{eq:Vpsi}.
The normalization factor $P_\text{EPR}$ is given by
\begin{align}
\label{eq:ideal_Pepr}
    P_\text{EPR} = \Tr[\Pi_{DD'} \ket{\tilde{\Phi}}\bra{\tilde{\Phi}}]
    =\frac{1}{d_A^2d_Bd_D}\figbox{0.4}{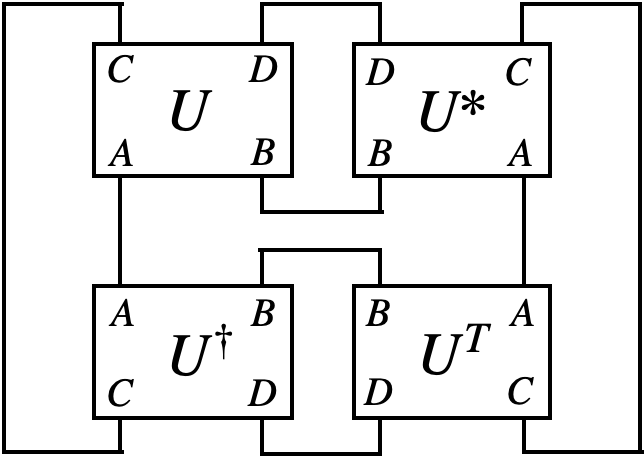}\, ,
\end{align}
which is the probability of the state $\ket{\tilde{\Phi}}$ being successfully projected onto $\ket{\text{EPR}}_{DD'}$.
Then, the fidelity, $F_\text{EPR}$, between $\ket{\Phi}$ and $\ket{\text{EPR}}_{RR'}$ quantifies the teleportation quality:
\begin{align}
\label{eq:ideal_Fepr}
    F_\text{EPR} =\Tr[\Pi_{RR'}\ket{\Phi}\bra{\Phi}]
    =\frac{1}{d_A^3d_Bd_DP_\text{EPR}}\figbox{0.4}{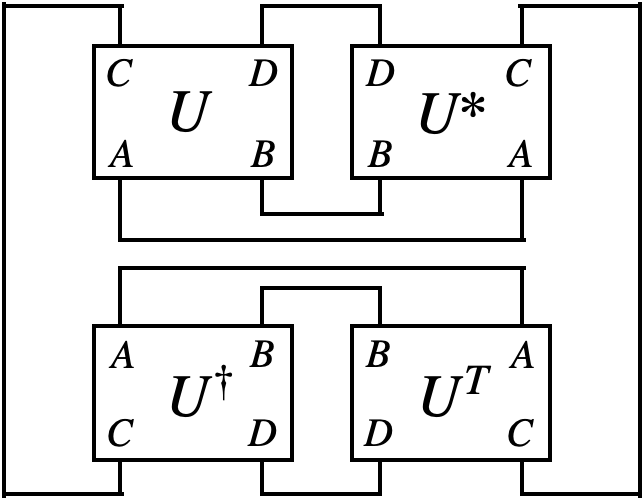}
    =\frac{1}{d_A^2P_\text{EPR}}.
\end{align}

\vspace{1em}

As discussed in Ref.~\cite{Yoshida:2017non}, the quantity $P_\text{EPR}$ is identified with an averaged OTOC,
\begin{align}
\label{eq:avOTOC}
    \overline{\langle\text{OTOC}\rangle}
    \coloneqq \int_\text{Haar}\diff O_A\diff O_D\Tr[O_AO_D(t)O_A^\dag O_D^\dag(t)].
\end{align}
Hence, the teleportation quality $F_\text{EPR}$ carries the information of the OTOC in the absence of noise effects:
\begin{align}
    F_\text{EPR} 
    =\frac{1}{d_A^2\overline{\langle\text{OTOC}\rangle}}.
\end{align}
The strong decay of averaged OTOC results in the high quality of teleportation in the absence of decoherence.

For the later purpose, we briefly discuss the results of $P_\text{EPR}$ and $F_\text{EPR}$ when the evolution operator $U$ is given by a Haar random evolution. In this case, the corresponding dynamics efficiently scrambles the information of input states. 
The projection probability $P_\text{EPR}$~\eqref{eq:ideal_Pepr} is reduced to (see Appendix~\ref{app:HP_Haar}):
\begin{align}
\label{eq:ideal_Pepr_Haar}
   P_\text{EPR}^\text{Haar} =\frac{1}{d^2-1}\left[d_B^2 + d_C^2
 -\frac{d_C^2}{d_A^2}-1\right]\approx \frac{1}{d_A^2}+\frac{1}{d_D^2}-\frac{1}{d_A^2d_D^2}.
\end{align}
The associated fidelity~\eqref{eq:ideal_Fepr} takes the form,
\begin{align}
\label{eq:ideal_Fepr_Haar}
   F_\text{EPR}^\text{Haar} =\frac{1}{d_A^2P_\text{EPR}^\text{Haar}}
   \approx \frac{1}{1+(d_A/d_D)^2}.
\end{align}
Therefore, the teleportation works almost perfectly when $d_D\gg d_A$ is satisfied, as implied by $N_D\ge N_A +\epsilon$.

\subsection{Noisy protocol}

OTOCs serve as proxies of scrambling or quantum chaos of a given quantum system. In particular, their decays imply that the corresponding quantum dynamics is scrambling.
It is, however, very challenging to measure the decay of OTOC stemming from the scrambling on a quantum computer when the noise effects of the quantum device are not negligible because such effects also induce the decay of OTOC.

On the other hand, $F_\text{EPR}$ increases when the system is scrambling while the noise effects likely to induce the decay of $F_\text{EPR}$. In particular, when quantum channels $U_{AB}$ and $U^\ast_{A'B'}$ are subject to the decoherence effect modeled by the depolarizing channel~\eqref{eq:depol_channel}, one can show~\cite{Yoshida:2018vly},
\begin{align}
\label{eq:noisy_Fepr}
    F_\text{EPR}^\text{decoh} = \frac{\delta}{d_A^2P_\text{EPR}^\text{decoh}}
    = \frac{2^{I^{(2)}(R,B'D)}}{d_A^2},
\end{align}
where the noise effect is characterized by $0\le\delta\le1$, satisfying $\delta = 2^{I^{(2)}(R,B'D)}P^\text{decoh}_\text{EPR}$. More importantly, $F^\text{decoh}_\text{EPR}$ is directly related to the R\'enyi-2 mutual information $I^{(2)}(R,B'D)$.
Therefore, it is easier to quantify the scrambling by using $F^\text{decoh}_\text{EPR}$ than directly measuring OTOCs.

We spell out the derivation of Eq.~\eqref{eq:noisy_Fepr}.
We replace the quantum channel $U$ (and $U^\ast$) with the following depolarizing channel~\cite{nielsen2002quantum},
\begin{align}
\label{eq:depol_channel}
    \calQ[\rho] = (1-p)(U_{AB}\otimes I_{\overline{AB}})\rho (U_{AB}^\dag\otimes I_{\overline{AB}}) + p\frac{I_{CD}}{d_Cd_D}\otimes \Tr_{AB}[\rho],
\end{align}
for an arbitrary density operator $\rho:\calH_A\otimes\calH_B\otimes\calH_{\overline{AB}}\to \calH_A\otimes\calH_B\otimes\calH_{\overline{AB}}$. $\overline{AB}$ stands for the spatial region complementary to $AB$. The error probability $p$ characterizes the strength of decoherence. This channel models the decoherence induced by an application of the Hamiltonian evolution operator $U$. Namely, the information of input state on $AB$ is lost with probability $p$. The channel is graphically represented by
\begin{align}
    \figbox{0.4}{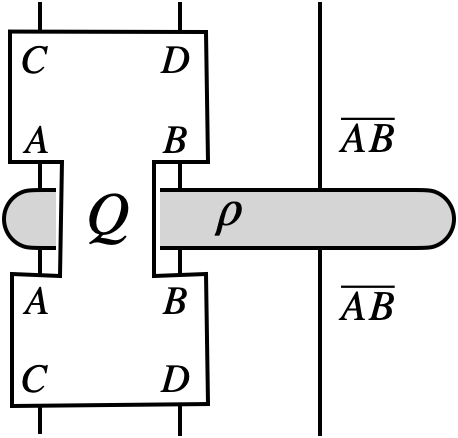}
    \ =\ 
    (1-p)\figbox{0.4}{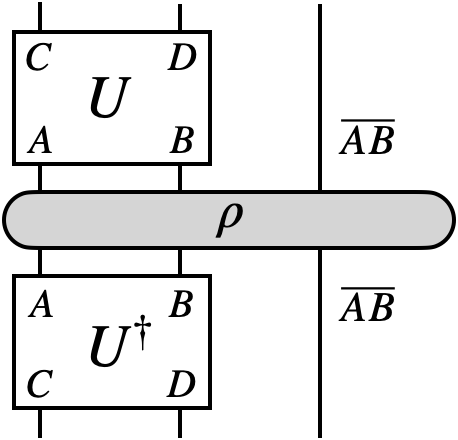}
    \ +\ \frac{p}{d_Cd_D}\figbox{0.4}{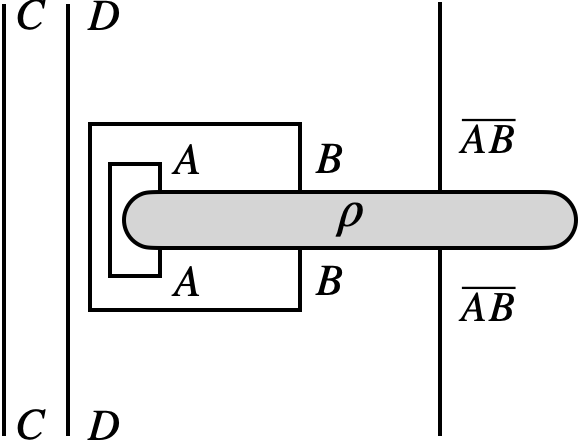}\,.
\end{align}

Replacing $U$ and $U^\ast$ with the corresponding depolarizing channels in Eq.~\eqref{eq:ideal_Pepr}, we find
\begin{align}
\label{eq:noisy_Pepr}
    P_\text{EPR}^\text{decoh}
    = \frac{1}{d_A^2d_Bd_D}\figbox{0.4}{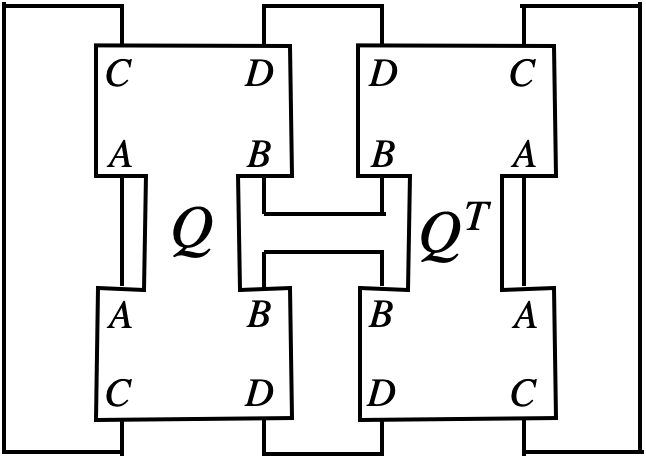}
    = (1-p)^2P_\text{EPR} + (2p-p^2)\frac{1}{d_D^2}.
\end{align}
By noting that $P_\text{EPR}$~\eqref{eq:ideal_Pepr} is lower-bounded by $\text{max}(1/d_A^{2},1/d_D^{2})$~\cite{Yoshida:2018vly}, we observe
\begin{align}
    P_\text{EPR} - P_\text{EPR}^\text{decoh} = (2p-p^2)\left(P_\text{EPR} -\frac{1}{d_D^2}\right)
    \ge(2p-p^2)\left(\text{max}\Big(\frac{1}{d_A^{2}},\frac{1}{d_D^{2}}\Big) -\frac{1}{d_D^2}\right) \ge 0,
\end{align}
{\it i.e.}, the depolarizing channel always suppresses $P_\text{EPR}$, or equivalently, the corresponding averaged OTOC~\eqref{eq:decoh_avOTOC}.
We remark that, even in the presence of decoherence, $P_\text{EPR}^\text{decoh}$ is identified with the averaged OTOC~\cite{Yoshida:2017non},
\begin{align}
\label{eq:decoh_avOTOC}
    P^\text{decoh}_\text{EPR} 
    = \overline{\langle\text{OTOC}\rangle}_\text{decoh}
    = \int_\text{Haar}\diff O_A\diff O_D\Tr[O_A\tilde{O}_D(t)O_A^\dag O_D^\dag(t)],
\end{align}
with
\begin{align}
    \tilde{O}_D(t) = (1-p)U^\dag O_D U + p\frac{I}{d}\Tr{O_D}.
\end{align}

The fidelity $F_\text{EPR}^\text{decoh}$ is computed in the same way:
\begin{align}
    F_\text{EPR}^\text{decoh} 
    = \frac{1}{d_A^3d_Bd_DP_\text{EPR}^\text{decoh}}\figbox{0.4}{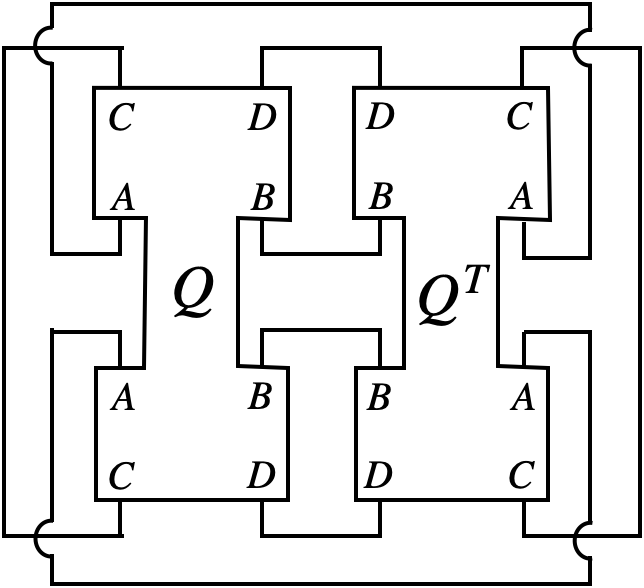}
    = \frac{1}{d_A^2P_\text{EPR}^\text{decoh}}\left[(1-p)^2 + (2p-p^2)\frac{1}{d_D^2}\right].
\end{align}

In order to make a connection to the R\'enyi-2 mutual information $I^{(2)}(R,B'D)$, we compute R\'enyi-2 entropies $S^{(2)}_{B'D}=-\log(\Tr[\rho_{B'D}^2])$ and $S^{(2)}_{RB'D}=-\log(\Tr[\rho_{RB'D}^2])$ associated with the state $\calQ_{AB}(\ket{\text{EPR}}\bra{\text{EPR}}_{RA}\otimes \ket{\text{EPR}}\bra{\text{EPR}}_{BB'})$.
We start with $S^{(2)}_{B'D}$:
\begin{align}
    2^{-S^{(2)}_{B'D}}=\Tr[\rho_{B'D}^2] = \frac{1}{d_A^2d_B^2}\figbox{0.35}{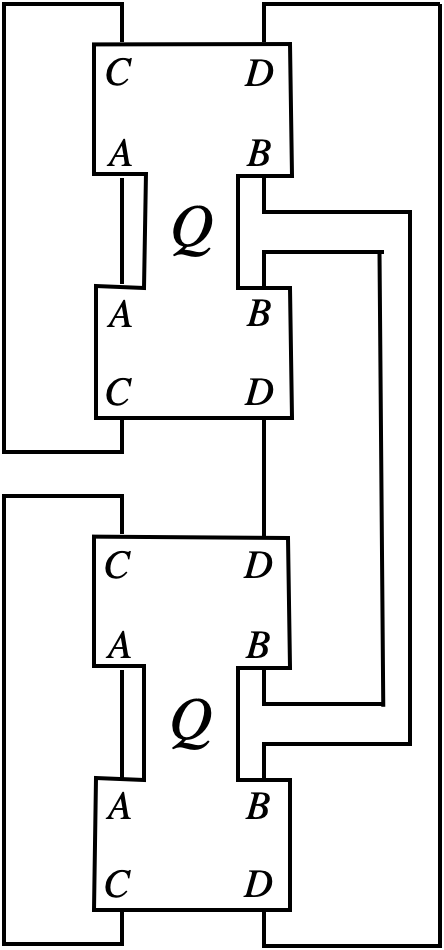}
    =\frac{d_D}{d_B}P_\text{EPR}^\text{decoh},
\end{align}
which leads to
\begin{align}
    P_\text{EPR}^\text{decoh} = \frac{d_C}{d_R}2^{-S^{(2)}_{B'D}} = 2^{S^{(2)}_{C}-S^{(2)}_{R}-S^{(2)}_{B'D}}.
\end{align}
Next, we compute $S^{(2)}_{RB'D}$:
\begin{align}
    2^{-S^{(2)}_{RB'D}}=\Tr[\rho_{RB'D}^2] = \frac{1}{d_A^2d_B^2}\figbox{0.35}{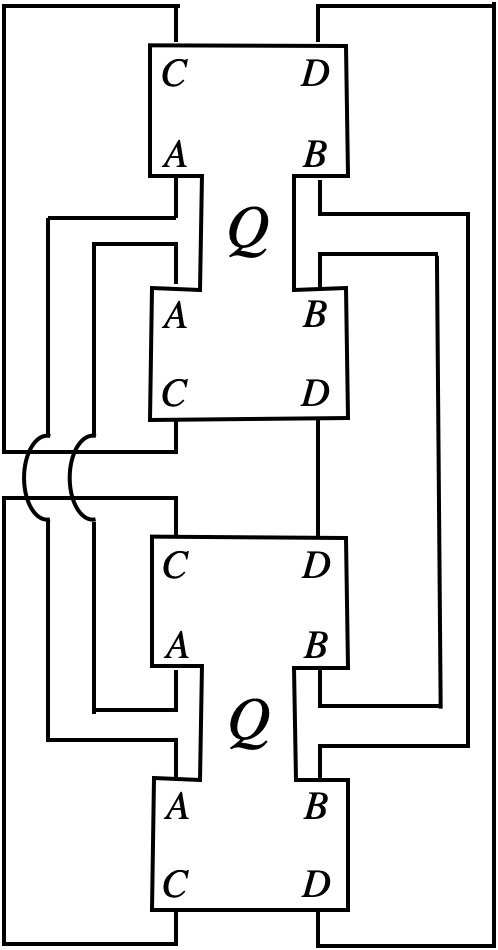}
    = \frac{d_Ad_D}{d_B}P_\text{EPR}^\text{decoh}F_\text{EPR}^\text{decoh}.
\end{align}
Hence, we find Eq.~\eqref{eq:noisy_Fepr},
\begin{align}
    F_\text{EPR}^\text{decoh} = \frac{2^{S^{(2)}_{C}-S^{(2)}_{RB'D}}}{d_A^2P_\text{EPR}^\text{decoh}}
    =\frac{2^{I^{(2)}(R,B'D)}}{d_A^2},
\end{align}
with $I^{(2)}(R,B'D) = S^{(2)}_{R}+S^{(2)}_{B'D}-S^{(2)}_{RB'D}$.

Interestingly, the decay of $P^\text{decoh}_\text{EPR}$ does not lead to the increase in $F^\text{decoh}_\text{EPR}$. Instead, the decoherence induces the decay of $F^\text{decoh}_\text{EPR}$ due to the decrease of $I^{(2)}(R,B'D)$, which is attributed to the leakage of mutual information to environment. Therefore, the effects of scrambling and decoherence induce the distinct consequences in the behavior of $F^\text{decoh}_\text{EPR}$. This is the reason why we favor $F^\text{decoh}_\text{EPR}$ to diagnose the information scrambling over $P^\text{decoh}_\text{EPR}$ or OTOCs.
In the following section, we demonstrate these observations by carrying out numerical simulations of the protocol in the Ising spin chain and the YM-Ising model.

\section{Diagnosis of scrambling in Ising spin chain}\label{sec:Diagnosis of scrambling in Ising spin chain}

\begin{figure}[th]
\centering
\includegraphics[width=.15\textwidth]{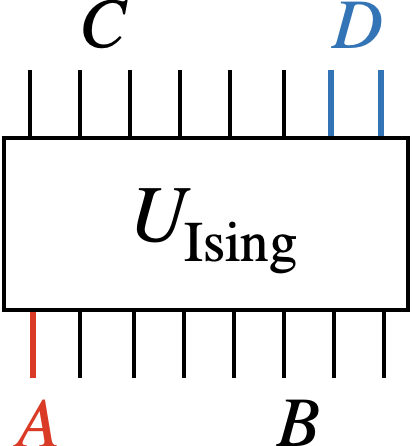}
\caption{The configuration of input and output Hilbert spaces of evolution operator $U_\text{Ising}$. The states on $\calH_A$ and $\calH_D$ are respectively represented by red and blue wires. The rest of wires in input and output wires are the states on $\calH_B$ and $\calH_C$, respectively. 
}
\label{fig:config_Ising}
\end{figure}

We demonstrate the ideas discussed in the previous section by applying the protocol to the Ising spin chain,
\begin{align}\label{eq:Ising}
    H_\text{Ising}
    = -\sum_{i=1}^{N-1}Z_iZ_{i+1}
    -h\sum_{i=1}^{N}X_i
    -m\sum_{i=1}^{N}Z_i,
\end{align}
where $(X_i,Y_i,Z_i)$ are the Pauli matrices $(\sigma_x,\sigma_y,\sigma_z)$ defined on the $i$-th site, and $N$ denotes the number of sites.
We consider the three sets of parameters:
(i) $h=m=0$,
(ii) $h=1$, $m=0$,
(iii) $h=-1.05$, $m=0.5$.
The first system is the classical Ising model, and the second one is the transverse-field Ising model at the critical point. Both of them are integrable systems. The third one is known to be a chaotic system~\cite{Banuls:2011}.
We numerically simulate the Hayden-Preskill protocol with $U$ given by the Hamiltonian evolution $U_\text{Ising}$~\eqref{eq:Ising_ST_formula} to compute $P_\text{EPR}$ ($P^\text{decoh}_\text{EPR}$) and $F_\text{EPR}$ ($F^\text{decoh}_\text{EPR}$) as functions of $t$. The configuration of input and output Hilbert spaces are shown in Fig.~\ref{fig:config_Ising} ($d=2^8$, $d_A=2$, and $d_D=2^2$).

The Hamiltonian evolution operator is implemented by using the Suzuki-Trotter product formula~\cite{Suzuki1991,Lloyd1073},
\begin{align}
\label{eq:Ising_ST_formula}
    U_\text{Ising}
    = \left(\e^{-\im \frac{t}{M}\left(-\sum_{i=1}^{N-1}Z_iZ_{i+1}
    -m\sum_{i=1}^{N}Z_i\right)}
    \e^{-\im \frac{t}{M}\left(-h\sum_{i=1}^{N}X_i\right)}\right)^M
    =e^{-\im H_\text{Ising}t}+\calO(t^2/M).
\end{align}
Their explicit implementations in terms of elementary quantum gates are delegated to Appendix~\ref{app:ciruits}.
For the noisy simulation, we replace all the CNOT gates appearing in the protocol with the depolarizing channels as follows,
\begin{align}
\label{eq:CNOTdepolarize}
    \text{CNOT}\rho \text{CNOT} \to (1-p)\text{CNOT}\rho \text{CNOT} + p\frac{I}{d}\otimes \Tr[\rho].
\end{align}
This replacement is motivated by the fact that multi-qubit entangling gates are the dominant source of errors inducing decoherence in near-term quantum hardware.
Note that it is not identical to the depolarizing channel introduced in the previous section. Nevertheless, it still reflects the essential feature that the Hamiltonian evolution is subject to the decoherence effect.\footnote{CNOT operations also appear in the preparation of initial EPR pairs and EPR measurements. However, the number of CNOTs in the Hamiltonian evolution scales linearly in $t$, and hence, the associated noise effects eventually dominate those caused by CNOTs for creating and measuring EPR pairs.} We remark that $2(N-1)$ CNOT operations are used in each Suzuki-Trotter step of Eq.~\eqref{eq:Ising_ST_formula} ($4(N-1)$ CNOT operations with $U_{AB}$ and $U_{A'B'}^\ast$ combined), and the noise effects accumulate as the Hamiltonian evolution proceeds.
\begin{figure}[h]
\centering
\begin{minipage}{.49\textwidth}
\subfloat[$P_\text{EPR}$]{
\includegraphics[width=.95\textwidth]{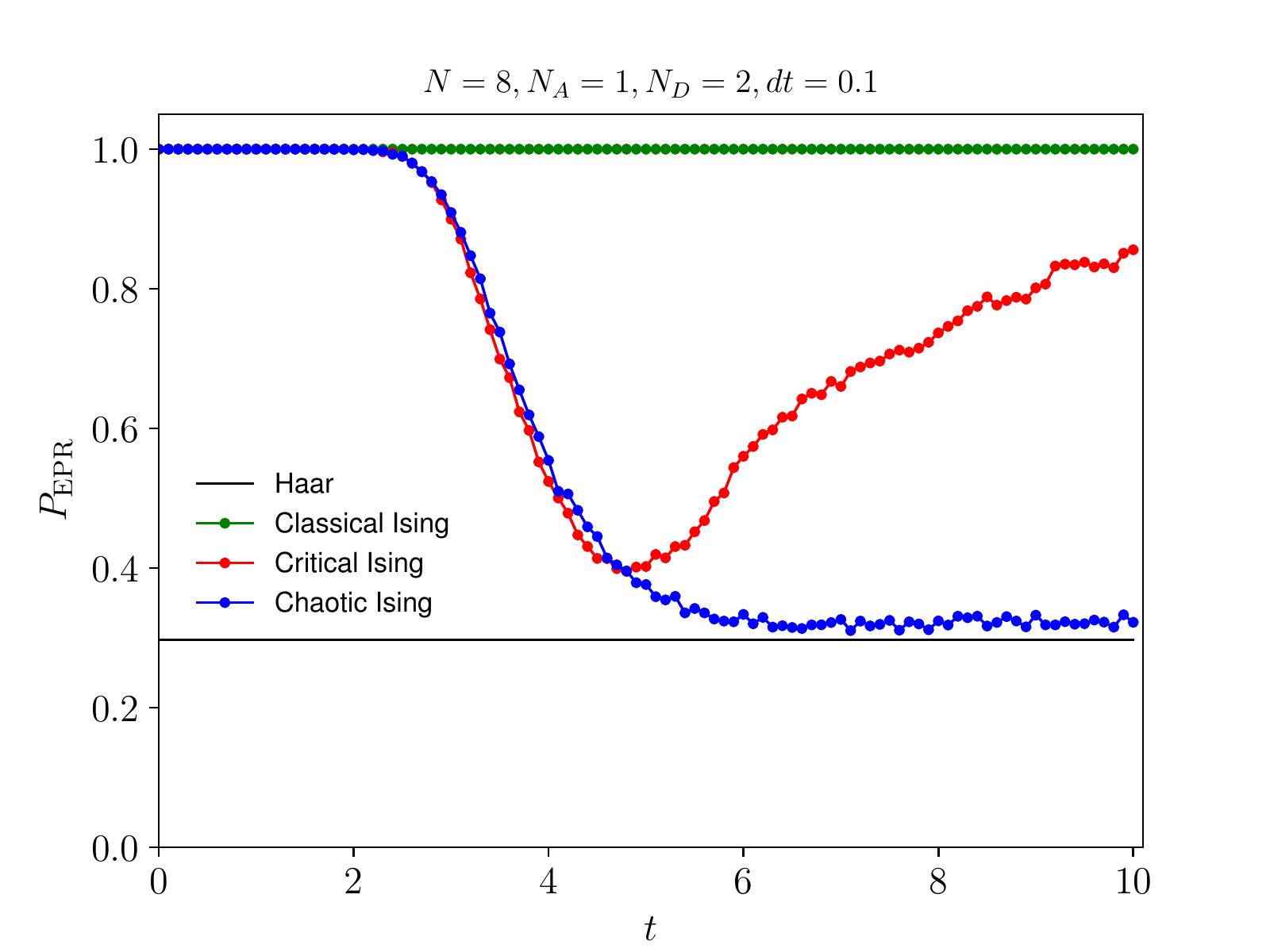}
}\end{minipage}\
\begin{minipage}{.49\textwidth}
\subfloat[$F_\text{EPR}$]{
\includegraphics[width=.95\textwidth]{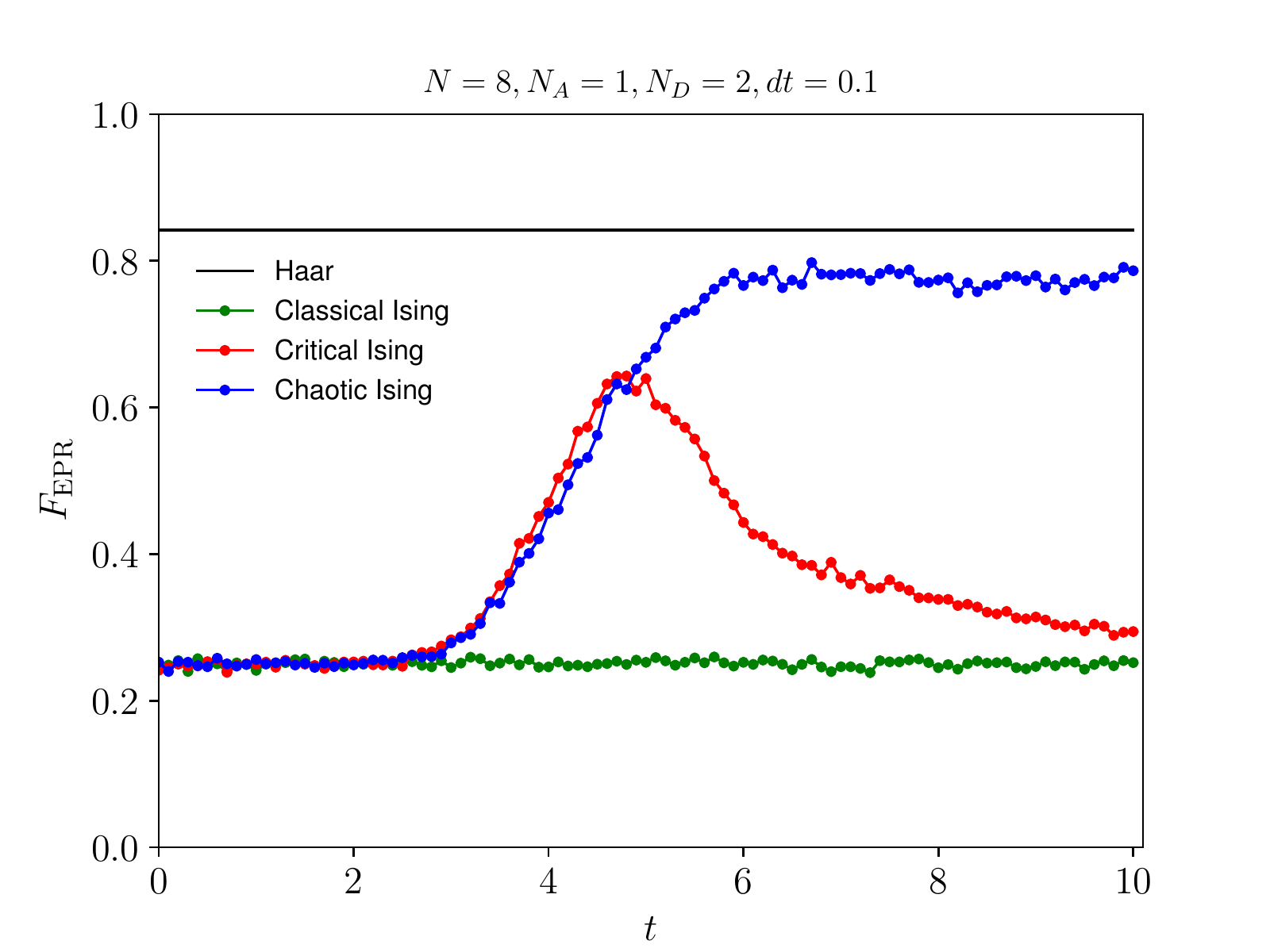}
}\end{minipage}
\caption{The ideal Hayden-Preskill recovery protocol with $U=U_\text{Ising}$. (a) The projection probability~\eqref{eq:ideal_Pepr}. (b) The fidelity~\eqref{eq:ideal_Fepr}. The parameters on each panel are $N\coloneqq\log d$, $N_A\coloneqq\log d_A$, $N_D\coloneqq\log d_D$, and $dt\coloneqq t/M$. The green, red, and blue points correspond to $(h,m)=(0,0)$ [classical], $(1.0,0)$ [critical], and $(-1.05,0.5)$ [chaotic], respectively. The black line represents the $P_\text{EPR}$~\eqref{eq:ideal_Pepr_Haar} and $F_\text{EPR}$~\eqref{eq:ideal_Fepr_Haar} computed with $U$ given by a Haar random unitary operator.}
\label{fig:ideal}
\end{figure}
\begin{figure}[h]
\centering
\begin{minipage}{.49\textwidth}
\subfloat[$P_\text{EPR}^\text{decoh}$]{
\includegraphics[width=.95\textwidth]{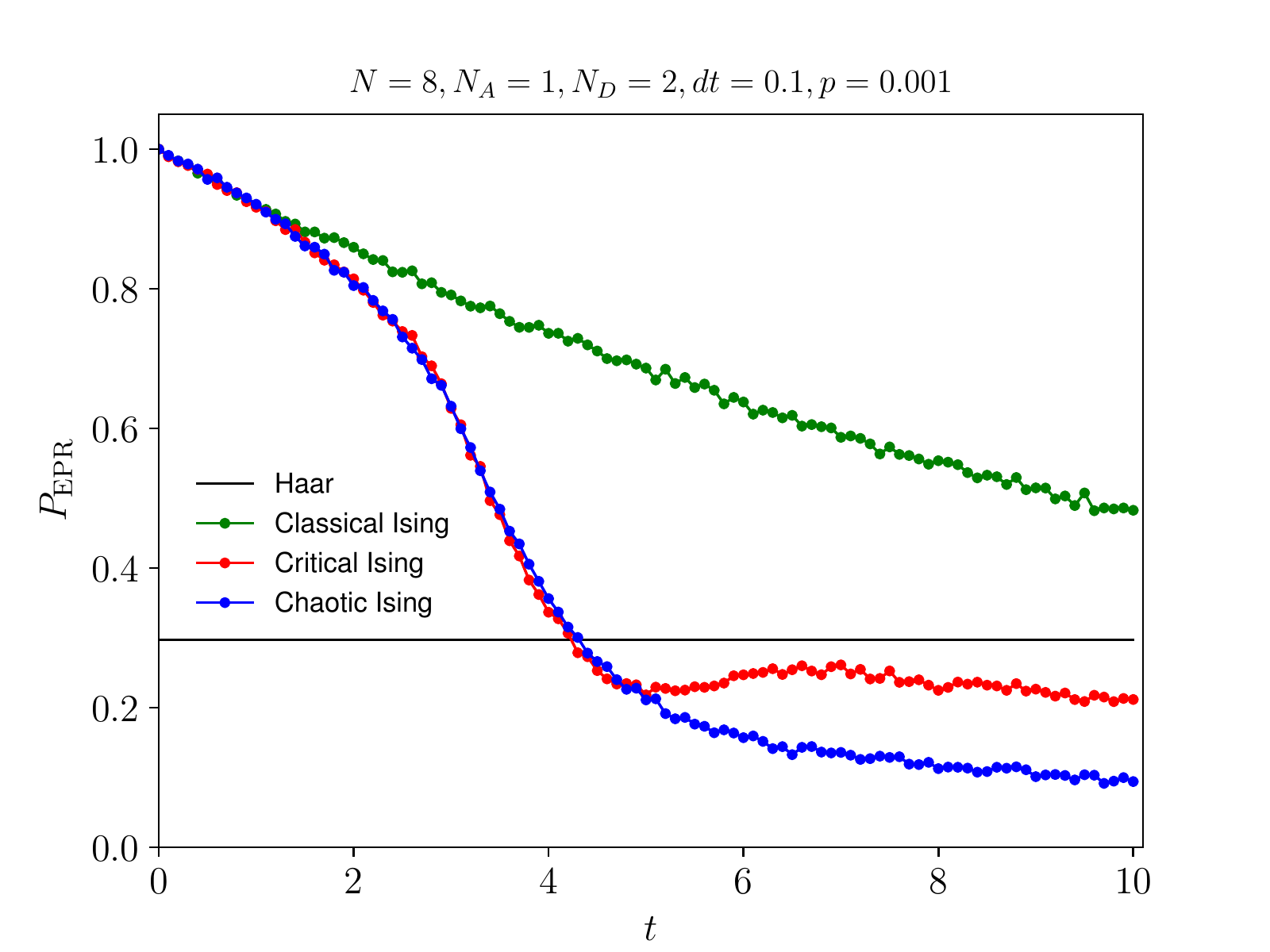}
}\end{minipage}\
\begin{minipage}{.49\textwidth}
\subfloat[$F_\text{EPR}^\text{decoh}$]{
\includegraphics[width=.95\textwidth]{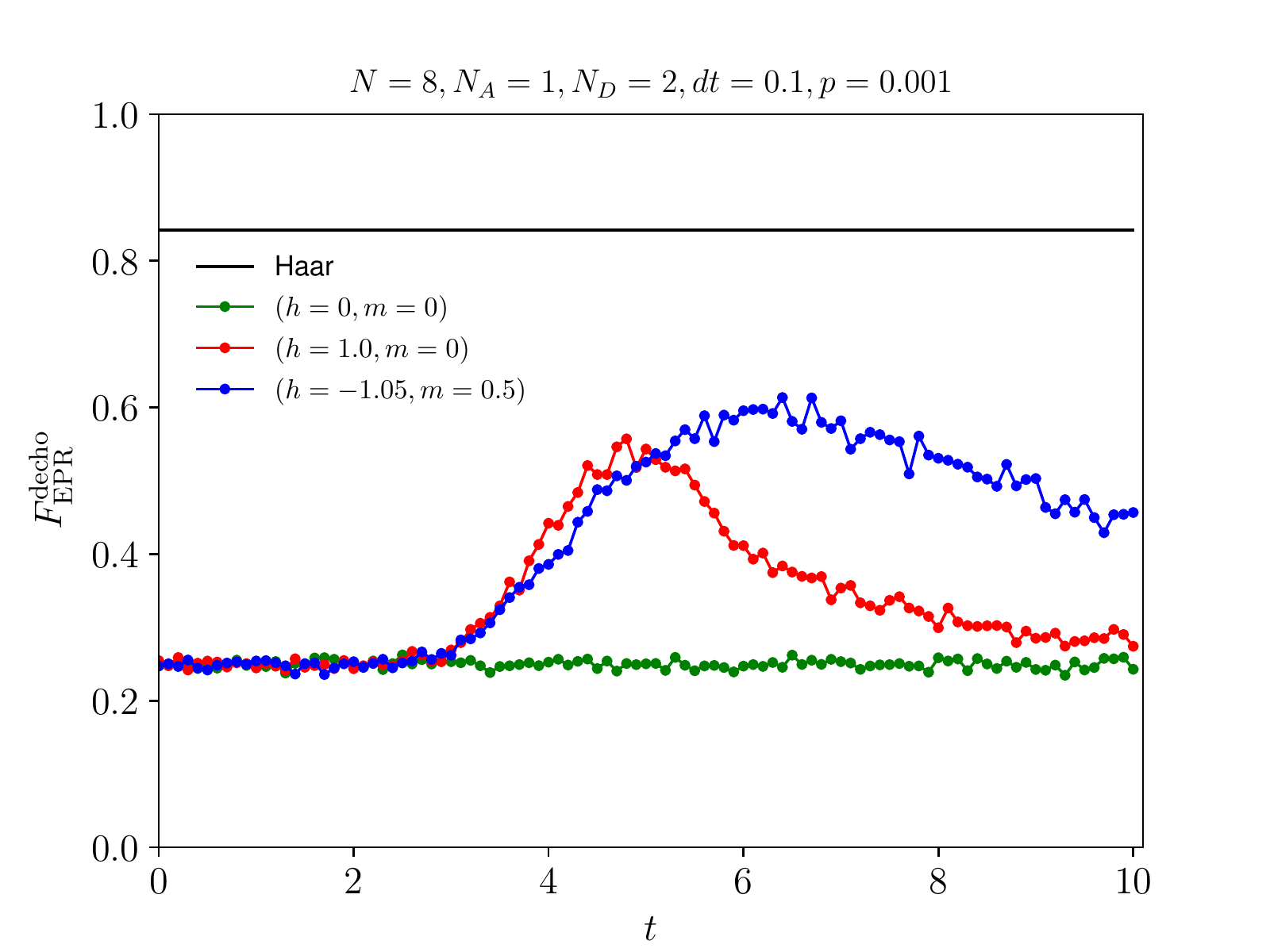}
}\end{minipage}
\caption{The Hayden-Preskill recovery protocol with $U=U_\text{Ising}$ in the presence of decoherence~\eqref{eq:CNOTdepolarize}. (a) The projection probability~\eqref{eq:noisy_Pepr}. (b) The fidelity~\eqref{eq:noisy_Fepr}. The parameters on each panel are $N\coloneqq\log d$, $N_A\coloneqq\log d_A$, $N_D\coloneqq\log d_D$, $dt\coloneqq t/M$, and $p$ is the error probability~\eqref{eq:CNOTdepolarize}. The green, red, and blue points correspond to $(h,m)=(0,0)$ [classical], $(1.0,0)$ [critical], and $(-1.05,0.5)$ [chaotic], respectively. The black line represents the $P_\text{EPR}$~\eqref{eq:ideal_Pepr_Haar} and $F_\text{EPR}$~\eqref{eq:ideal_Fepr_Haar} computed with $U$ given by a Haar random unitary operator without decoherence effects.}
\label{fig:noisy}
\end{figure}

Figure~\ref{fig:ideal} shows the $t$-dependence of $P_\text{EPR}$ and $F_\text{EPR}$ in the absence of noise effects. Remember that the former is equivalent to the averaged OTOC~\eqref{eq:avOTOC}. In the chaotic spin chain, $P_\text{EPR}$ decays to almost saturate and remains at the value of Haar random unitary case. This is in contrast to the other integrable spin chains. The consistent behaviors are observed in the computation of $F_\text{EPR}$. We emphasize that the relation~\eqref{eq:ideal_Fepr} in the ideal case implies that $P_\text{EPR}$ and $F_\text{EPR}$ carry exactly the same information.
Namely, the decay of $P_\text{EPR}$ indicates the growth of $F_\text{EPR}$ and vice versa.
\red{See Appendix~\ref{app:numerial_data} for system-size dependence
of $P_\text{EPR}$/$F_\text{EPR}$.}

Figure~\ref{fig:noisy} shows the computations of $P^\text{decoh}_\text{EPR}$ and $F^\text{decoh}_\text{EPR}$ in the presence of decoherence effects.
Here, we introduce the decoherence by replacing all the CNOT gates with the depolarizing channels~\eqref{eq:CNOTdepolarize}.
In the left panel, the decay of $P^\text{decoh}_\text{EPR}$ is observed even in the integrable spin chains. This computation manifestly demonstrates that it is hard to discriminate the scrambling effects from the noise effects because both of them induce the decrease of $P^\text{decoh}_\text{EPR}$, or equivalently averaged OTOC~\eqref{eq:decoh_avOTOC}.
On the other hand, the effects of decoherence and scrambling are distinguishable when $F^\text{decoh}_\text{EPR}$ is used as shown in the right panel. The scrambling raises $F^\text{decoh}_\text{EPR}$ while the decoherence suppresses it. This is because the decoherence is caused by the entanglement between the principle system and environment, which reduces the entanglement, and hence the mutual information, between $R$ and $B'D$.

\section{Diagnosis of scrambling in Yang-Mills-Ising model}\label{sec:Diagnosis of scrambling in Yang-Mills-Ising model}

Having seen the application of the Hayden-Preskill protocol to the transverse Ising model, we turn to the problem of more physical interest, SU(2) Yang-Mills theory.
We give a brief recap of the Hamiltonian formulation of the lattice SU($2$) Yang-Mills (YM) theory based on the Kogut-Susskind formulation~\cite{Kogut:1974ag}.

The link variable $U_\mu(\bm x)$, a $2\times2$ matrix-valued operator, is defined on the link from a site $\bm x$ to $\bm x+e_{\mu}$, 
where $e_{\mu}$ is the unit vector along the $\mu=x,y$ direction.
We introduce the generators of left and right transformation, $E^{a}_L(\bm x,\mu)$, and $E^a_R(\bm x,\mu)$ [$a=x,y,z$],
which satisfy
\begin{align}
  [E^{a}_L(\bm x,\mu),U_\nu(\bm y)] &=-T^aU_\mu(\bm x)\delta_{\mu\nu}\delta_{\bm x,\bm y} , \label{eq:[E_L,U]}
  \\
  [E^{a}_R(\bm x,\mu),U_\nu(\bm y)] &=U_\mu(\bm x) T^a\delta_{\mu\nu}\delta_{\bm x,\bm y},\\
[E^{a}_{L}(\bm x,\mu),E^{b}_{L}(\bm x,\nu)] &=\ri\epsilon^{abc}E^{c}_{L}(\bm x,\mu)\delta_{\mu\nu}\delta_{\bm x,\bm y},     \\
[E^{a}_{R}(\bm x,\mu),E^{b}_{R}(\bm x,\nu)] &=\ri\epsilon^{abc}E^{c}_{R}(\bm x,\mu)\delta_{\mu\nu}\delta_{\bm x,\bm y},     \label{eq:[E_R,E_R]}
\end{align}
where $T^a$ is the generator of $SU(2)$ that satisfies $[T^a,T^b]=\ri\epsilon^{abc}T^c$ with the Levi-Civita symbol $\epsilon^{abc}$.
The generators $E^{a}_{R}(\bm x,\mu)$ and $E^{b}_{L}(\bm x,\nu)$ are not independent, but related to the parallel transport as $-U^\dag_\nu(\bm x) E_{L}(\bm x,\nu)U_\nu(\bm x) = E_{R}(\bm x,\nu)$, where $E_{R/L}(\bm x,\nu)= \sum_aT^a E^a_{R/L}(\bm x,\nu)$. 
This leads to the constraint,
\bea
\begin{split}
E^2(\bm x,\mu) =  \sum_aE^{a}_L(\bm x,\mu)E^{a}_L(\bm x,\mu)=\sum_a E^{a}_R(\bm x,\mu)E^{a}_R(\bm x,\mu).
\end{split}
\label{eq:constraint}
\eea
The Hamiltonian is given as the sum of electric and magnetic parts,
\begin{align}
  &H=H_E+H_B
  \label{eq:H_YM}
  \\
  &H_E=\sum_{\bm x,\mu}\frac{1}{2}E^2(\bm x,\mu),
  \label{eq:electric}
  \\
  &H_B=
   -\frac{K}{2}\sum_{\bm{x}}(\mathrm{tr}U_{\square}(\bm{x})+\mathrm{tr}U^\dag_{\square}(\bm{x}) ),
  \label{eq:magnetic}    
\end{align}
where
$U_{\square}(\bm{x})=\tr[U_{x}(\bm x)U_{y}(\bm x+e_x)U_{x}^{\dagger}(\bm x+e_y)U_{y}^{\dagger}(\bm x)]$.
$K$ is the coupling constant, which is inversely proportional to the square of the gauge coupling $g$. 
$H_E$ is the electric part of the Hamiltonian, while $H_B$ is the plaquette Hamiltonian; it involves the Wilson loop operator.
In addition, the theory is restricted to the physical Hilbert space $|\Psi\rangle$ by the Gauss law constraints: 
\bea
\sum_\mu\left( E^a_L(\bm x,\mu)+E^{a}_R(\bm x-e_{\mu},\mu) \right)|\Psi\rangle=0 .
\label{eq:gauss}
\eea
The Gauss law constraints represent the conservation law of electric flux at a site $\bm x$.
\begin{figure}[th]
\centering
\includegraphics[scale=0.5]{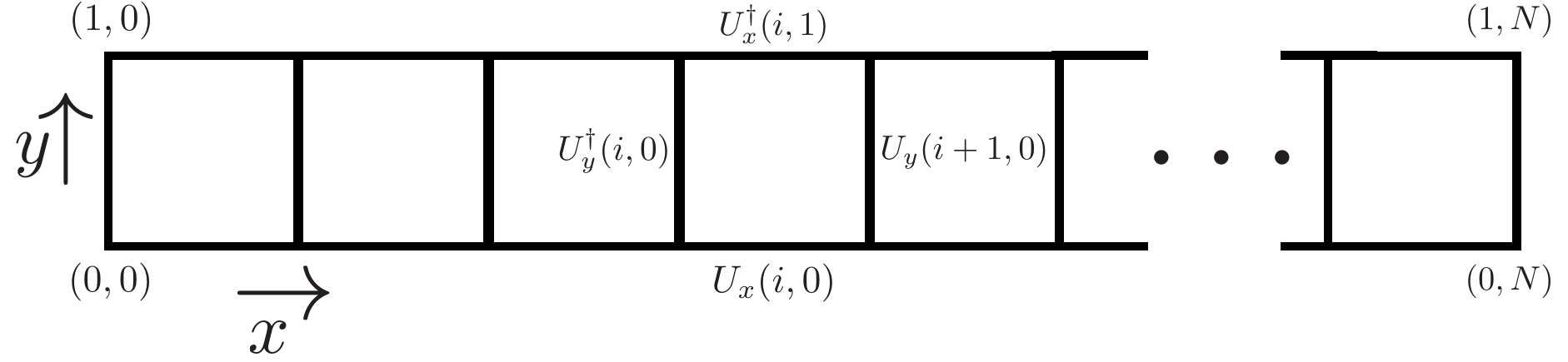}
\caption{Two-leg ladder model of $SU(2)$ Yang-Mills theory}
\label{fig:ladderLattice}
\end{figure}
In what follows, we will consider a two-leg ladder geometry shown in Fig.~\ref{fig:ladderLattice}.

\subsection{Yang-Mills-Ising model}

With the suitable truncation scheme detailed in Appendix~\ref{app:truncatedYM}, the SU(2) lattice YM theory in a two-leg ladder geometry can be mapped to the spin chain, which we call Yang-Mills-Ising (YM-Ising) model, and the resultant Hamiltonian of the spin chain has the form,
\begin{align}
\label{eq:YM_Hspin}
  H &=\sum_{i=0}^N\frac{3}{16}( 3-2Z_i-Z_{i-1}Z_{i})  -K\sum_{i=0}^{N-1} \frac{1}{16}X_i(1+3Z_{i-1})(1+3Z_{i+1}),
\end{align}
with $Z_{-1}=Z_{N}=1$ representing open boundary conditions.
This serves as an effective model of the SU(2) lattice gauge theory in the sense that the model has the SU(2) gauge symmetry, and satisfies the nonabelian Gauss law constraints. However, the dimension of the representation of SU($2$) (equivalent to the dimension of the Hilbert space) is significantly reduced; we use only the spin 0 and $1/2$ representations {\it i.e}., up to the first nontrivial representation. Although an extrapolation to the infinite-dimensional representation is needed to discuss the original lattice YM theory, we anticipate that the essential features of the nonabelian gauge symmetry and Gauss law constraints are incorporated into the truncated theory. 

\subsection{Numerical simulation}
\begin{figure}[th]
\centering
\includegraphics[width=.15\textwidth]{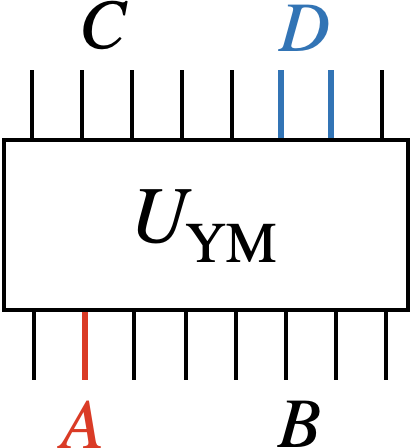}
\caption{The configuration of input and output Hilbert spaces of evolution operator $U_\text{YM}$. The states on $\calH_A$ and $\calH_D$ are respectively represented by red and blue wires. The rest of wires in input and output wires are the states on $\calH_B$ and $\calH_C$, respectively. $A$ and $D$ are kept away from the boundaries to reduce possible boundary effects.
}
\label{fig:config_YM}
\end{figure}

We study information scrambling on the YM-Ising model on the basis of the Hayden-Preskill protocol with the configuration shown in~Fig.~\ref{fig:config_YM} ($d=2^8$, $d_A=2$, and $d_D=2^2$).

The unitary evolution operator $U_{\rm YM}$ with the aforementioned Hamiltonian is implemented using the Suzuki-Trotter decomposition formula:
\begin{align}
\label{eq:STformula_YM}
\begin{split}
    U_\text{YM}
    &= \Big(\e^{-\im \frac{t}{M}\left(-\frac{3}{16}\sum_{i=0}^{N}(2Z_i+Z_{i-1}Z_{i})\right)}
    \\
    &\times
    \prod_{i=0}^{N-1}
    \e^{-\im \frac{t}{M}\left(-\frac{K}{16}X_i\right)}
    \e^{-\im \frac{t}{M}\left(-\frac{3K}{16}X_iZ_{i-1}\right)}
    \e^{-\im \frac{t}{M}\left(-\frac{3K}{16}X_iZ_{i+1}\right)}
    \e^{-\im \frac{t}{M}\left(-\frac{9K}{16}X_iZ_{i-1}Z_{i+1}\right)}\Big)^M
    \\
    &=e^{-\im H_\text{YM}t}+\calO(t^2/M).
\end{split}
\end{align}
The $10N-14$ CNOT operations are required in each Suzuki-Trotter step.
Since it is five times as large as that in the Ising spin chain, the YM-Ising model suffers from stronger decoherence effects when computing the Trotter evolution on a noisy quantum simulator.

\begin{figure}[h]
\centering
\begin{minipage}{.49\textwidth}
\subfloat[$P_\text{EPR}$]{
\includegraphics[width=.95\textwidth]{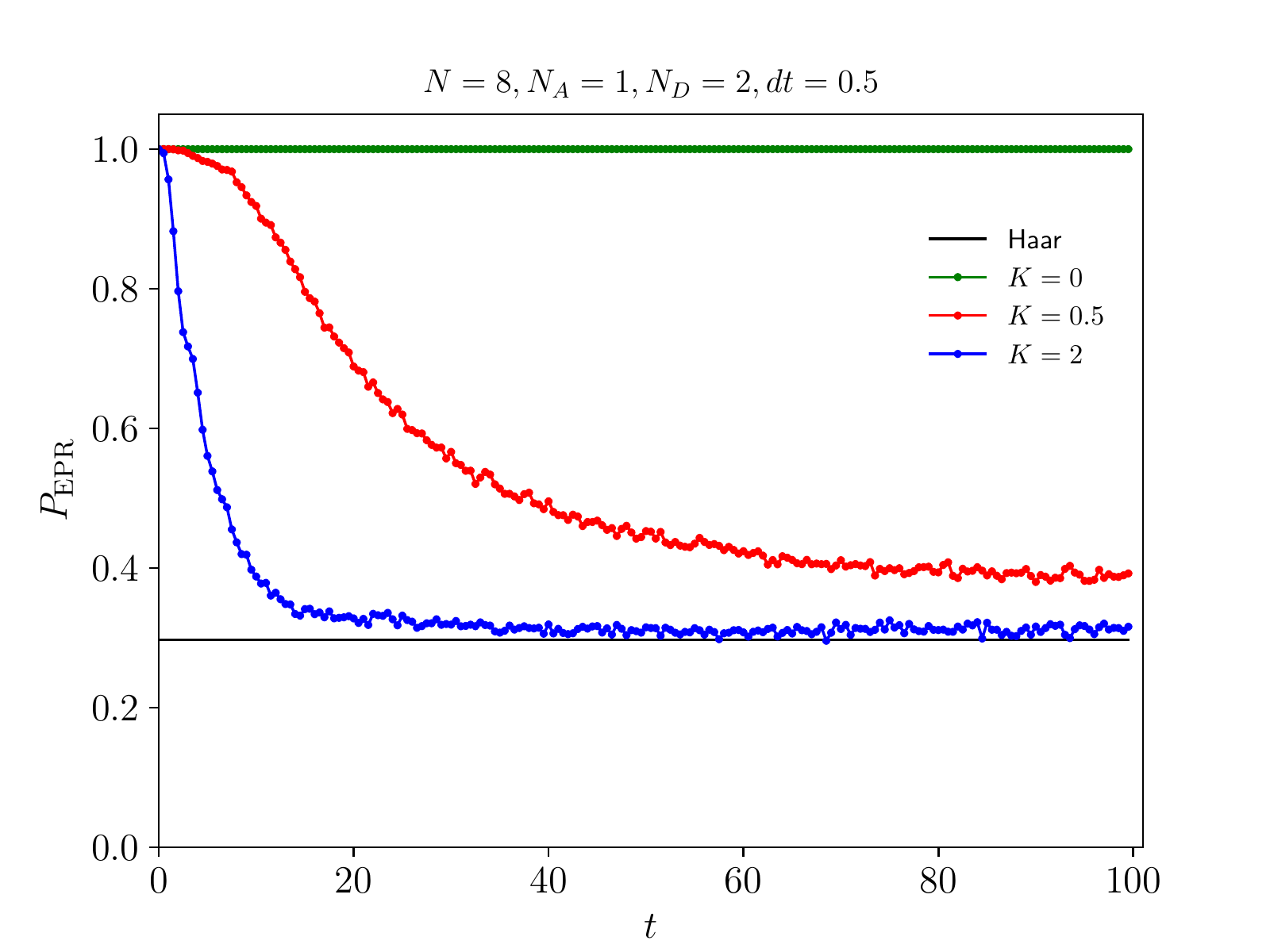}
\label{fig:Ferasure_ym}
}\end{minipage}\
\begin{minipage}{.49\textwidth}
\subfloat[$F_\text{EPR}$]{
\includegraphics[width=.95\textwidth]{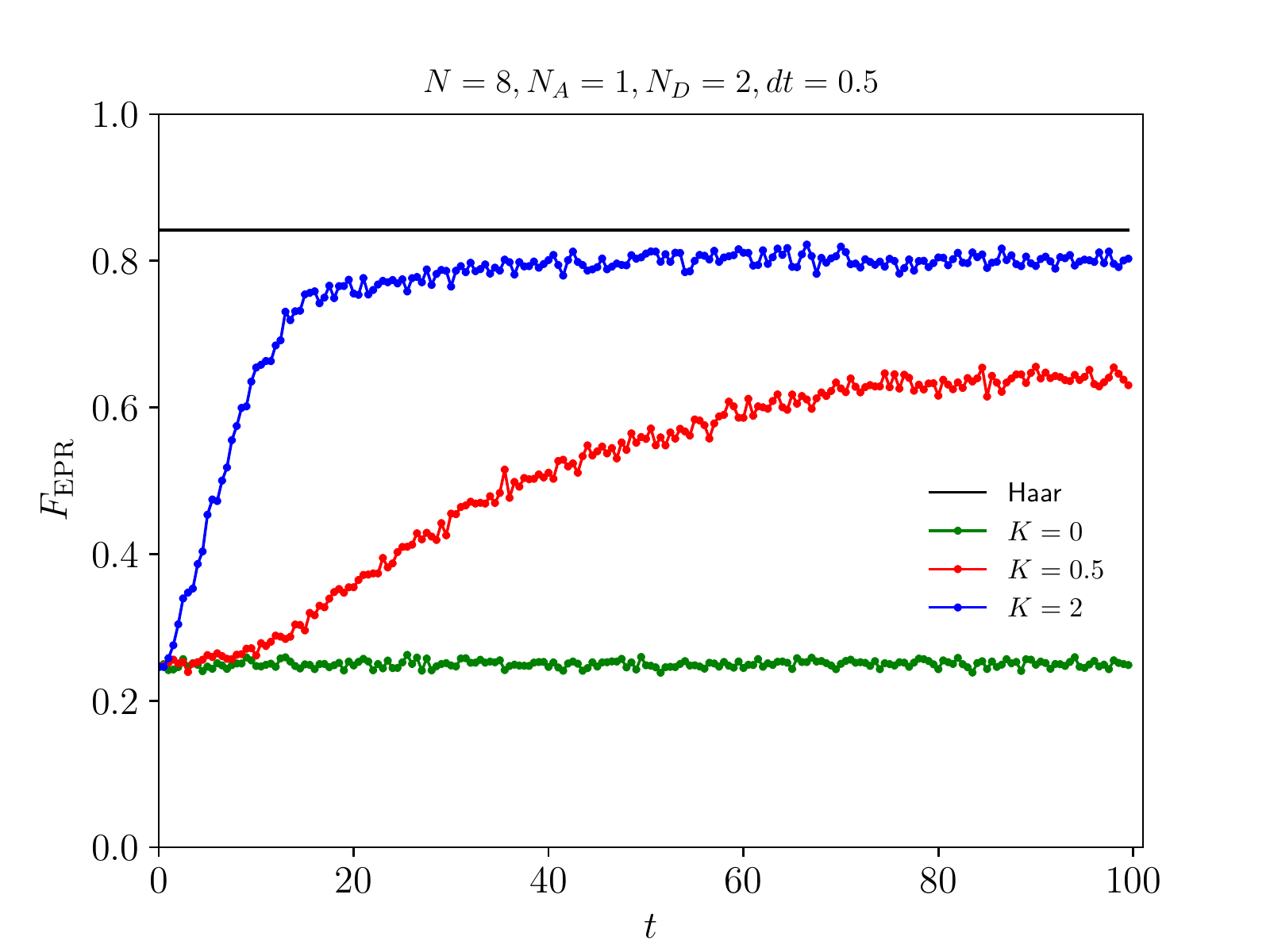}
\label{fig:Fdecoh_ym}
}\end{minipage}
\caption{The ideal Hayden-Preskill recovery protocol with $U=U_{\rm YM}$. (a) The projection probability~\eqref{eq:ideal_Pepr}. (b) The recovery fidelity~\eqref{eq:ideal_Fepr}. The parameters on each panel are $N\coloneqq\log d$, $N_A\coloneqq \log d_A$, $N_D\coloneqq\log d_D$, and $dt\coloneqq t/M$. The green, red, and blue points correspond to $K=0$, $K=0.5$, and $K=2$, respectively. The black line represents the $P_\text{EPR}$~\eqref{eq:ideal_Pepr_Haar} and $F_\text{EPR}$~\eqref{eq:ideal_Fepr_Haar} computed with $U$ given by a Haar random unitary operator.}
\label{fig:ideal_ym}
\end{figure}
\begin{figure}[h]
\centering
\begin{minipage}{.49\textwidth}
\subfloat[$P_\text{EPR}$]{
\includegraphics[width=.9\textwidth]{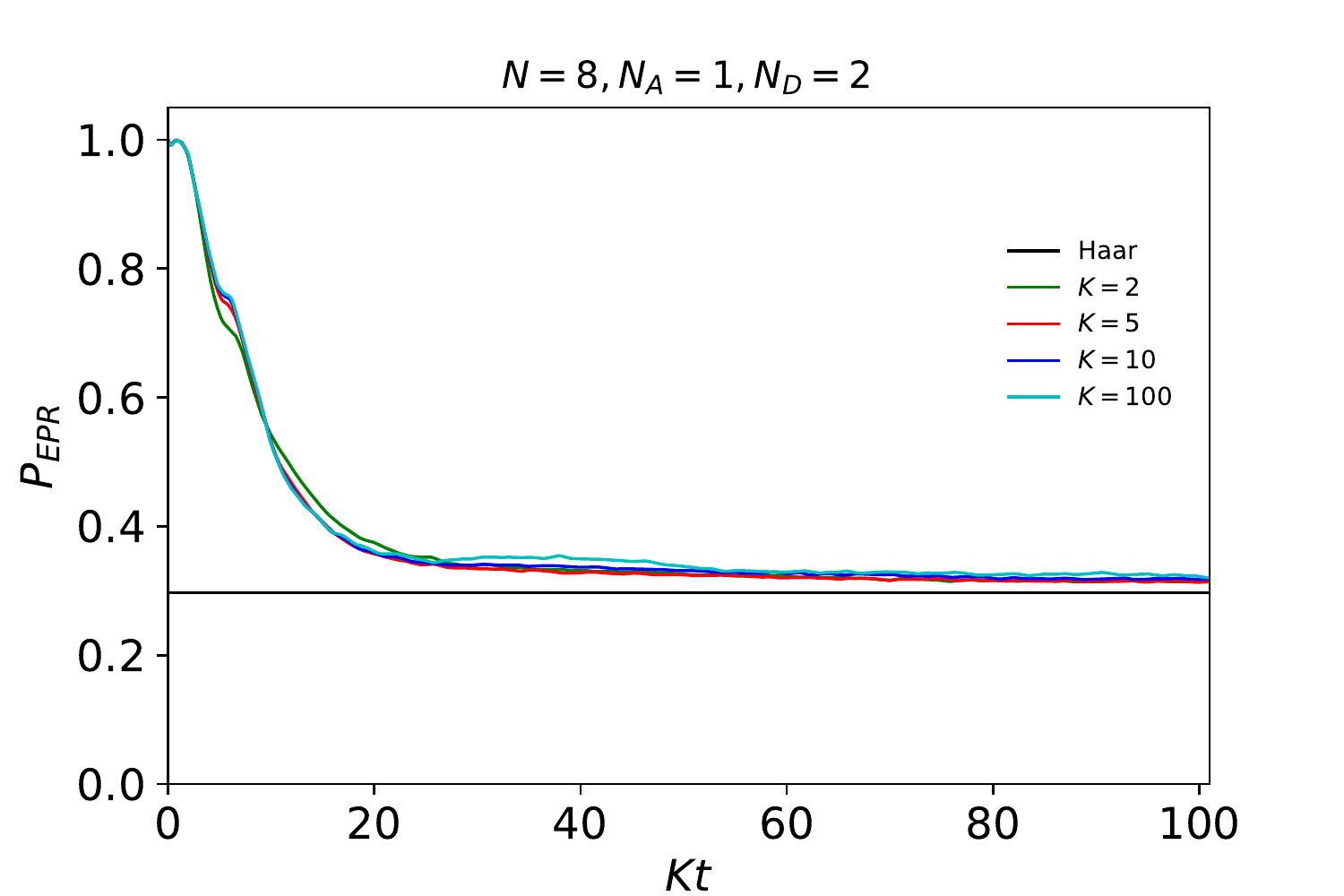}
}\end{minipage}\
\begin{minipage}{.49\textwidth}
\subfloat[$F_\text{EPR}$]{
\includegraphics[width=.9\textwidth]{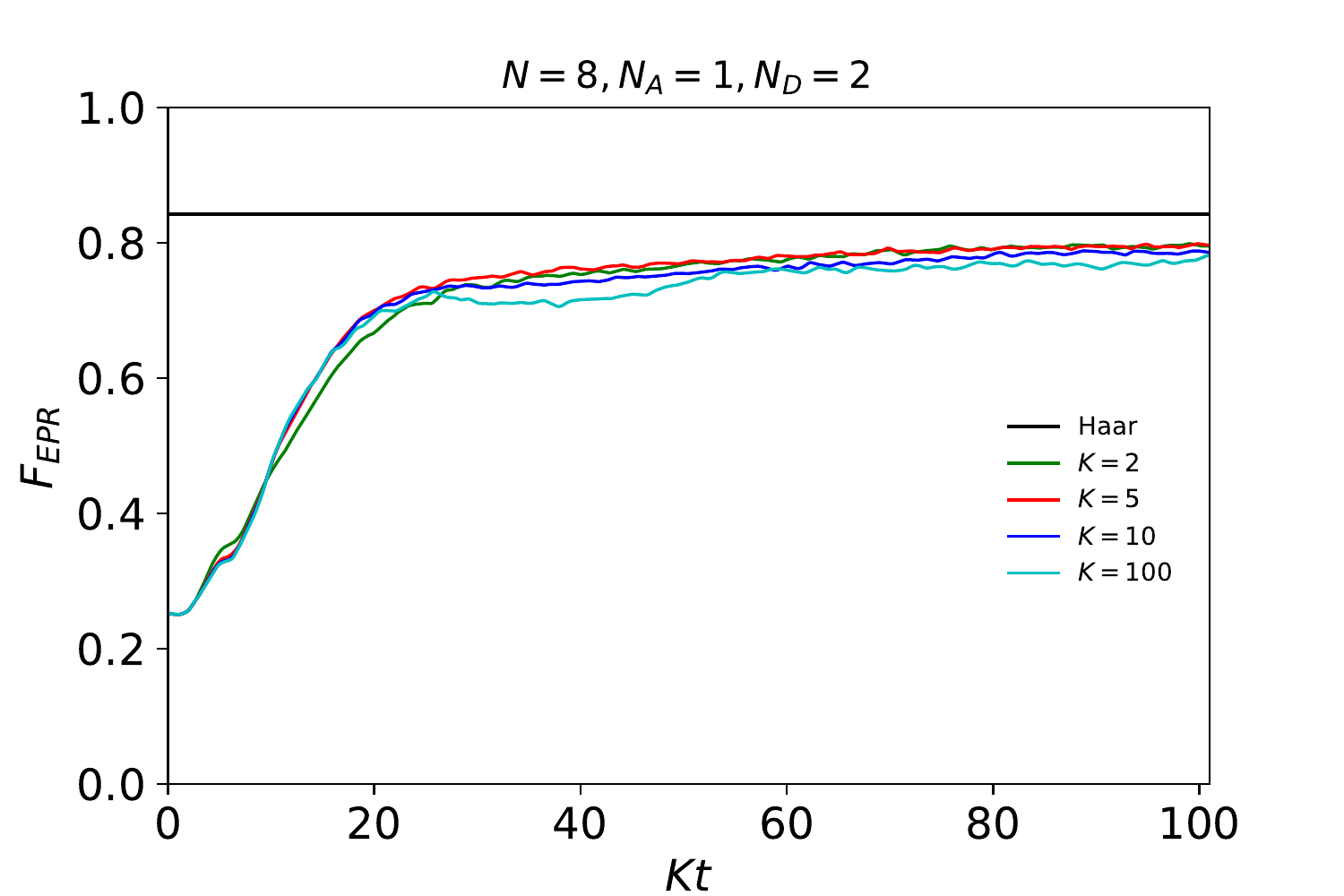}
}\end{minipage}
\begin{minipage}{.49\textwidth}
\subfloat[$P_\text{EPR}$ at late time]{
\includegraphics[width=.9\textwidth]{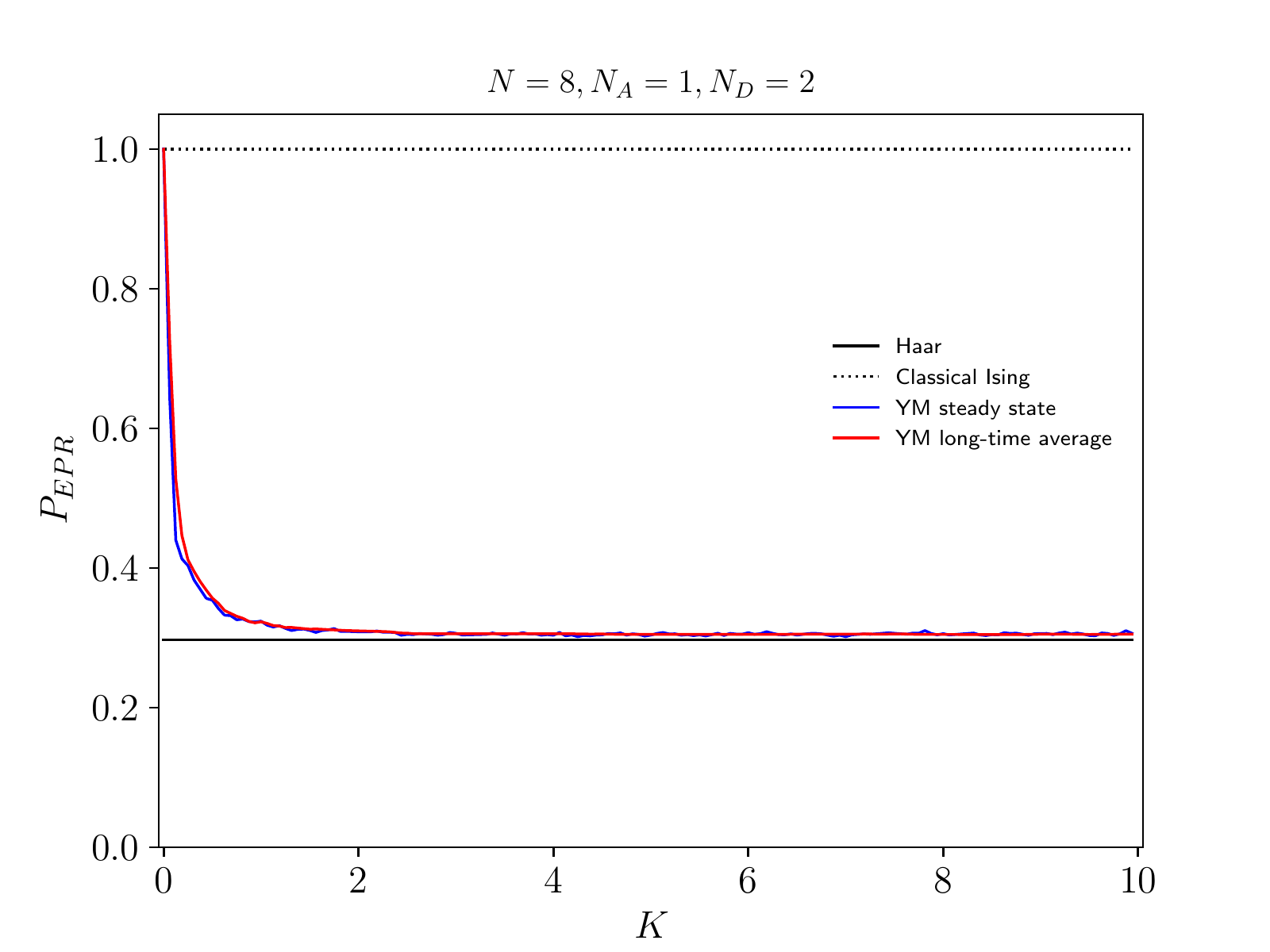}
}\end{minipage}\
\begin{minipage}{.49\textwidth}
\subfloat[$F_\text{EPR}$ at late time]{
\includegraphics[width=.9\textwidth]{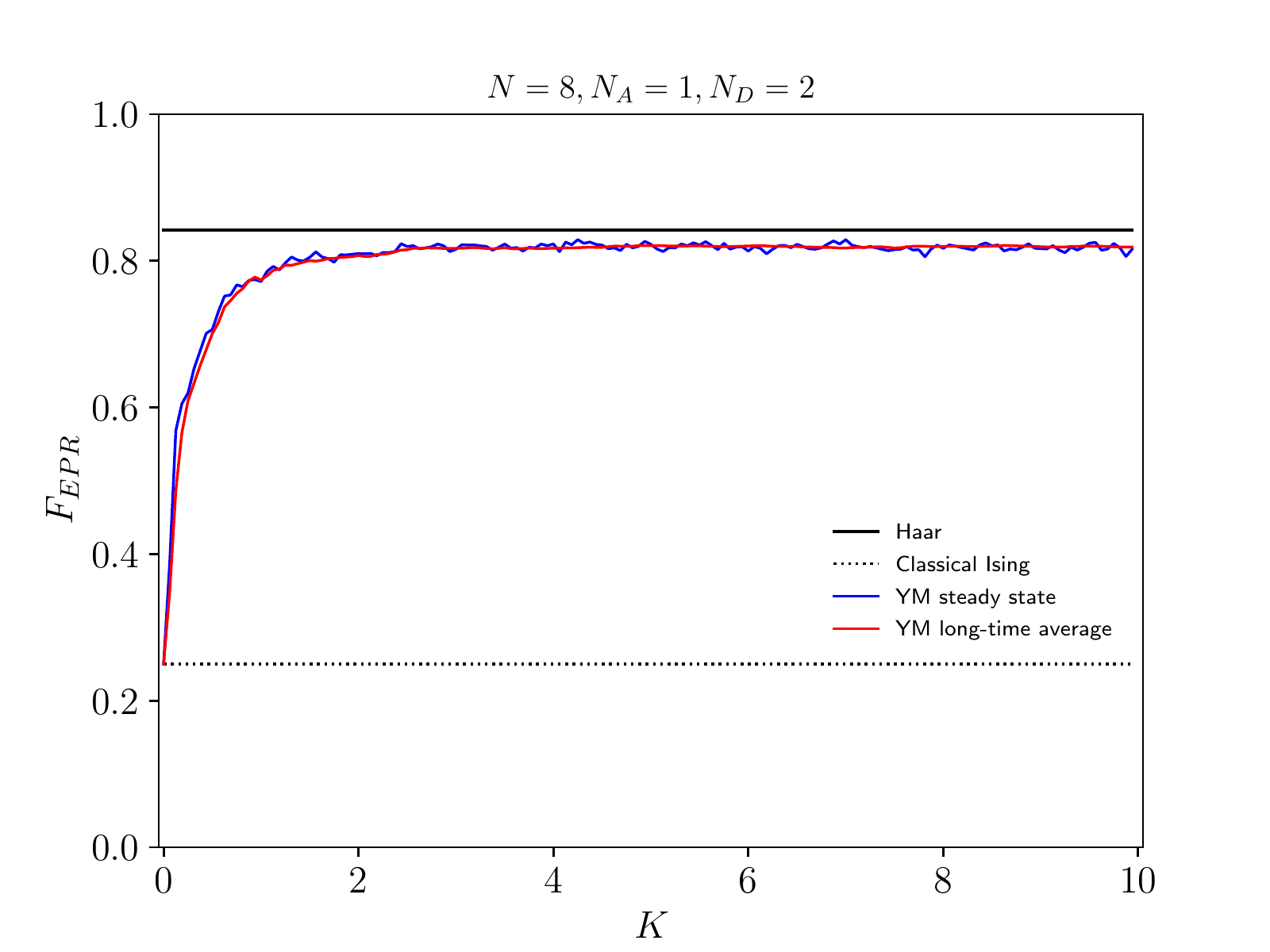}
}\end{minipage}
\caption{The Hayden-Preskill recovery protocol with $U=U_\text{YM}$ in the absence of decoherence. The projection probability (a) and the recovery fidelity (b) against the rescaled time $Kt$. The lower panels show the $K$-dependence of the late-time values of $P_\text{EPR}$ (c) and $F_\text{EPR}$ (d). The blue data are the values of $P_\text{EPR}$ ($F_\text{EPR}$) at $t=1000$, and the red data are obtained by averaging $P_\text{EPR}$ ($F_\text{EPR}$) over the time duration $100\le t\le 1000$. The solid and dotted black lines represent those by the Haar random evolution and classical Ising Hamiltonian (YM-Ising model at $K=0$).
}
\label{fig:ladder_Kdep}
\end{figure}

\begin{figure}[h]
\centering
\begin{minipage}{.49\textwidth}
\subfloat[$P_\text{EPR}^\text{decoh}$]{
\includegraphics[width=.95\textwidth]{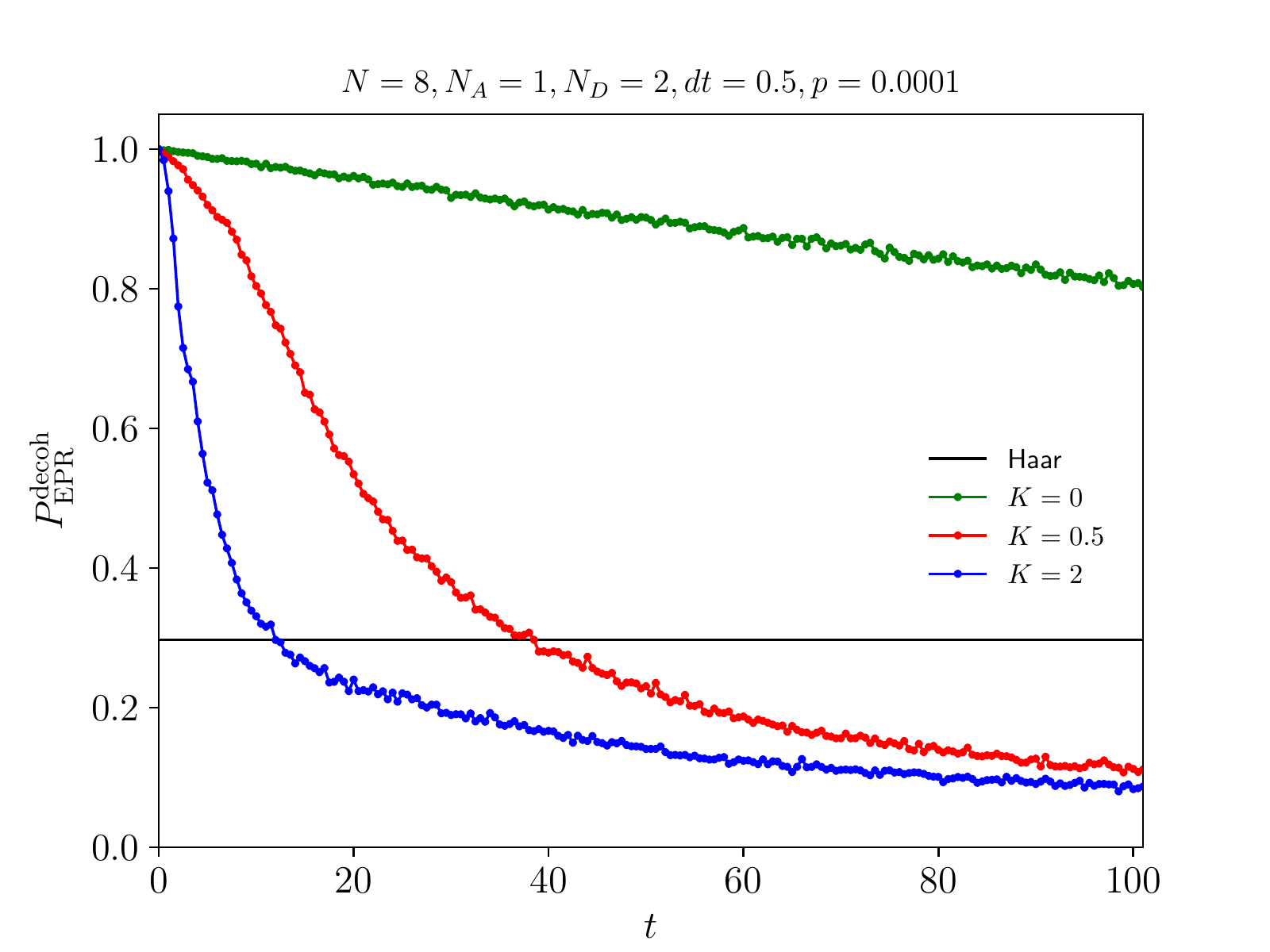}
\label{fig:Ferasure}
}\end{minipage}\
\begin{minipage}{.49\textwidth}
\subfloat[$F_\text{EPR}^\text{decoh}$]{
\includegraphics[width=.95\textwidth]{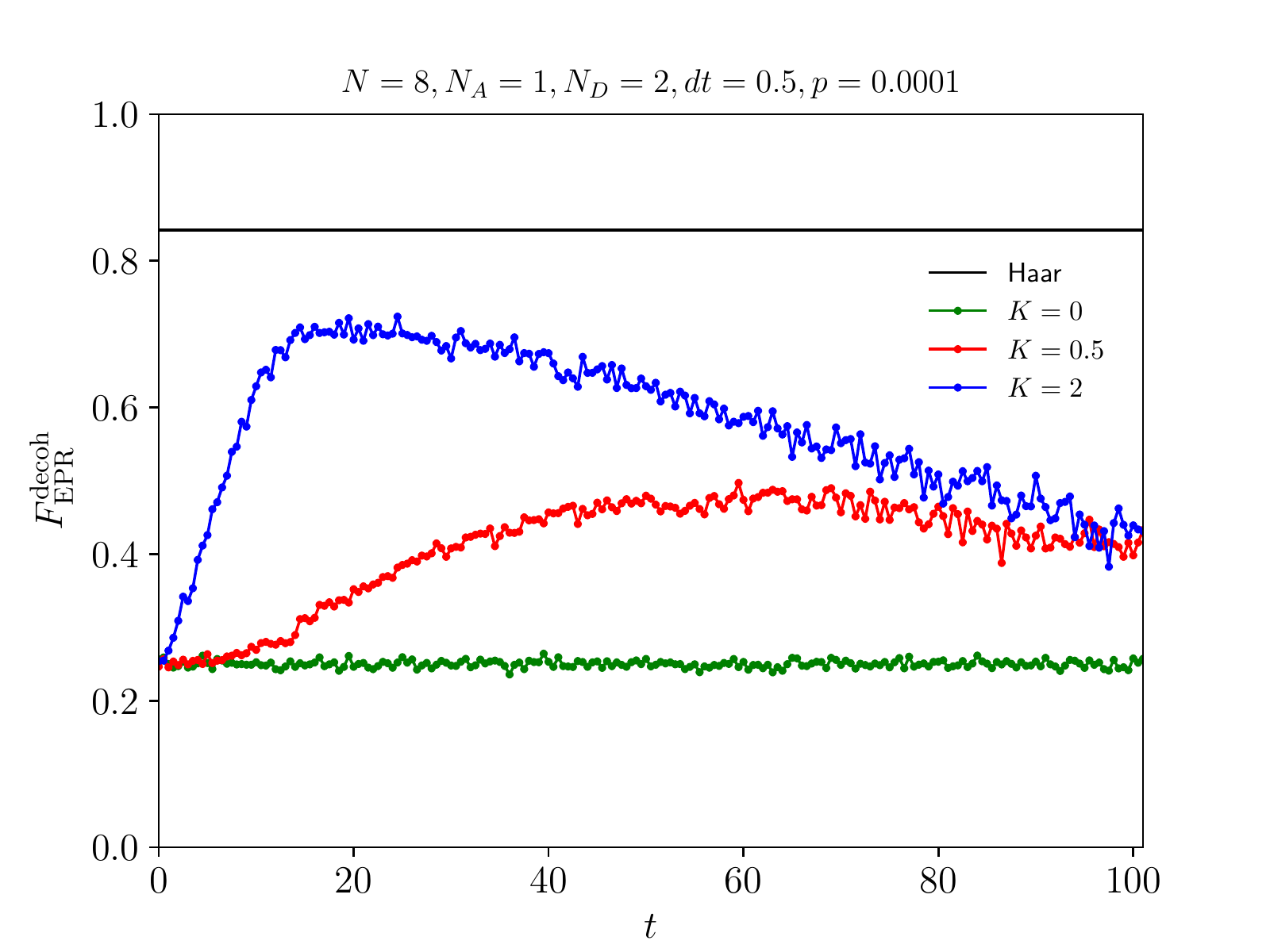}
\label{fig:Fdecoh}
}\end{minipage}
\caption{The Hayden-Preskill recovery protocol with $U=U_{\rm YM}$ in the presence of decoherence~\eqref{eq:CNOTdepolarize}. (a) The projection probability~\eqref{eq:noisy_Pepr}. (b) The recovery fidelity~\eqref{eq:noisy_Fepr}. The parameters on each panel are $N\coloneqq\log d$, $N_A\coloneqq\log d_A$, $N_D\coloneqq\log d_D$, $dt\coloneqq t/M$, and $p$ is the error probability~\eqref{eq:CNOTdepolarize}. The green, red, and blue points correspond to $K=0$, $K=0.5$, and $K=2$, respectively. The black line represents the $P_\text{EPR}$~\eqref{eq:ideal_Pepr_Haar} and $F_\text{EPR}$~\eqref{eq:ideal_Fepr_Haar} computed with $U$ given by a Haar random unitary operator without decoherence effects.}
\label{fig:noisy_ym}
\end{figure}

Figure~\ref{fig:ideal_ym} shows the $t$-dependence of $P_\text{EPR}$ and $F_\text{EPR}$ in the absence of noise effects. 
We see that $P_\text{EPR}$ decays to the value computed with Haar random evolutions for $K>1$. On the other hand, the decay becomes weak for $K<1$, and $P_\text{EPR}$ ceases to decay for $K\ll 1$, as is evident from the $K=0$ limit where our model is reduced to the classical Ising model with the longitudinal field. In view of the original link variables, the $K=0$ limit becomes the free theory of rigid rotators that should show no signal of the information scrambling. 
In Fig.~\ref{fig:ladder_Kdep}, $K$-dependences of $P_\text{EPR}$ and $F_\text{EPR}$ are shown in more detail. Figures~\ref{fig:ladder_Kdep}(a) and (b) are plotted for the time rescaled by $K$, in which the curves with different values of $K$ mostly overlap.
This is because the chaotic dynamics is dominated by the term proportional to $K$, that is, the magnetic term. Then, the results of different values of $K$ coincide by rescaling the time by each $K$. Without the rescaling, a larger $K$ leads to larger initial growth rate of $F_\text{EPR}$. Furthermore, as shown in Figs.~\ref{fig:ladder_Kdep}(c) and (d), $P_\text{EPR}$ ($F_\text{EPR}$) at late times take the values slightly above (below) the Haar random value for $K\gtrsim 1$. While their late-time behaviors do not depend on the systems size $N$, the decay (growth) of $P_\text{EPR}$ ($F_\text{EPR}$) start earlier for a smaller system size due to the shorter distance between the subsystems $A$ and $D$. See Appendix~\ref{app:numerial_data} for more on the system-size dependence and later time behavior of $P_\text{EPR}$/$F_\text{EPR}$.

Although more quantitative analyses with larger system sizes and larger cutoff of representations are certainly demanded, the high teleportation quality observed in the YM-Ising model implies that the scrambling also occurs in the original lattice YM theory since the number of qubits participating in the local interaction increases and the spreading of information becomes more prominent as cutoff increases.

Finally, we turn to the computation of $P^\text{decoh}_\text{EPR}$ and $F^\text{decoh}_\text{EPR}$ in the presence of decoherence effects (Fig.~\ref{fig:noisy_ym}).
Here, we introduce the decoherence by the same procedure as the case of Ising spin chain~\eqref{eq:CNOTdepolarize}.
In the left panel, the decay of $P^\text{decoh}_\text{EPR}$ is observed even in $K=0$, that is, the classical spin chain.
On the other hand, the effects of decoherence and scrambling are distinguishable in the case of $F^\text{decoh}_\text{EPR}$ as shown in the right panel. 
These features are the same as those observed in the Ising spin chain.

\section{Summary and discussion}\label{sec:Summary and discussion}

We demonstrated how the Hayden-Preskill protocol can be applied to diagnose the information scrambling of a given Hamiltonian dynamics under decoherence. To this end, we performed the numerical simulations with the Hamiltonian evolution of the Ising spin chain and the YM-Ising model.

The real-time simulations of the Ising spin chain show that the fidelity $F_\text{EPR}$ is well suited to extract the signal of information scrambling even in the presence of decoherence. This is because the scrambling results in the growth of $F_\text{EPR}$ while the decoherence modeled by the depolarizing channel leads to its decay. The behavior of $F_\text{EPR}$ should be contrasted to those of $P_\text{EPR}$ or OTOCs, which can decay both due to the scrambling and decoherence effects.
Hence, our study suggests that the measurement of $F_\text{EPR}$ in the Hayden-Preskill protocol is favorable to quantify the information scrambling on a near-term quantum device.

The protocol has also been applied to the SU(2) lattice Yang-Mills theory in a two-leg ladder geometry with a suitable truncation of the Hilbert space that retains the exact SU(2) gauge symmetry.
The Yang-Mills theory was formulated in a way that its real-time dynamics is efficiently simulated with the limited computational resource; gauge-redundant states are eliminated upon application of the Gauss' law constraint, and the Hilbert space is truncated after mapped to the spin basis. In the present work, we retained the gauge-invariant Hilbert space on each plaquette with $j\le j_\text{max}=1/2$ in the spin basis ($j_\text{max}$ is cutoff of the spin quantum number). Then, the Yang-Mills Hamiltonian on a ladder geometry has been mapped to the simple spin Hamiltonian, that is, the YM-Ising model~\eqref{eq:YM_Hspin}. The model serves as an effective model that would reproduce the original lattice Yang-Mills theory as we set $j_\text{max}$ to a sufficiently large value.
Employing the evolution operator with the YM-Ising Hamiltonian, we numerically performed the Hayden-Preskill protocol.
As a result, we observed the growth and late-time saturation in $F_\text{EPR}$, implying the information scrambling of the YM-Ising model.
While the scrambling in the original lattice Yang-Mills theory is yet to be confirmed from our study, it is anticipated that it also shows the signal of scrambling because the spread of information can be more prominent as the cutoff $j_\text{max}$ taken to be large.
This can be studied by the same numerical analysis with increasing $j_\text{max}$, that is left to a future study.

Finally, we remark that we formulated our entire protocols so that they can be directly implemented on digital quantum computers, and the numerical simulations were indeed performed on classical simulators provided by IBM Qiskit.
We anticipate that the simulations presented here will be implemented on a near-term quantum device in the near future.

\section*{Acknowledgements}
The numerical calculations were carried out on Yukawa-21 at YITP in Kyoto University, and on cluster computers at iTHEMS
in RIKEN with use of the QasmSimulator in IBM’s open source SDK Qiskit~\cite{Qiskit}.
This work was supported by JSPS KAKENHI Grant Numbers~17H06462, 18H01211, and the U.S. Department of Energy, Office of Science, National Quantum Information Science Research Centers, Co-design Center for Quantum Advantage under contract number DE-SC0012704.

\appendix

\section{Hayden-Preskill protocol with Haar random unitary evolution}
\label{app:HP_Haar}

We present the computations of $\Tr[\rho_{RD}^2]$ used for~Eq.~\eqref{eq:I(RBD)} and $P_\text{EPR}$~\eqref{eq:ideal_Pepr} provided the evolution operator $U$ is a Haar random unitary operator, which is obtained by integrating $U$ over rank-$d$ unitary operators with the Haar measure. $d$ is the dimension of the Hilbert space that $U$ acts on.
For an arbitrary function $f$ and a unitary operator $V$, the integral over the Haar measure satisfies the following properties:
\begin{align}
 &\int\diff U =1,
 \qquad
 \int\diff f(U) = \int\diff f(UV) = \int\diff f(VU).
\end{align}
In this Appendix, we will repeatedly use the following Haar integral formulas:
\begin{align}
 \int\diff U\ U_{i_1j_1}U^\ast_{i_2j_2}
 &=\frac{\delta_{i_1i_2}\delta_{j_2j_1}}{d},
\\
 \int\diff U\  U_{i_1j_1}U_{i_2j_2}U^\ast_{i_3j_3}U^\ast_{i_4j_4}
 &=\frac{\delta_{i_1i_3}\delta_{i_2i_4}\delta_{j_1j_3}\delta_{j_2j_4} 
 + \delta_{i_1i_4}\delta_{i_2i_3}\delta_{j_1j_4}\delta_{j_2j_3}}{d^2-1}
 \nonumber\\
 &-\frac{\delta_{i_1i_3}\delta_{i_2i_4}\delta_{j_1j_4}\delta_{j_2j_3} 
 + \delta_{i_1i_4}\delta_{i_2i_3}\delta_{j_1j_3}\delta_{j_2j_4}}{d(d^2-1)}.
\end{align}

We first calculate the purity $\Tr[\rho_{RD}^2]$ of the reduced density operator $\rho_{RD}=\Tr_{B'C}[\ket{\Psi}\bra{\Psi}]$,
\begin{align}
    \rho_{RD} = \frac{1}{d_Ad_B}\figbox{0.4}{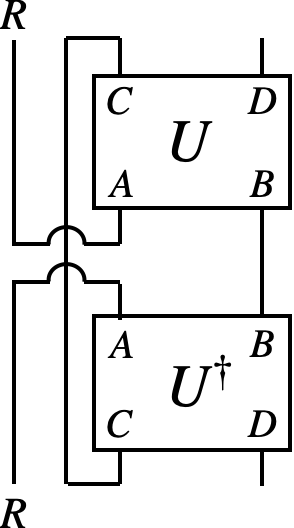}\, ,
\end{align}
which is used to compute $I^{(2)}(R,D)$~\eqref{eq:I(RBD)}.
Then, the associated purity is
\begin{align}
    \Tr[\rho_{RD}^2]
    &=\frac{1}{d_A^2d_B^2}\figbox{0.4}{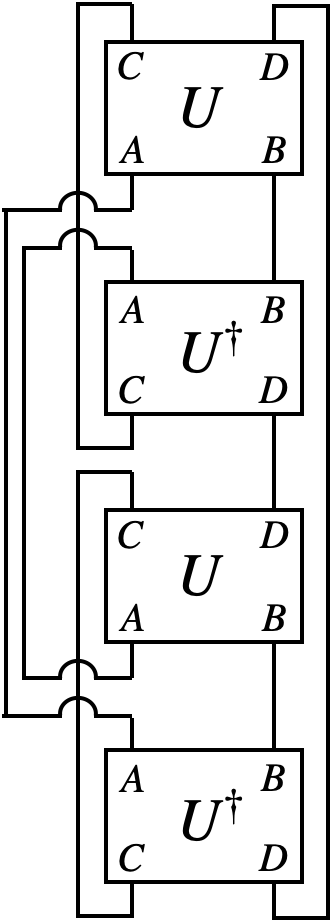}
    =\frac{1}{d_A^2d_B^2}U_{a_1b_1c_1d_1}U_{a_2b_2c_2d_2}U^\ast_{a_2b_1c_1d_2}U^\ast_{a_1b_2c_2d_1}.
\end{align}
Its Haar integral is
\begin{align}
\begin{split}
 &\int\diff U\ \Tr[\rho_{RD}^2]
 \\
 &=\frac{1}{d_A^2d_B^2}\int\diff U\  U_{a_1b_1c_1d_1}U_{a_2b_2c_2d_2}U^\ast_{a_2b_1c_1d_2}U^\ast_{a_1b_2c_2d_1}
 \\
 &=\frac{1}{d_A^2d_B^2}\Big[\frac{\delta_{a_1a_2}\delta_{b_1b_1}\delta_{a_2a_1}\delta_{b_2b_2}\ \delta_{c_1c_1}\delta_{d_1d_2}\delta_{c_2c_2}\delta_{d_2d_1} 
 + \delta_{a_1a_1}\delta_{b_1b_2}\delta_{a_2a_2}\delta_{b_2b_1}\ \delta_{c_1c_2}\delta_{d_1d_1}\delta_{c_1c_2}\delta_{d_2d_2}}{d^2-1}
 \\
 &-\frac{\delta_{a_1a_2}\delta_{b_1b_1}\delta_{a_2a_1}\delta_{b_2b_2}\ \delta_{c_1c_2}\delta_{d_1d_1}\delta_{c_2c_1}\delta_{d_2d_2} 
 + \delta_{a_1a_1}\delta_{b_1b_2}\delta_{a_2a_2}\delta_{b_2b_1}\ \delta_{c_1c_1}\delta_{d_1d_2}\delta_{c_2c_2}\delta_{d_2d_1}}{d(d^2-1)}\Big]
 \\
 &=\frac{1}{d_A^2d_B^2}\Big[\frac{d_Ad_B^2d_C^2d_D + d_A^2d_Bd_Cd_D^2}{d^2-1}
 -\frac{d_Ad_B^2d_Cd_D^2+ d_A^2d_Bd_C^2d_D}{d(d^2-1)}\Big]
 \\
 &=\frac{d_Bd_C+d_Ad_D}{d^2-1}-\frac{d_Bd_D+d_Ad_C}{d(d^2-1)}
 \\
 &\approx \frac{1}{d_Ad_D} + \frac{1}{d_Bd_C},
\end{split}
\end{align}
where we assumed $d\gg1$ in the last line.

We next compute $P_\text{EPR}$~\eqref{eq:ideal_Pepr}:
\begin{align}
    P_\text{EPR} 
    &=\frac{1}{d_A^2d_Bd_D}\figbox{0.4}{fig_Pepr}
    =\frac{1}{d_A^2d_Bd_D}U_{a_1b_1c_1d_1}U_{a_2b_2c_2d_2}U^\ast_{a_1b_2c_1d_2}U^\ast_{a_2b_1c_2d_1}.
\end{align}
Its Haar integral is computed as
\begin{align}
\begin{split}
 &\int\diff U\ P_\text{EPR}
 \\
 &=\frac{1}{d_A^2d_Bd_D}\int\diff U\  U_{a_1b_1c_1d_1}U_{a_2b_2c_2d_2}U^\ast_{a_1b_2c_1d_2}U^\ast_{a_2b_1c_2d_1}
 \\
 &=\frac{1}{d_A^2d_Bd_D}\Big[\frac{\delta_{a_1a_1}\delta_{b_1b_2}\delta_{a_2a_2}\delta_{b_2b_1}\ \delta_{c_1c_1}\delta_{d_1d_2}\delta_{c_2c_2}\delta_{d_2d_1} 
 + \delta_{a_1a_2}\delta_{b_1b_1}\delta_{a_2a_1}\delta_{b_2b_2}\ \delta_{c_1c_2}\delta_{d_1d_1}\delta_{c_2c_1}\delta_{d_2d_2}}{d^2-1}
 \\
  &-\frac{\delta_{a_1a_1}\delta_{b_1b_2}\delta_{a_2a_2}\delta_{b_2b_1}\ \delta_{c_1c_2}\delta_{d_1d_1}\delta_{c_2c_1}\delta_{d_2d_2} 
 + \delta_{a_1a_2}\delta_{b_1b_1}\delta_{a_2a_1}\delta_{b_2b_2}\ \delta_{c_1c_1}\delta_{d_1d_2}\delta_{c_2c_2}\delta_{d_2d_1}}{d(d^2-1)}\Big]
 \\
 &=\frac{1}{d_A^2d_Bd_D}\Big[\frac{d_A^2d_Bd_C^2d_D + d_Ad_B^2d_Cd_D^2}{d^2-1}
 -\frac{d_A^2d_Bd_Cd_D^2 + d_Ad_B^2d_C^2d_D}{d(d^2-1)}\Big]
 \\
 &=\frac{1}{d^2-1}\left[d_B^2 + d_C^2
 -\frac{d_C^2}{d_A^2}-1\right]\approx \frac{1}{d_A^2}+\frac{1}{d_D^2}-\frac{1}{d_A^2}\frac{1}{d_D^2},
\end{split}
\end{align}
where we assumed $d\gg1$ in the last line.

\section{State teleportation}
\label{app:stateTel}

Given a quantum channel $U_{AB}:\calH_A\otimes\calH_B\to \calH_C\otimes\calH_D$ and a state $\ket{\psi}_A\in\calH_A$, we consider the following quantum state~\cite{Hayden:2007cs}:
\begin{align}
    (I_{C}\otimes V_{DB'})(U_{AB}\otimes I_{B'})(\ket{\psi}_A\otimes \ket{\text{EPR}}_{BB'})
    =\figbox{0.4}{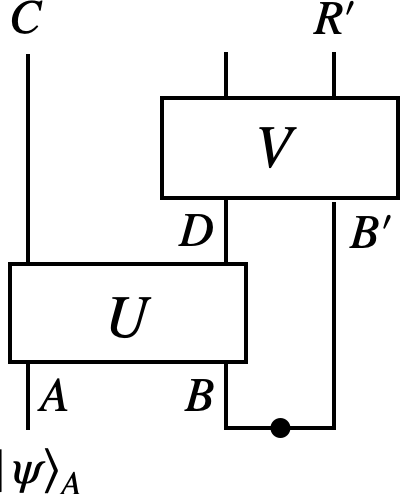}.
\end{align}
Here, we consider transmitting some pure state $\ket{\psi}_A$ on $\calH_{A}$ to $\calH_{R'}$.
A concrete teleportation protocol $V_{DB'}$ is as follows~\cite{Yoshida:2017non,Yoshida:2018vly}:
\begin{enumerate}
    \item Extend the Hilbert space by attaching $\calH_{A'}\otimes\calH_{R'}$ and prepare an EPR state, $\ket{\text{EPR}}_{A'R'}$.
    \item Apply a backward evolution $U^\ast_{A'B'}$ on $\calH_{A'}\otimes\calH_{B'}$ to find a state $\ket{\tilde{\Phi}^\psi}$.
    \item Project the state $\ket{\tilde{\Phi}^\psi}$ onto $\ket{\text{EPR}}_{DD'}$.
\end{enumerate}
Then, the resultant quantum state $\ket{\Phi^\psi}$ is given by
\begin{align}
\begin{split}
    \ket{\tilde{\Phi}^\psi} 
    &= (U_{AB}\otimes U^\ast_{A'B'}\otimes I_{R'})(\ket{\psi}_A\otimes \ket{\text{EPR}}_{BB'}\otimes \ket{\text{EPR}}_{A'R'}),
    \\
    \ket{\Phi^\psi} 
    &= \frac{\Pi_{DD'} \ket{\tilde{\Phi}^\psi}}{\sqrt{P_\text{EPR}^\psi}}
    =\frac{1}{\sqrt{P_\text{EPR}^\psi}}\figbox{0.4}{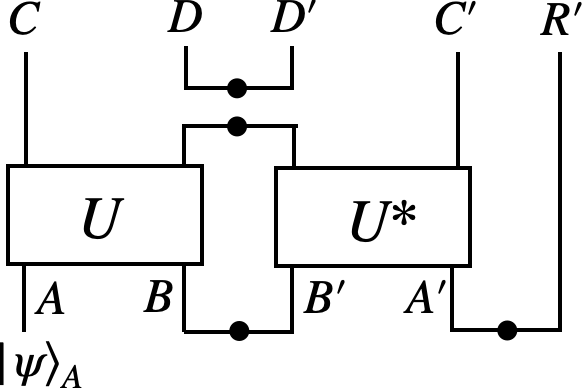}\,,
\end{split}
\end{align}
where $\Pi_{DD'}\coloneqq\ket{\text{EPR}}\bra{\text{EPR}}_{DD'}$ is the projection operator on $\ket{\text{EPR}}_{DD'}$.
The normalization factor $P_\text{EPR}^{\psi}$ is given by
\begin{align}
    P_\text{EPR}^\psi = \Tr[\Pi_{DD'} \ket{\tilde{\Phi}^\psi}\bra{\tilde{\Phi}^\psi}]
    =\frac{1}{d_Ad_Bd_D}\figbox{0.4}{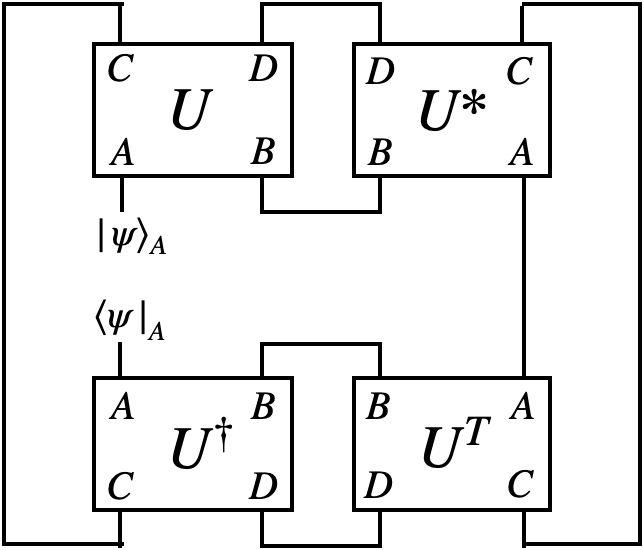}\,.
\end{align}
Let us evaluate the quality of teleportation by computing the following fidelity,
\begin{align}
\label{eq:Fepr_psi}
    F_\text{EPR}^\psi =\Tr[\Pi^\psi_{R'}\ket{\Phi^\psi}\bra{\Phi^\psi}]
    =\frac{1}{d_Ad_Bd_DP_\text{EPR}^\psi}\figbox{0.4}{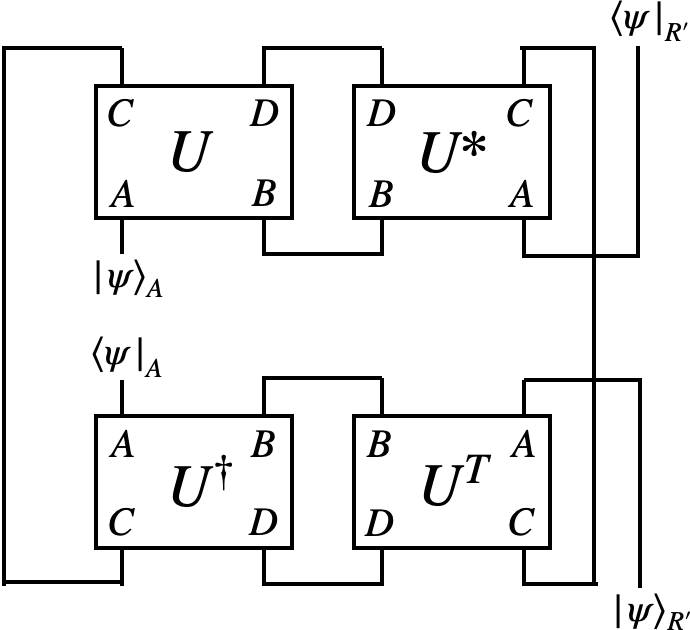}
\end{align}
with $\Pi^\psi_{R'}=\ket{\psi}\bra{\psi}_{R'}$.

Finally, we make a connection between $(P^\psi_\text{EPR},F^\psi_\text{EPR})$ and $(P_\text{EPR},F_\text{EPR})$. Taking a Haar average of $\ket{\psi}$ in Ref.~\eqref{eq:Fepr_psi}, we find
\begin{align}
\begin{split}
    \int_\text{Haar}\diff\psi P^\psi_\text{EPR}F^\psi_\text{EPR}
    &=\frac{1}{d_Ad_Bd_D(d_A^2-1)}\left(1-\frac{1}{d_A}\right)[d_A^3d_Bd_DP_\text{EPR}F_\text{EPR} + d_A^2d_Bd_DP_\text{EPR}]
    \\
    &=\frac{d_A}{d_A+1}P_\text{EPR}\left(F_\text{EPR} +\frac{1}{d_A}\right).
\end{split}
\end{align}
In the absence of decoherence effects (ideal case), where $P_\text{EPR}F_\text{EPR}=d_A^{-2}$~\eqref{eq:ideal_Fepr} holds, it is reduced to $\int_\text{Haar}\diff\psi P^\psi_\text{EPR}F^\psi_\text{EPR}=(P_\text{EPR}+d_A^{-1})/(d_A+1)$.

\section{Quantum circuits}
\label{app:ciruits}

We present the quantum circuits used in our simulations. All the quantum operations consist of the following three elementary gates.
Note that the time flows from left to right in the circuit diagrams presented in this Appendix.
\begin{itemize}
\item Hadamard gate:
\begin{align}
\label{eq:gate1}
&\begin{array}{c}
\Qcircuit @C=0.5cm @R=.3cm {
&  \gate{H} & \qw
}
\end{array}=\frac{1}{\sqrt{2}}
\begin{pmatrix}
 1 & 1 \\ 1 & -1
\end{pmatrix}.
\end{align}

\item Phase (Z-rotation) gate:
\begin{align}
\label{eq:gate2}
&\begin{array}{c}
\Qcircuit @C=0.5cm @R=.3cm {
&  \gate{R_Z(\theta)} & \qw
}
\end{array}
= \e^{-\im\frac{\theta}{2}Z}=
\begin{pmatrix}
 \e^{-\im\frac{\theta}{2}} & 0 \\ 0 & \e^{\im\frac{\theta}{2}}
\end{pmatrix}.
\end{align}

\item CNOT gate: 
\begin{align}
\label{eq:gate3}
\begin{array}{c}
\Qcircuit @C=1em @R=.7em {
& \ctrl{1} & \qw
\\
& \targ{} & \qw
}
\end{array}
= \begin{pmatrix} 
1 & 0  & 0 & 0 \cr  
0 & 1  & 0 & 0 \cr  
0 & 0  & 0 & 1 \cr  
0 & 0  & 1 & 0 \cr  
\end{pmatrix}.
\end{align}
\end{itemize}
The matrix representation of a single-qubit operation is given on the basis $(\ket{0}\, \ket{1})$, and that of two-qubit operation is given on the basis $(\ket{00}\, \ket{01}\,\ket{10}\,\ket{11})$, where $\ket{0}$ ($\ket{1}$) is an eigenstate of $Z$ operator, $Z\ket{0}=\ket{0}$ ($Z\ket{1}=-\ket{1}$).

A $2N$-qubit EPR state, $\ket{\text{EPR}}=\frac{1}{2^{N/2}}\sum_{i=0}^{2^N-1}\ket{i_\text{bin}}_A\otimes\ket{i_\text{bin}}_B$ with $i_\text{bin}$ being a $N$-digit binary representation of $i$, is created as follows e.g., for $N=3$:
\begin{align}
\label{eq:EPRprep}
 \ket{\text{EPR}} &= 
 \hspace{3em}
\begin{array}{c}
\Qcircuit @C=0.4cm @R=.1cm {
\lstick{\ket{0}_A}&\gate{H}&\ctrl{3}&\qw&\qw&\qw
\\
\lstick{\ket{0}_A}&\gate{H}&\qw&\ctrl{3}&\qw&\qw
\\
\lstick{\ket{0}_A}&\gate{H}&\qw&\qw&\ctrl{3}&\qw
\\
\lstick{\ket{0}_B}&\qw&\targ{}&\qw&\qw&\qw
\\
\lstick{\ket{0}_B}&\qw&\qw&\targ{}&\qw&\qw
\\
\lstick{\ket{0}_B}&\qw&\qw&\qw&\targ{}&\qw
}
\end{array}\,.
\end{align}
In general, $N$ CNOT gates are required to create a $2N$-qubit EPR state.

Next, we consider an $2N$-qubit EPR measurement/projection. Given an arbitrary $2N$-qubit input state, the measurement circuit e.g., for $N=3$ is as follows:
\begin{align}
\label{eq:EPRmeas}
\begin{array}{c}
\Qcircuit @C=0.4cm @R=.1cm {
&\ctrl{3}&\qw&\qw&\gate{H}&\meter{}
\\
&\qw&\ctrl{3}&\qw&\gate{H}&\meter{}
\\
&\qw&\qw&\ctrl{3}&\gate{H}&\meter{}
\\
&\targ{}&\qw&\qw&\qw&\meter{}
\\
&\qw&\targ{}&\qw&\qw&\meter{}
\\
&\qw&\qw&\targ{}&\qw&\meter{}
}
\end{array}\,.
\end{align}
The input state is successfully projected onto an EPR state when the measurement result on the right end of circuit reads ``000000'' on $Z$-basis.
A $2N$-qubit EPR measurement uses $N$ CNOT gates.
In the rest of this Appendix, we show how to implement the time evolution operators used in the main text.

\subsection{Ising spin chain}

Upon application of the Suzuki-Trotter decomposition, the time evolution operator of the Ising spin chain is given by
\begin{align}
    U_\text{Ising}
    = \left(\e^{-\im \frac{t}{M}\left(-\sum_{i=1}^{N-1}Z_iZ_{i+1}
    -m\sum_{i=1}^{N}Z_i\right)}
    \e^{-\im \frac{t}{M}\left(-h\sum_{i=1}^{N}X_i\right)}\right)^M.
\end{align}
It is implemented with the following two quantum gates:
\begin{itemize}

\item Ising-coupling gate: 
\begin{align}
\e^{-\im \frac{\theta}{2}Z\otimes Z}=
\begin{array}{c}
\Qcircuit @C=1em @R=.7em {
& \ctrl{1} & \qw & \ctrl{1}& \qw
\\
& \targ{} & \gate{R_Z(\theta)} & \targ{} & \qw
}
\end{array}\,.
\end{align}
\item X-rotation gate:
\begin{align}
\e^{-\im\frac{\theta}{2}X}=
&\begin{array}{c}
\Qcircuit @C=0.5cm @R=.3cm {
&\gate{H}&  \gate{R_Z(\theta)} &\gate{H}& \qw
}
\end{array}.
\end{align}
\end{itemize}
Thus, implementing $U_\text{Ising}$ at each Suzuki-Trotter step requires $2(N-1)$ CNOT gates.

\subsection{YM-Ising model}

Upon application of the Suzuki-Trotter decomposition, the time evolution operator of the Yang-Mills-Ising model~\eqref{eq:YM_Hspin} is given by
\begin{align}
\label{eq:ST_YM}
    U_\text{YM}
    &= \Big(\e^{-\im \frac{t}{M}\left(-\frac{3}{16}\sum_{i=0}^{N}(2Z_i+Z_{i-1}Z_{i})\right)}
    \nonumber\\
    &\times
    \prod_{i=0}^{N-1}
    \e^{-\im \frac{t}{M}\left(-\frac{K}{16}X_i\right)}
    \e^{-\im \frac{t}{M}\left(-\frac{3K}{16}X_iZ_{i-1}\right)}
    \e^{-\im \frac{t}{M}\left(-\frac{3K}{16}X_iZ_{i+1}\right)}
    \e^{-\im \frac{t}{M}\left(-\frac{9K}{16}X_iZ_{i-1}Z_{i+1}\right)}\Big)^M
\end{align}
with $Z_{-1}=Z_N=1$.
It is implemented with the following gates in addition to those presented so far:
\begin{itemize}

\item $XZ$ gate: 
\begin{align}
\e^{-\im \frac{\theta}{2}X\otimes Z}=
\begin{array}{c}
\Qcircuit @C=1em @R=.7em {
&\gate{H} & \ctrl{1} & \qw & \ctrl{1}&\gate{H} & \qw
\\
&\qw & \targ{} & \gate{R_Z(\theta)} & \targ{} &\qw & \qw
}
\end{array}\,.
\end{align}

\item $ZXZ$ gate:
\begin{align}
\e^{-\im \frac{\theta}{2}Z\otimes X\otimes Z}=
\begin{array}{c}
\Qcircuit @C=1em @R=.7em {
&\qw & \ctrl{2} &\qw & \qw & \qw & \ctrl{2}&\qw & \qw
\\
&\gate{H} & \qw & \ctrl{1} & \qw & \ctrl{1}& \qw&\gate{H} & \qw
\\
&\qw & \targ{} &\targ{} & \gate{R_Z(\theta)} & \targ{} &\targ{} &\qw & \qw
}
\end{array}\,.
\end{align}

\end{itemize}
Thus, implementing $U_\text{YM}$ at each Suzuki-Trotter step requires $10N-14$ CNOT gates.
\section{More numerical analyses of $P_\text{EPR}$ and $F_\text{EPR}$}
\label{app:numerial_data}
\subsection{Suzuki-Trotter decomposition versus Exact evolution}
\label{app:ST_dep}

\begin{figure}[h]
\centering
\begin{minipage}{.49\textwidth}
\subfloat[$P_\text{EPR}^\text{decoh}$]{
\includegraphics[width=.95\textwidth]{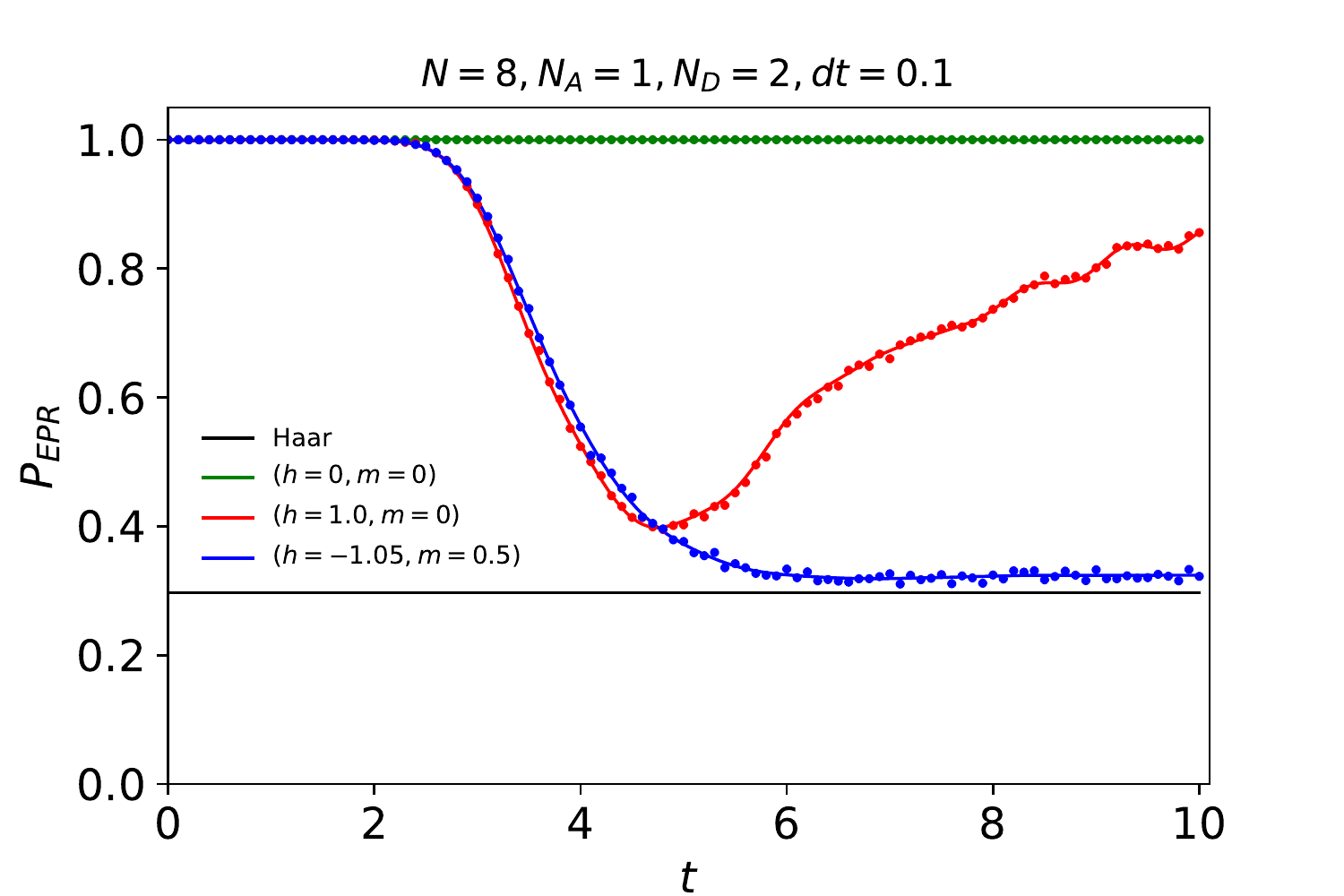}
}\end{minipage}\
\begin{minipage}{.49\textwidth}
\subfloat[$F_\text{EPR}^\text{decoh}$]{
\includegraphics[width=.95\textwidth]{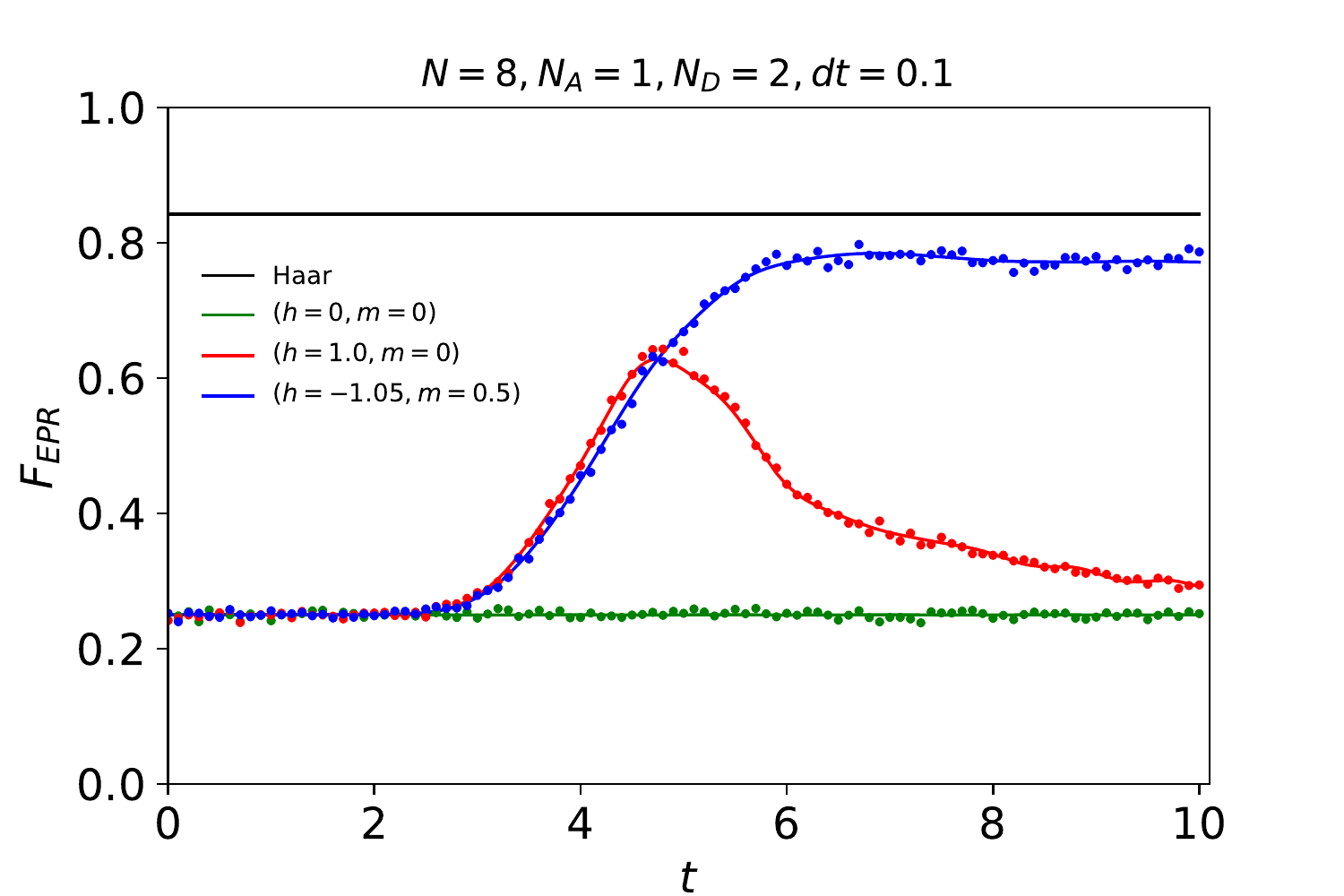}
}\end{minipage}
\caption{The Hayden-Preskill recovery protocol for the Ising spin chain in the absence of decoherence. (a) The projection probability. (b) The recovery fidelity. The data points are obtaind by the computation with the Suzuki-Trotter decomposition applided to the evolution operator, while the curves are obtaind by the exact Hamiltonian evolution.}
\label{fig:exact_Ising}
\end{figure}

\begin{figure}[h]
\centering
\begin{minipage}{.49\textwidth}
\subfloat[$P_\text{EPR}^\text{decoh}$]{
\includegraphics[width=.95\textwidth]{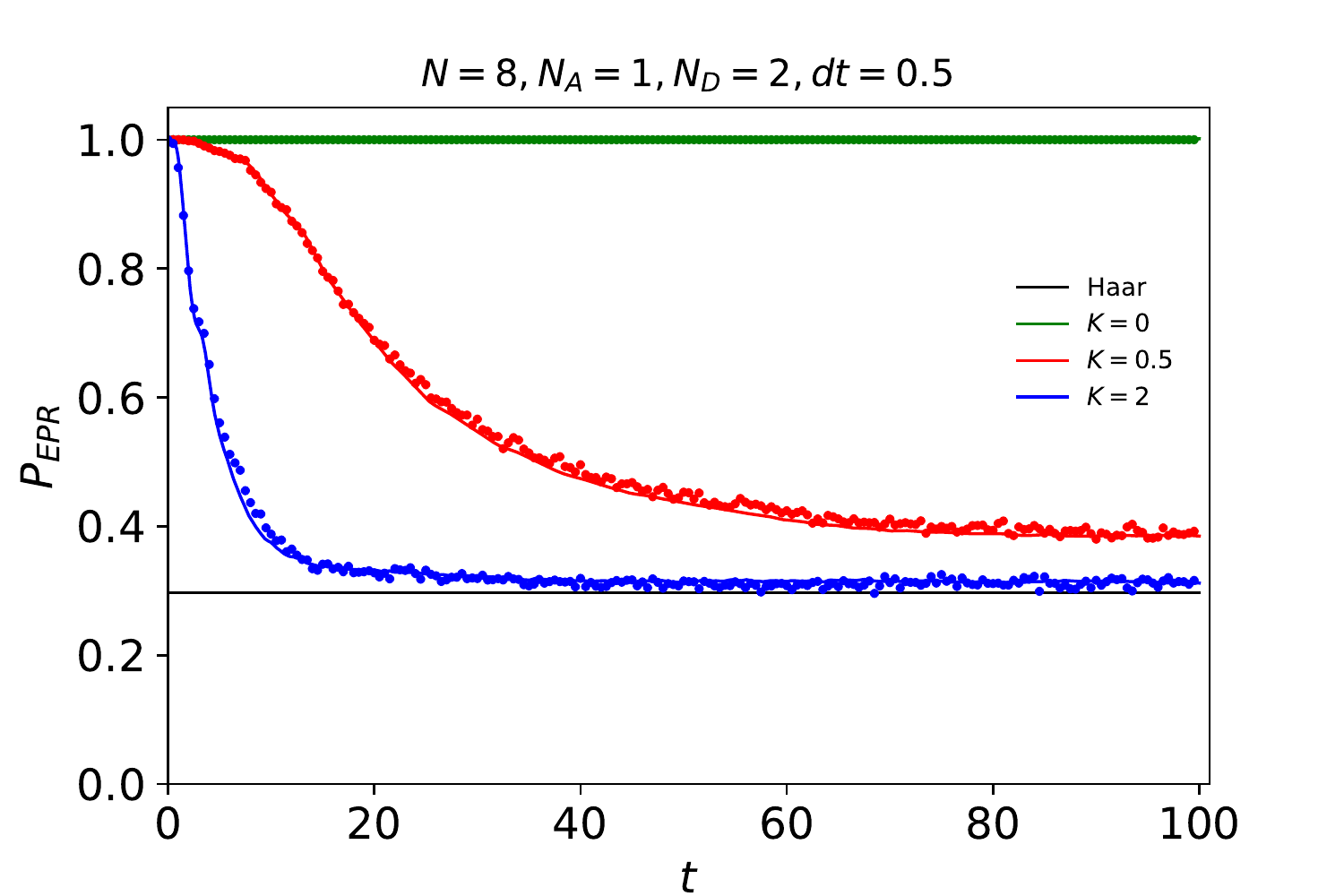}
}\end{minipage}\
\begin{minipage}{.49\textwidth}
\subfloat[$F_\text{EPR}^\text{decoh}$]{
\includegraphics[width=.95\textwidth]{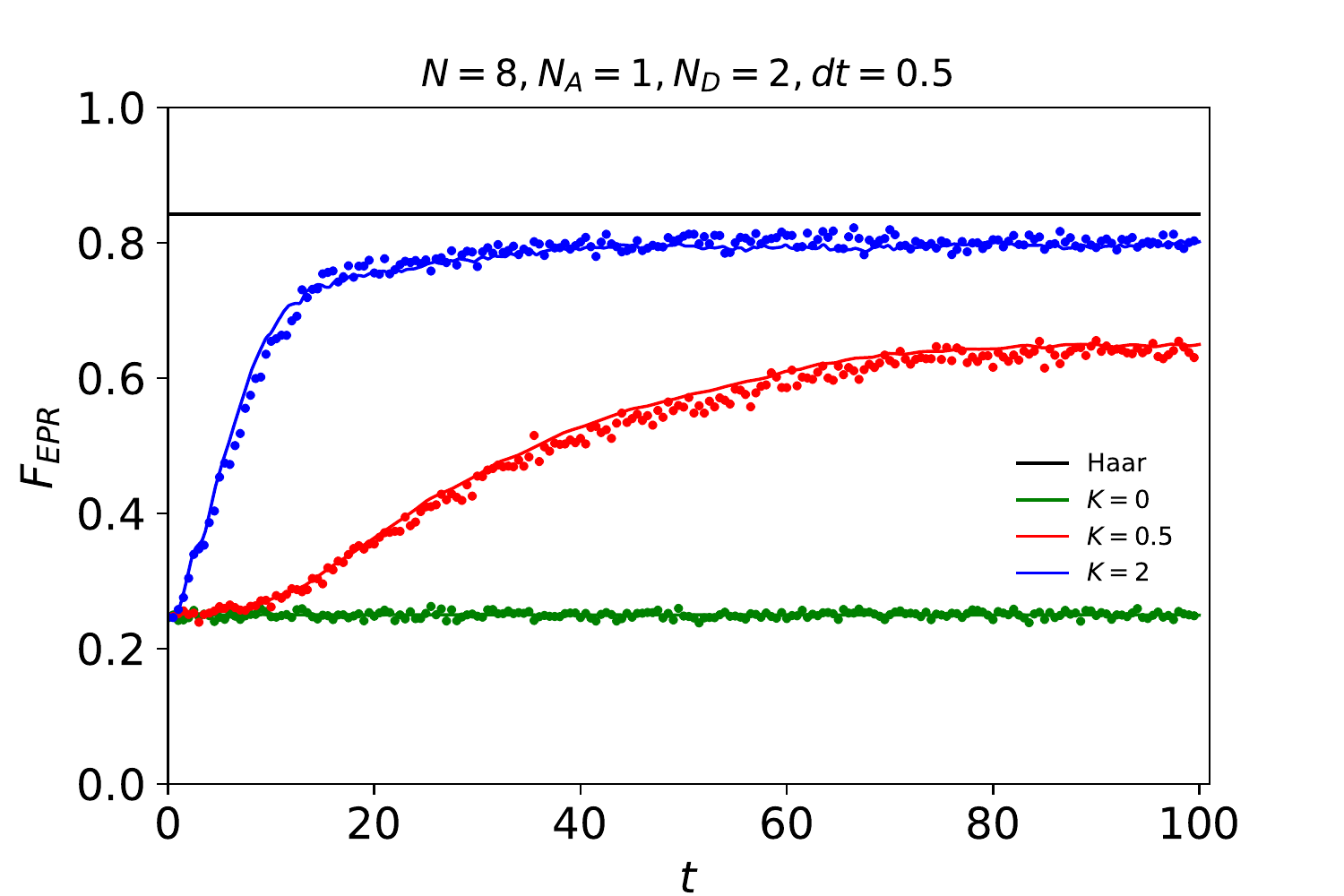}
}\end{minipage}
\caption{The Hayden-Preskill recovery protocol for the YM-Ising model in the absence of decoherence. (a) The projection probability. (b) The recovery fidelity. The data points are obtaind by the computation with the Suzuki-Trotter decomposition applided to the evolution operator, while the curves are obtaind by the exact Hamiltonian evolution.}
\label{fig:exact_ladder}
\end{figure}

The Suzuki-Trotter decomposition introduces a systematic error due to its temporal discretization of the Hamiltonian evolution. Here, we show that the Suzuki-Trotter error is well suppressed in our simulations. Figures \ref{fig:exact_Ising} and \ref{fig:exact_ladder} are the numerical results of $P_\mathrm{EPR}$ and $F_\mathrm{EPR}$ with the Suzuki-Trotter decomposition and exact Hamiltonian evolution. Indeed, there are no significant deviation between the results using the decomposition and exact evolution.

\subsection{System-size dependence}
\label{app:N_dep}

\begin{figure}[h]
\centering
\begin{minipage}{.49\textwidth}
\subfloat[$P_\text{EPR}$]{
\includegraphics[width=.95\textwidth]{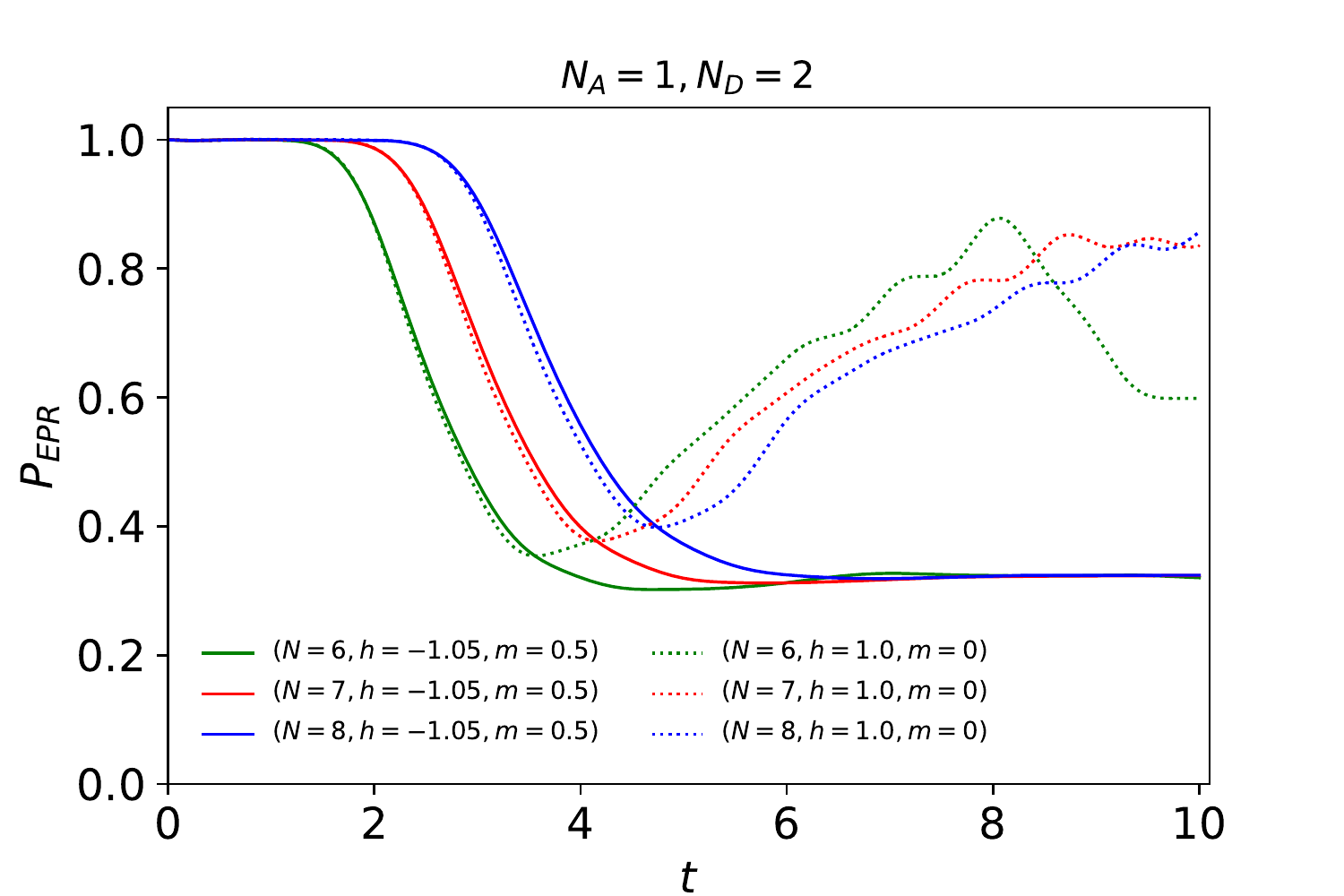}
}\end{minipage}\
\begin{minipage}{.49\textwidth}
\subfloat[$F_\text{EPR}$]{
\includegraphics[width=.95\textwidth]{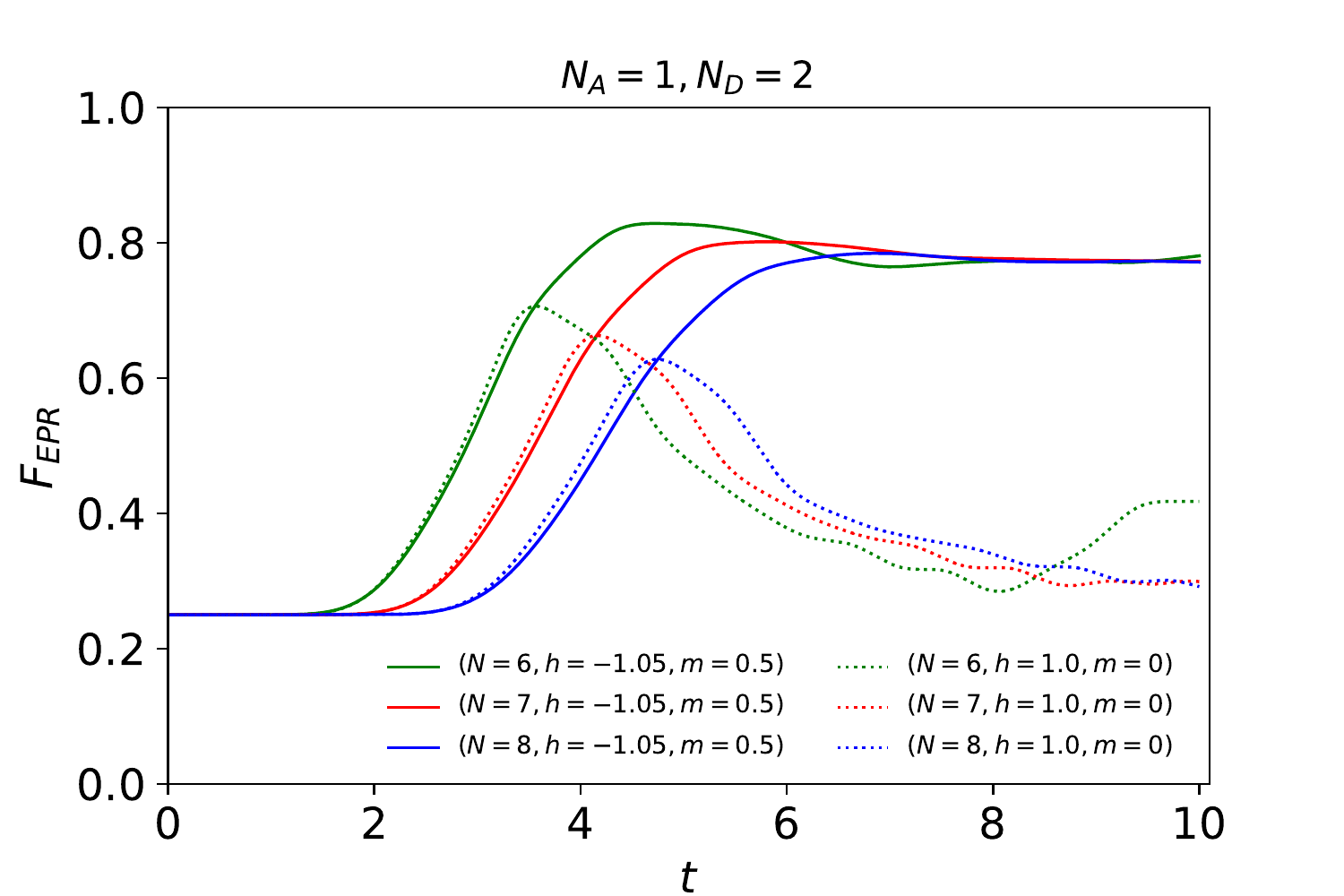}
}\end{minipage}
\caption{The Hayden-Preskill recovery protocol for the Ising spin chain in the absence of decoherence. The projection probability (a), and the recovery fidelity (b) are computed with system sizes $N=6,7,8$. 
}
\label{fig:Ising_size}
\end{figure}

\begin{figure}[h]
\centering
\begin{minipage}{.49\textwidth}
\subfloat[$P_\text{EPR}^\text{decoh}$]{
\includegraphics[width=.95\textwidth]{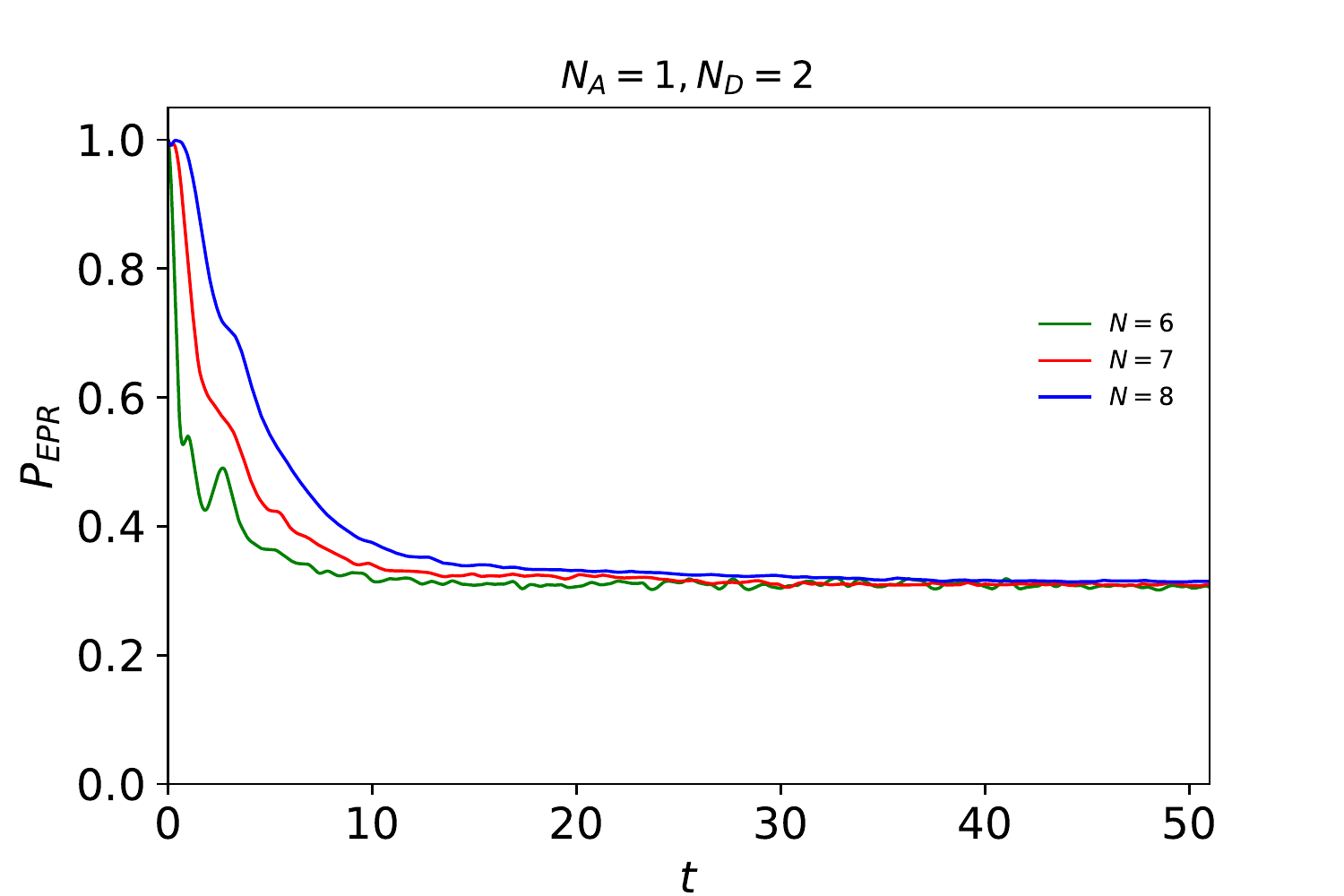}
}\end{minipage}\
\begin{minipage}{.45\textwidth}
\subfloat[$F_\text{EPR}^\text{decoh}$]{
\includegraphics[width=.95\textwidth]{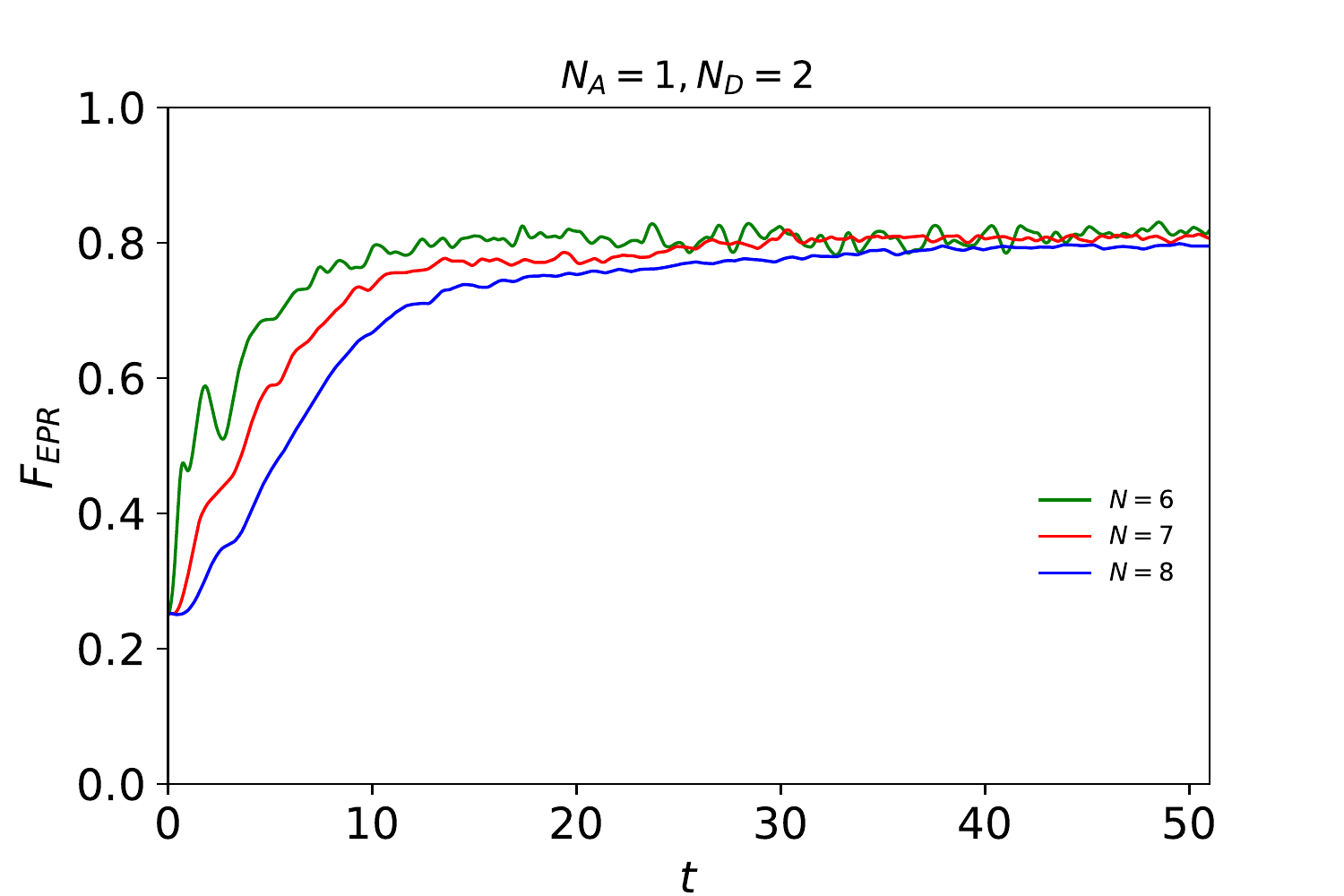}
}\end{minipage}
\caption{The Hayden-Preskill recovery protocol for the YM-Ising model with $K=2$ in the absence of decoherence. The projection probability (a), and the recovery fidelity (b) are computed with system sizes $N=6,7,8$.}
\label{fig:ladder_size}
\end{figure}

We here study how the projection probability $P_\text{EPR}$ and the recovery fidelity $F_\text{EPR}$ change as the system size is varied.
We computed them for the critical and chaotic Ising models with the number of lattice sites $N=6,7$, and $8$ (Fig.~\ref{fig:Ising_size}). The smaller the system size is, the earlier the decay of $P_\text{EPR}$ and growth of $F_\text{EPR}$ start because the distance between the subsystems $A$ and $D$ are smaller.
The qualitatively same behavior is observed in case of the YM-Ising model (Fig.~\ref{fig:ladder_size}).
\subsection{Late-time behaviors of $P_\text{EPR}$ and $F_\text{EPR}$ in YM-Ising model}
\label{app:K_dep}%
\begin{figure}[t]
\centering
\begin{minipage}{.49\textwidth}
\subfloat[$P_\text{EPR}$]{
\includegraphics[width=.95\textwidth]{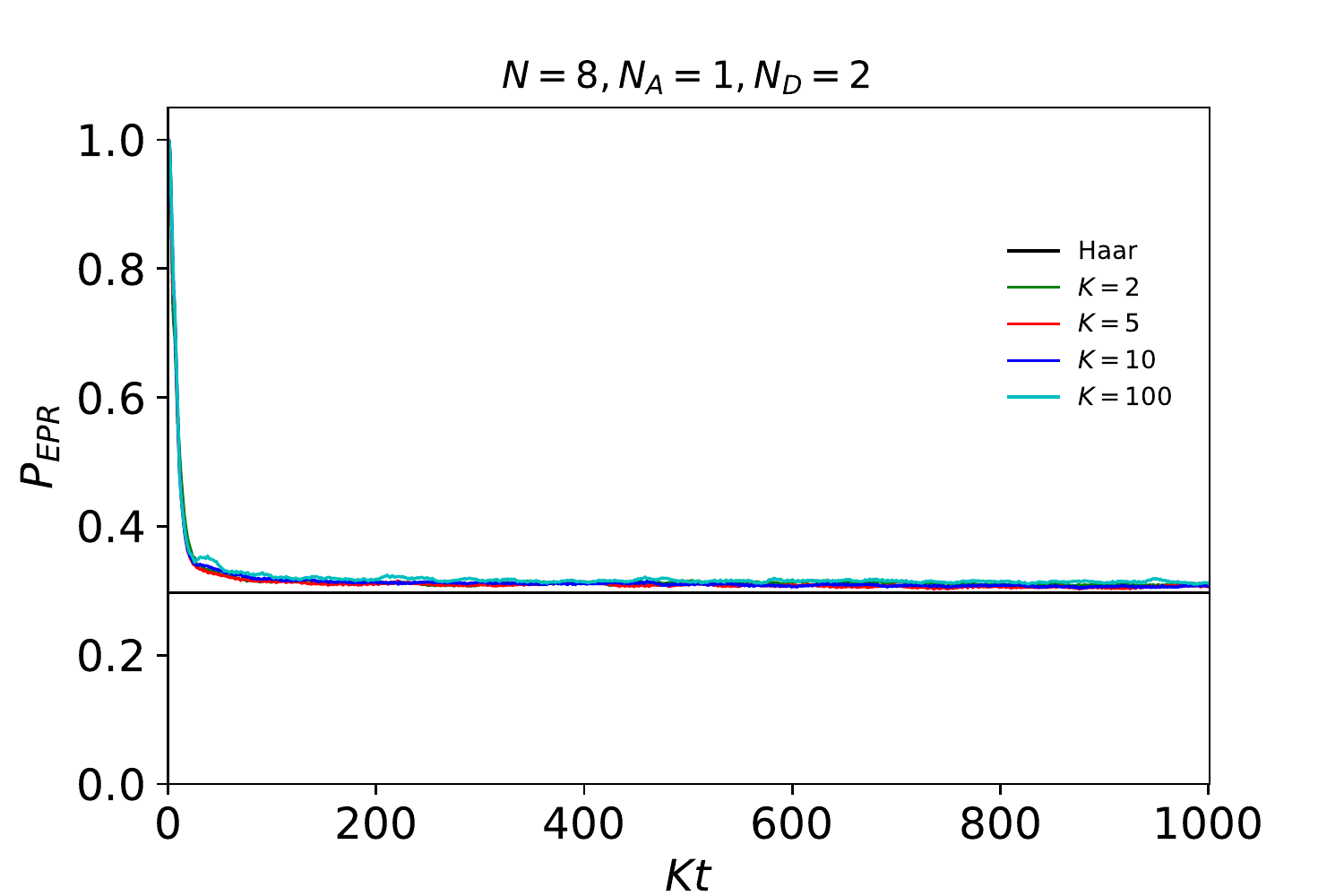}
}\end{minipage}\
\begin{minipage}{.45\textwidth}
\subfloat[$F_\text{EPR}$]{
\includegraphics[width=.95\textwidth]{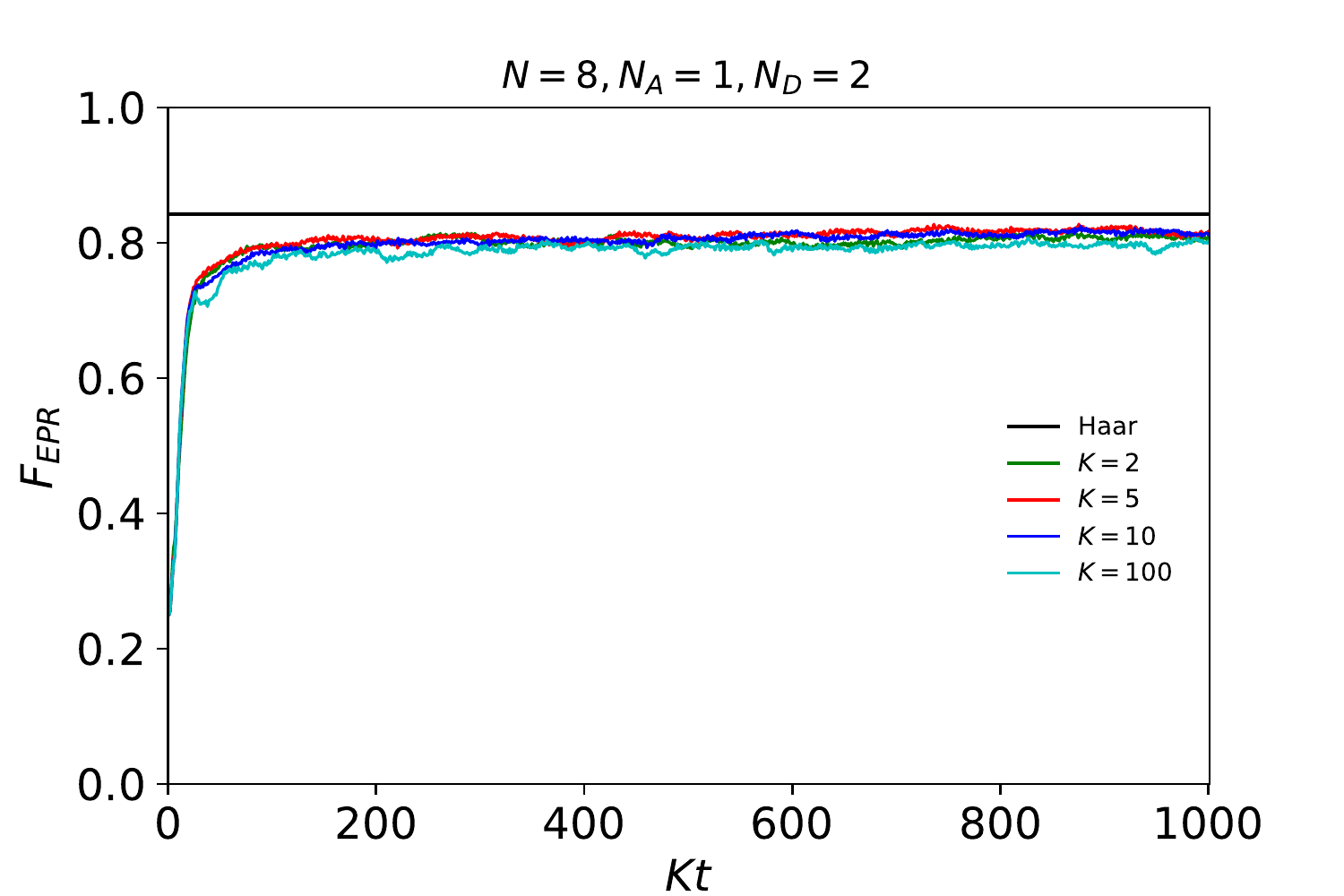}
}\end{minipage}
\caption{The Hayden-Preskill recovery protocol for the YM-Ising model in the absence of decoherence. The projection probability (a), and the recovery fidelity (b) against the rescaled time $Kt$.
}
\label{fig:ladder_Kt}
\end{figure}
In order to confirm that $P_\text{EPR}$ and $F_\text{EPR}$ in the YM-Ising model saturate at certain values close to the one from the Haar random evolution at late times, we computed their evolutions up to $Kt=1000$ as shown in Fig.~\ref{fig:ladder_Kt} (the plots of their evolutions up to $Kt=100$ are shown in Fig.~\ref{fig:ladder_Kdep}).

\section{Formulation of YM-Ising model}
\label{app:truncatedYM}

We present the reduction of the lattice Yang-Mills Hamiltonian~\eqref{eq:H_YM} on a ladder geometry to the spin Hamiltonian, the Yang-Mills-Ising model~\eqref{eq:YM_Hspin}, by employing a suitable truncation of the Hilbert space.

\subsection{Schwinger boson representation of YM theory}

We compute the matrix elements of the lattice Yang-Mills Hamiltonian~\eqref{eq:H_YM} in terms of the Schwinger bosons~\cite{Mathur:2004kr}.
Although we will consider a two-leg ladder geometry shown in Fig.~\ref{fig:ladderLattice},
we here formulate a general two-dimensional cartesian lattice system.

First, we write the chromo-electric fields using the Schwinger bosons.
Since the chromo-electric fields are generators of SU($2$), they can be identified as the angular momentum operator, and written by using the creation and annihilation operators of the spin doublet bosons (Schwinger bosons) as
\begin{align}
  E^a_{L}(\bm x,\mu) =a^\dagger(\bm x,\mu)T^a a(\bm x,\mu), 
  \quad
  E^a_{R}(\bm x,\mu)=b^\dagger(\bm x,\mu)T^a b(\bm x,\mu) ,
  \label{eq:E_right}
\end{align}
where $a=(a_\up,a_\down)$,  $a^\dagger=(a^\dagger_\up,a^\dagger_\down)$ [$b=(b_\up,b_\down)$,  $b^\dagger=(b^\dagger_\up,b^\dagger_\down)$] are the two component annihilation and creation operators of the Schwinger bosons, which are defined on the left (right) end of a link.
In the Schwinger-boson formalism, the constraint~\eqref{eq:constraint} implies
\be
N_L(\bm x,\mu)|\Psi\rangle=N_R(\bm x,\mu) |\Psi\rangle,
\ee
where $N_L= a^\dagger a=a_\up^\dagger a_\up+a_\down^\dagger a_\down$, and $N_R= b^\dagger b=b_\up^\dagger b_\up+b_\down^\dagger b_\down$ are the number operators of the Schwinger bosons.
Namely, the number of Schwinger bosons living on each end of a link must be the same.

Next, we write the link operator using the Schwinger bosons.
Using the Schwinger bosons, the link operator is written as
$U=U_L(a)U_R(b)$ with
 \begin{align}
  U_L(a) &=
  \frac{1}{\sqrt{N_L+1}}
  \begin{pmatrix}
  a_\down^\dagger & a_\up \\
  -a_\up^\dagger & a_\down
  \end{pmatrix} , \quad
  U_R(b) =
  \begin{pmatrix}
  b_\up^\dagger & b_\down^\dagger \\
  -b_\down & b_\up
  \end{pmatrix}
  \frac{1}{\sqrt{N_R+1}} .
  \label{eq:link}     
 \end{align}
Using the commutation relations between creation and annihilation operators, Eqs.~\eqref{eq:E_right}, and \eqref{eq:link}
reproduces those between the chromo-electric fields, and link operators in Eqs.~\eqref{eq:[E_L,U]}-\eqref{eq:[E_R,E_R]}.
\subsection{Physical states}
\begin{figure}[th]
\centering
\includegraphics[scale=0.25]{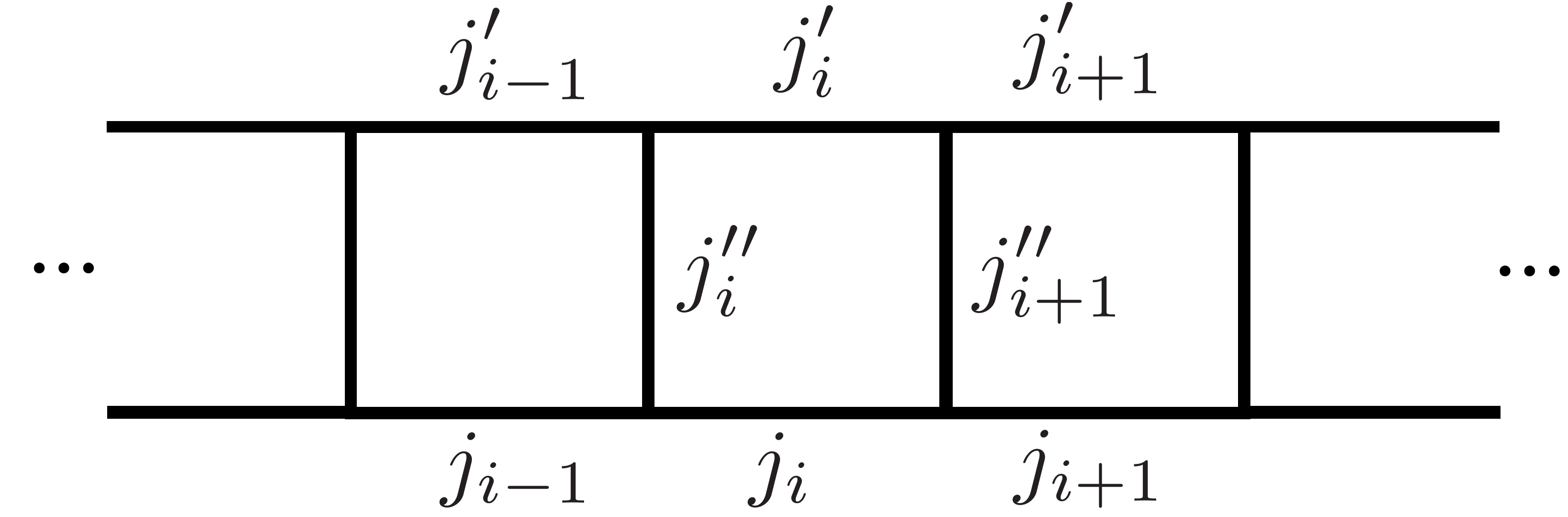}
\caption{Label of a physical state of $SU(2)$ Yang-Mills theory in a two-leg ladder geometry.}
\label{fig:ladder model}
\end{figure}
Here, we explicitly construct the physical states of the two-leg ladder system.
To this end, we solve the Gauss law constraint 
  $\sum_\mu\left( E^a_L(\bm x,\mu)+E^{a}_R(\bm x-\hat{e}_\mu,\mu) \right)|\Psi\rangle=0$
imposed at each vertex.
Let us focus on a vertex connecting to links $(i,j,k)$:
\begin{equation}
    \parbox{3cm}{\includegraphics[scale=0.3]{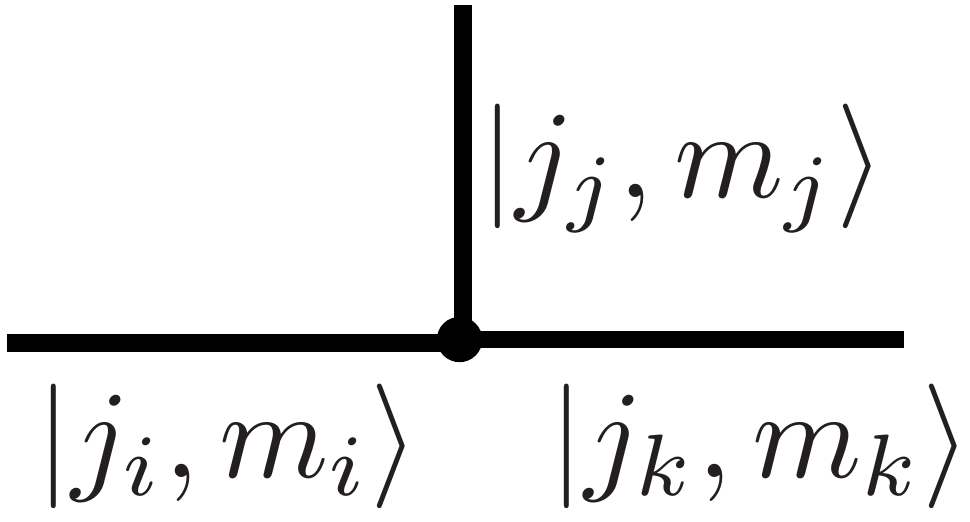}},
\end{equation}
in which the local state is expressed by 
$|j_i \, m_i\rangle|j_j \, m_j\rangle|j_k \, m_k\rangle=|N_{i\uparrow} \, N_{i\downarrow}\rangle|N_{j\uparrow} \, N_{j\downarrow}\rangle|N_{k\uparrow} \, N_{k\downarrow}\rangle$
with $j_a=(N_{a\uparrow}+N_{a\downarrow})/2$ and 
$m_a=(N_{a\uparrow}-N_{a\downarrow})/2$ ($a=i,j,k$).
$|N_{a\uparrow} \, N_{a\downarrow}\rangle$ is the eigenstate of the number operators $a^\dag_\uparrow a_\uparrow$ ($b^\dag_\uparrow b_\uparrow$) and $a^\dag_\downarrow a_\downarrow$ ($b^\dag_\downarrow b_\downarrow$) with the eigenvalues $N_{a\uparrow}$ and $N_{a\downarrow}$.
We have introduced the spin basis, which is useful for solving the Gauss law constraint.
The Gauss law constraint forces the local spin states contacted at a vertex to be the spin singlet.
The singlet resulting from the composition of the three spins is 
\begin{equation}
|j_i,j_j,j_k\rangle \coloneqq  \sum_{m_i=-j_i}^{j_i}\sum_{m_j=-j_j}^{j_j}\sum_{m_k=-j_k}^{j_k} 
\begin{pmatrix}
  j_i & j_j & j_k\\
  m_i & m_j & m_k
\end{pmatrix}
|j_i,m_i\rangle|j_j,m_j\rangle|j_k,m_k\rangle,
\end{equation}
where $\begin{psmallmatrix}
  j_i & j_j & j_k\\
  m_i & m_j & m_k
\end{psmallmatrix}$ is the Wigner $3$-$j$ symbols.
By construction, this satsifies $\langle j_i,j_j,j_k|j_{l},j_{m},j_{n}\rangle=\delta_{il}\delta_{jm}\delta_{kn}$.
The states $|j_i,j_j,j_k\rangle$ exists only if the triangle conditions $|j_i-j_j|\leq j_k\leq j_i+j_j$ and $j_i+j_j+j_k\in \mathbb{Z}$ are satisfied.
By using the creation and annihilation operators, one can directly check that the electric fields vanish on this state, i.e., $(E_R^a(i)+E_L^a(j)+E_L^a(k))|j_i,j_j,j_k\rangle=0$.
Using the singlet, the gauge invariant local states defined on vertices in lower and upper legs can be labeled by three indices on the links connecting the vertices:
\begin{align}\label{eq:SU(2)basis}
    \parbox{2.cm}{\includegraphics[scale=0.35]{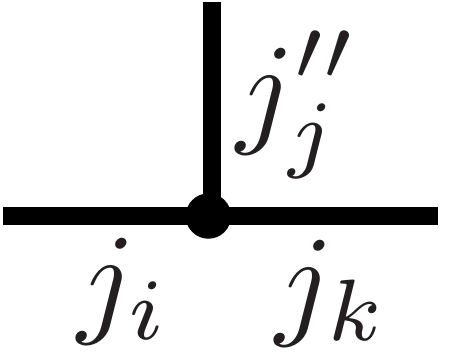}}&=|j_i,j''_j,j_k\rangle,\qquad
    \parbox{2.cm}{\includegraphics[scale=0.35]{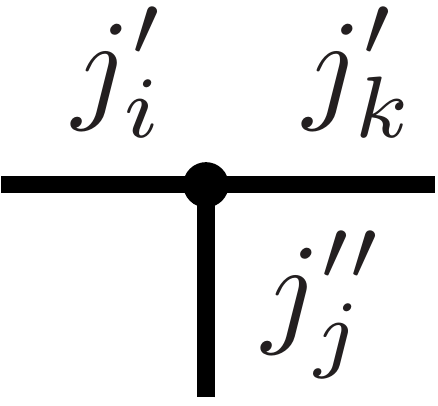}}=|j'_i,j_k',j''_j\rangle.
\end{align}
Here, we choose the order of $j_i's$ clockwise.

The $U(1)$ constraint $N_L(\bm{x},\mu)|\Psi\rangle=  N_R(\bm{x},\mu)|\Psi\rangle$ implies the vertices connecting to a link share the same spin $j$. 
We label the spins by $j_i$ and $j_i'$ for horizontal links on the $y=0$ and $y=1$, respectively.
We also label the spins on the vertical links of $x=i$ by $j_{i}''$ (See Fig.~\ref{fig:ladder model}).
We can express a physical state by using the spins of links, $\bm{j}=(j_{0},\cdots j_{N-1};j'_0,\cdots j'_{N-1};j''_{0},\cdots, j''_{N} )$ as
\begin{equation}
  \begin{split}
  |\bm{j}\rangle
  &\coloneqq \prod_{i=0}^{N} |j_{i-1},j''_{i},j_{i}\rangle |j'_{i-1},j'_{i},j''_{i}\rangle.
  \label{eq:physical state}
  \end{split}
\end{equation}
Here $j_{-1}=j'_{-1}=j_{N}=j'_{N}=0$ from the open boundary condition.
We note that not all $j_i$'s are independent. They must satisfy the triangle conditions.
In numerical simulations, we truncate the spin's maximum value, $j_\text{max}$, to make the dimension of the Hilbert space finite.
In this paper, we will consider a model with $j_{\rm max}=1/2$.

\subsection{Matrix elements of Hamiltonian}
We here evaluate the matrix elements of the Hamiltonian $H=H_E+H_M$.
Because $|\bm{j}\rangle$ is an eigenstate of $H_E$, the matrix elements of $H_E$ are the sum of all spins:
\begin{equation}
  \langle {\bm{k}}|H_E|\bm{j}\rangle = \sum_{i=0}^{N}\Bigl(\frac{j_i(j_i+1)}{2}+\frac{j'_i(j'_i+1)}{2}+\frac{j''_{i}(j''_{i}+1)}{2} \Bigr)\delta_{\bm{k},\bm{j}}.
\end{equation}
The magnetic part needs more consideration. Since $H_M$ consists of plaquettes, let us first evaluate the single plaquette, 
\begin{equation}
    \parbox{3.cm}{\includegraphics[scale=0.3]{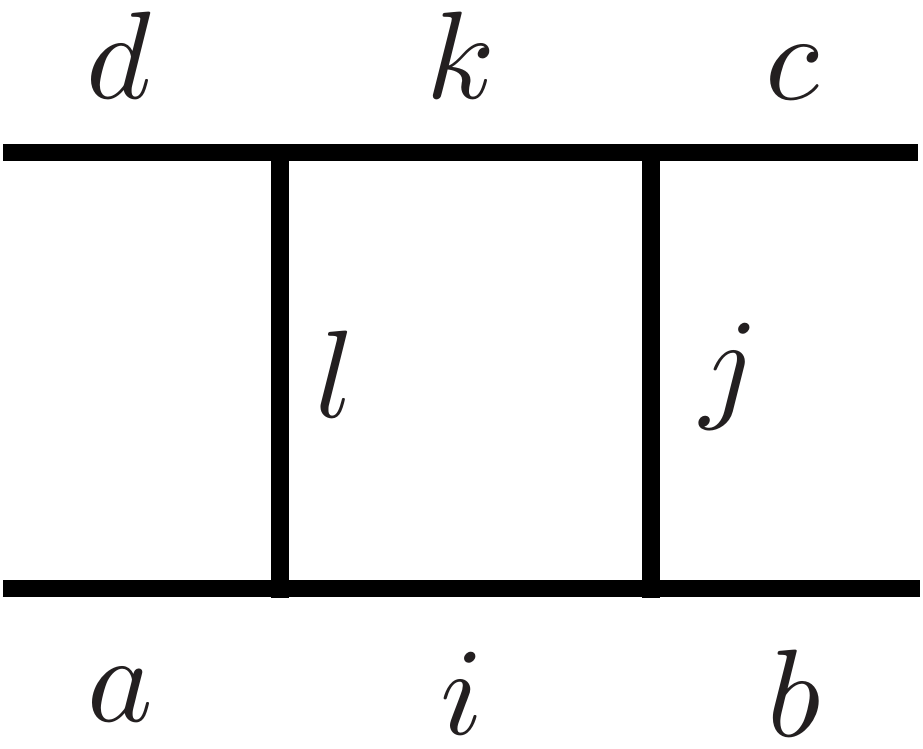}}\;\;\;\;\;,
\end{equation}
in which the plaquette operator $\mathrm{tr}(U_{i}U_{j}U_{k}^\dag U_{l}^\dag)$ acts on a state 
$|j_a,j_l,j_i\rangle |j_i,j_j,j_b\rangle|j_d,j_k,j_l\rangle|j_k,j_c,j_j\rangle$.

Noting $U_{i}=U_L(a_i)U_R(b_i)$, the plaquette operator can be decomposed into local operators on the vertices:
\begin{equation}
  \mathrm{tr}(U_{i}U_{j}U_{k}^\dag U_{l}^\dag)=
  \mathrm{tr}\bigl([U_R(b_i)U_L(a_j)][U_R(b_j)U_R^\dag(b_k)][U_L^\dag(a_k)U_R^\dag(b_l)][U_L^\dag(a_l)U_L(a_i)]\bigr),
\end{equation}
where, we used the cyclic property of the trace.
In addition, introducing
\begin{align}
  \mathcal{L}^{++}(b,a)&=
  \frac{1}{\sqrt{a_{\uparrow}^\dag a_{\uparrow}+a_{\downarrow}^\dag a_{\downarrow}+1}}
  (b_{\uparrow}^\dag a_{\downarrow}^\dag -b_{\downarrow}^\dag a_{\uparrow}^\dag)
  \frac{1}{\sqrt{b_{\uparrow}^\dag b_{\uparrow}+b_{\downarrow}^\dag b_{\downarrow}+1}},
  \\
  \mathcal{L}^{-+}(b,a)&=
  \frac{1}{\sqrt{a_{\uparrow}^\dag a_{\uparrow}+a_{\downarrow}^\dag a_{\downarrow}+1}}
  (-b_\downarrow a_{\downarrow}^\dag-b_{\uparrow} a_{\uparrow}^\dag)
  \frac{1}{\sqrt{b_{\uparrow}^\dag b_{\uparrow}+b_{\downarrow}^\dag b_{\downarrow}+1}},
  \\
\mathcal{L}^{+-}(b,a)&=
\frac{1}{\sqrt{a_{\uparrow}^\dag a_{\uparrow}+a_{\downarrow}^\dag a_{\downarrow}+1}}
(b_{\uparrow}^\dag a_{\uparrow}+b_\downarrow^\dag a_{\downarrow})
\frac{1}{\sqrt{b_{\uparrow}^\dag b_{\uparrow}+b_{\downarrow}^\dag b_{\downarrow}+1}},
\\
\mathcal{L}^{--}(b,a)&=
\frac{1}{\sqrt{a_{\uparrow}^\dag a_{\uparrow}+a_{\downarrow}^\dag a_{\downarrow}+1}}
(-b_{\downarrow} a_{\uparrow}+b_{\uparrow} a_{\downarrow})
\frac{1}{\sqrt{b_{\uparrow}^\dag b_{\uparrow}+b_{\downarrow}^\dag b_{\downarrow}+1}},
\end{align}
we can write the local operations as
\begin{align}
  U_R(b_i)U_L(a_j)
&=\begin{pmatrix}
\mathcal{L}^{++}(b_i,a_j) & \mathcal{L}^{+-}(b_i,a_j)\\
\mathcal{L}^{-+}(b_i,a_j) & \mathcal{L}^{--}(b_i,a_j)
\end{pmatrix},
\quad  U_R(b_j) U_R^\dag(b_k)
=
\begin{pmatrix}
\mathcal{L}^{+-}(b_j,b_k) & -\mathcal{L}^{++}(b_j,b_k)\\
\mathcal{L}^{--}(b_j,b_k) & -\mathcal{L}^{-+}(b_j,b_k)
\end{pmatrix},\notag\\
U_L^\dag(a_k) U_R^\dag(b_l)
&=
\begin{pmatrix}
-\mathcal{L}^{--}(a_k,b_l) & \mathcal{L}^{-+}(a_k,b_l)\\
\mathcal{L}^{+-}(a_k,b_l) & -\mathcal{L}^{++}(a_k,b_l)
\end{pmatrix},\quad
U_L^\dag(a_l) U_L(a_i)
=
\begin{pmatrix}
-\mathcal{L}^{-+}(a_l,a_i) & -\mathcal{L}^{--}(a_l,a_i)\\
\mathcal{L}^{++}(a_l,a_i) & \mathcal{L}^{+-}(a_l,a_i)
\end{pmatrix}.
\end{align}
Using these, we can express the plaquette operator as the sum of $\mathcal{L}$'s
\begin{equation}
  \mathrm{tr}(U_{i}U_{j}U_{k}^\dag U_{l}^\dag)
  =\sum_{s_i,s_j,s_k,s_l=\pm1}{\mathcal{L}}^{s_is_j}(b_i,a_j) {\mathcal{L}}^{s_js_k}(b_j,b_k) 
{\mathcal{L}}^{s_ks_l}(a_k,b_l) {\mathcal{L}}^{s_ls_i}(a_l,a_i).
\end{equation}
Each ${\mathcal{L}}^{s's}(b,a)$ locally acts on the Hilbert space defined on a single vertex, so that it is enough to consider the action of ${\mathcal{L}}^{s's}(b,a)$ on it.
Let us consider the local state $| j_i,j_j,j_{b}\rangle$.
Since ${\mathcal{L}}^{s's}(b,a)$ is written by the creation and annihilation operators of the Schwinger bosons,
we can calculate the action of ${\mathcal{L}}^{s's}(b,a)$.
Recall $|j_i,m_i\rangle=|N_{i\uparrow},N_{i\downarrow}\rangle$ with $j_i=(N_{i\uparrow}+N_{i\downarrow})/2$ and $m_i=(N_{i\uparrow}-N_{i\downarrow})/2$; we can calculate
$a_{i\uparrow}^{\dag}|j_i,m_i\rangle$ as
\begin{equation}
  a_{i\uparrow}^{\dag}|j_i,m_i\rangle
  =a_{i\uparrow}^{\dag}|N_{i\uparrow},N_{i\downarrow}\rangle
  =\sqrt{N_{i\uparrow}+1}|N_{i\uparrow}+1,N_{i\downarrow}\rangle
  =\sqrt{j_i+m_i+1}|j_i+\frac{1}{2},m_i+\frac{1}{2}\rangle.
\end{equation}
Similarly, we have
\begin{align}
  a_{i\uparrow}|j_i,m_i\rangle
  &=\sqrt{j_i+m_i}|j_i-\frac{1}{2},m_i-\frac{1}{2}\rangle,\\
  a_{i\downarrow}^{\dag}|j_i,m_i\rangle
  &=\sqrt{j_i-m_i+1}|j_i+\frac{1}{2},m_i-\frac{1}{2}\rangle,\\
  a_{i\downarrow}|j_i,m_i\rangle
  &=\sqrt{j_i-m_i}|j_i-\frac{1}{2},m_i+\frac{1}{2}\rangle.
\end{align}
Using these relations, we find the action of $\mathcal{L}^{s_is_j}(b_i,a_j)$ on $|j_i,j_j,j_b\rangle$:
\begin{equation}
  \begin{split}
  &\mathcal{L}^{s_is_j}(b_i,a_j)
    |j_i,j_j,j_b\rangle\\
     &=  \sum_{m_i=-j_i}^{j_i}\sum_{m_j=-j_j}^{j_j}\sum_{m_b=-j_b}^{j_b} 
    \begin{pmatrix}
      j_i & j_j & j_b\\
      m_i & m_j & m_b
    \end{pmatrix}\\
    &\quad\times
    \Bigl(
      s_i\sqrt{\frac{(j_i+s_im_i+\frac{1+s_i}{2})(j_j-s_jm_j+\frac{1+s_j}{2})}{(2j_i+1)(2j_j+1+s_j)}}|j_i+\frac{s_i}{2},m_i+\frac{1}{2}\rangle|j_j+\frac{s_j}{2},m_j-\frac{1}{2}\rangle\\
      &\quad-s_j\sqrt{\frac{(j_i-s_im_i+\frac{1+s_i}{2})(j_j+s_jm_j+\frac{1+s_j}{2})}{(2j_i+1)(2j_j+1+s_j)}}|j_i+\frac{s_i}{2},m_i-\frac{1}{2}\rangle|j_j+\frac{s_j}{2},m_j+\frac{1}{2}\rangle
    \Bigr)|j_b,m_b\rangle.
  \end{split}
  \label{eq:L}
\end{equation}
Since $\mathcal{L}^{s_is_j}(b_i,a_j)$ commutes with the Gauss law constraints, the right hand side of Eq.~\eqref{eq:L} must be proportional to $| j_i+{s_i}/{2},j_j+{s_j}/{2},j_{b}\rangle$, {\it i.e.},
$\mathcal{L}^{s_is_j}(b_i,a_j)| j_i,j_j,j_{b}\rangle=\lambda_{s_is_j}(j_i,j_j,j_b)| j_i+{s_i}/{2},j_j+{s_j}/{2},j_{b}\rangle$.
Comparing the wave functions of $\mathcal{L}^{s_is_j}(b_i,a_j)| j_i,j_j,j_{b}\rangle$ and $| j_i+{s_i}/{2},j_j+{s_j}/{2},j_{b}\rangle$, we obtain
\begin{align}
  \lambda_{++}(j_i,j_j,j_b)&=\sqrt{\frac{(2+j_b+j_i+j_j)(1-j_b+j_i+j_j)}{(2j_i+1)(2j_j+2)}},\label{eq:lambda++}\\
  \lambda_{--}(j_i,j_j,j_b)
&=\sqrt{ \frac{(j_b-j_i-j_j)(-1-j_i-j_j-j_b)}{(2j_i+1)(2j_j)} },\label{eq:lambda--}\\
\lambda_{+-}(j_i,j_j,j_b)
&=-\sqrt{\frac{(1+j_b+j_i-j_j)(j_b-j_i+j_j)}{(2j_i+1)(2j_j)}},\label{eq:lambda+-}\\
\lambda_{-+}(j_i,j_j,j_b)
&= \sqrt{\frac{(1+j_b-j_i+j_j)(j_b+j_i-j_j)}{(2j_i+1)(2j_j+2)} }.\label{eq:lambda-+}
\end{align}
Similar calculations work for other vertices, $|j_a,j_l,j_i\rangle$,  $|j_d,j_k,j_l\rangle$ and  $|j_k,j_c,j_j\rangle$.
Combining these results, we get
\begin{equation}
  \begin{split}
  &\mathrm{tr}(U_{i}U_{j}U_{k}^\dag U_{l}^\dag)
  |j_i,j_j,j_b\rangle
  |j_k,j_c,j_j\rangle
  |j_d,j_k,j_l\rangle
  |j_a,j_l,j_i\rangle \\
  &=\sum_{s_i,s_j,s_k,s_l=\pm 1}
  {\lambda}_{s_is_j}(j_i,j_j,j_{b})
  {\lambda}_{s_js_k}(j_j,j_k,j_{c})
  {\lambda}_{s_ks_l}(j_k,j_l,j_{d})
  {\lambda}_{s_ls_i}(j_l,j_i,j_{a}) 
  \\
  &\quad\times
  |j_i+\frac{s_i}{2},j_j+\frac{s_j}{2},j_b\rangle
  |j_k+\frac{s_k}{2},j_c,j_j+\frac{s_j}{2}\rangle
  |j_d,j_k+\frac{s_k}{2},j_l+\frac{s_l}{2}\rangle
  |j_a,j_l+\frac{s_l}{2},j_i+\frac{s_i}{2}\rangle .
  \end{split}
\end{equation}
In the same way, we can show the action of the conjugate plaquette is the same, that is, $\mathrm{tr}(U_{i}U_{j}U_{k}^\dag U_{l}^\dag)|\bm{j}\rangle=\mathrm{tr}(U_{l}U_{k}U_{k}^\dag U_{i}^\dag)|\bm{j}\rangle$.
Eventually, the plaquette Hamiltonian is written as
\begin{equation}\label{eq:magnetic part}
 \langle \bm{k}|H_M|\bm{j}\rangle
 = -K\sum_{i=0}^{N-1}\sum_{s_i,s'_i,s''_{i},s''_{i+1}=\pm1}\Lambda(\bm{s},\bm{j})
 \delta_{\bm{k},\bm{j}+\frac{\bm{s}}{2}},
\end{equation}
where we define $\bm{s}=(0,\cdots,0,s_i,0,\cdots;0,\cdots,0,s_i',0,\cdots;0,\cdots,s_i'',s_{i+1}'',0,\cdots)$,
and
\begin{equation}
\Lambda(\bm{s},\bm{j})\coloneqq\lambda_{s_i,s''_{i+1}}(j_i,j''_{i+1},j_{i+1})\lambda_{s_{i+1}'',s'_i}(j''_{i+1},j'_{i},j'_{i+1})\lambda_{s'_i,s''_{i}}(j_i',j''_{i},j'_{i-1})\lambda_{s''_{i},s_i}(j''_{i},j_i,j_{i-1}).
\end{equation}

\subsection{$j_{\rm max}=1/2$ model}
When the maximal value of spins is restricted to $1/2$, our model is significantly simplified and reduces to a spin chain model. We call the reduced model Yang-Mills-Ising model because it is converted to the Ising spin chain~\eqref{eq:YM_Hspin}.
In this case, the local Hilbert space on a vertex is four dimensional, whose basis is graphically expressed as
\begin{align}
    \parbox{2.cm}{\includegraphics[scale=0.25]{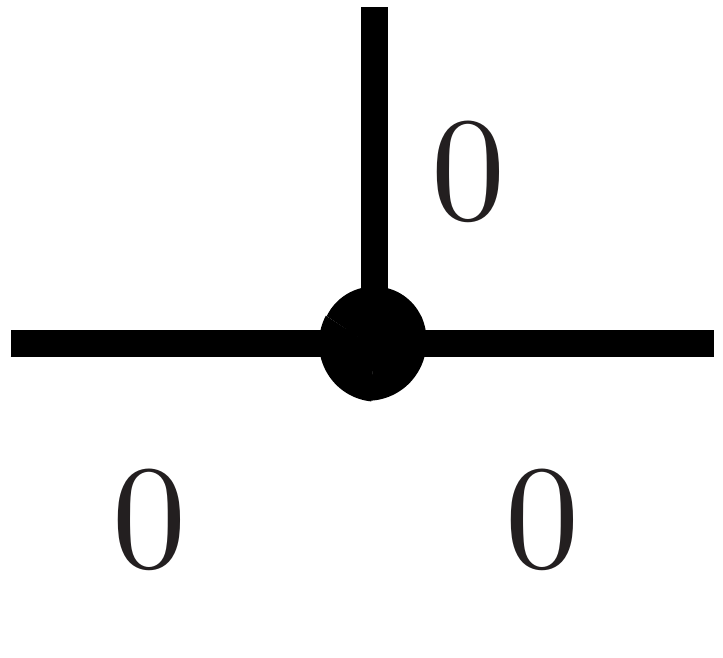}}&=\Bigl|0,0,0\Bigr\rangle,\qquad
    \parbox{2.cm}{\includegraphics[scale=0.25]{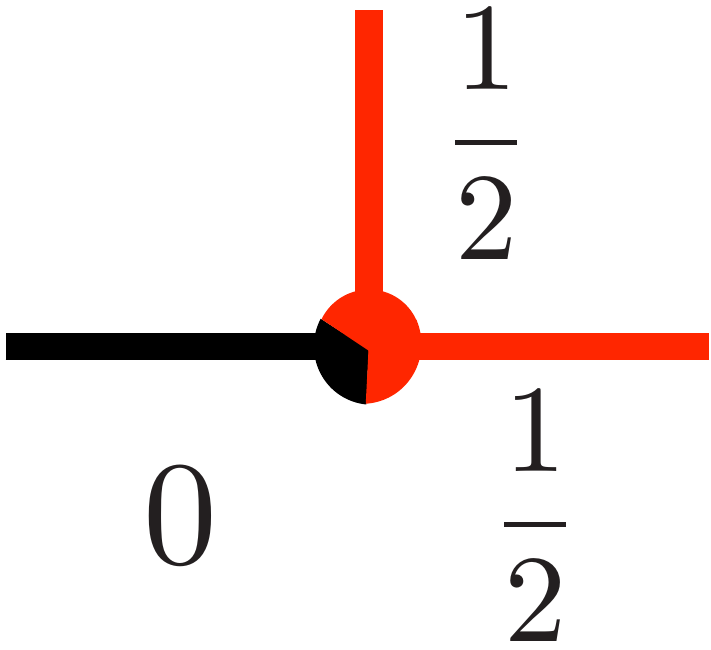}}=\Bigl|0,\frac{1}{2},\frac{1}{2}\Bigr\rangle,\\
    \parbox{2.cm}{\includegraphics[scale=0.25]{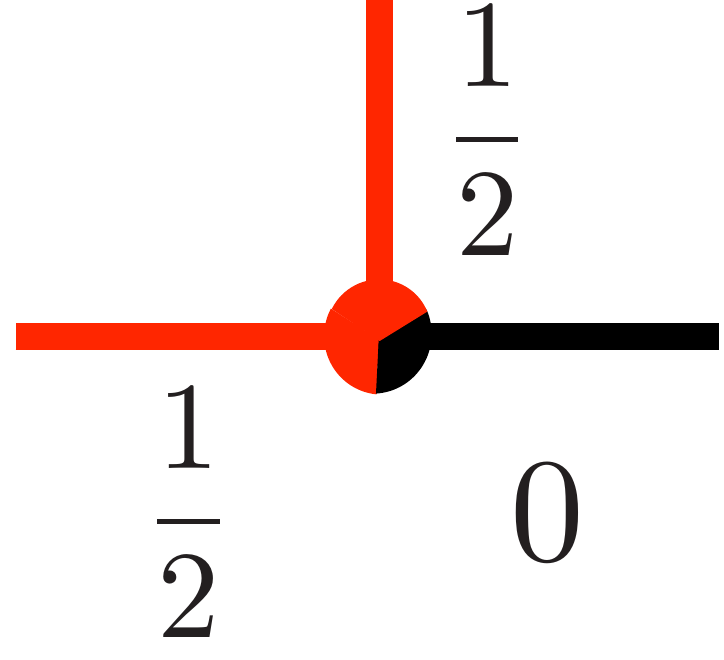}}&=\Bigl|\frac{1}{2},\frac{1}{2},0\Bigr\rangle,\qquad
    \parbox{2.cm}{\includegraphics[scale=0.25]{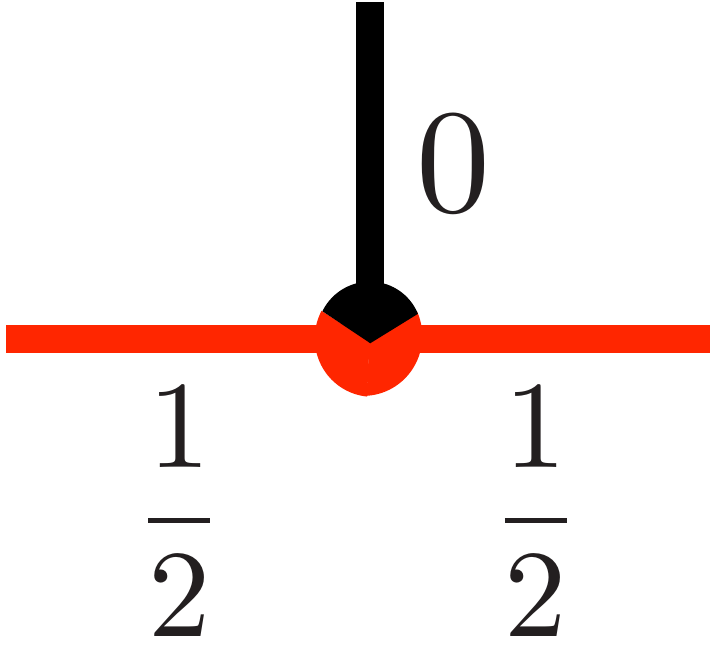}}=\Bigl|\frac{1}{2},0,\frac{1}{2}\Bigr\rangle.
\end{align}
The U(1) constraints imply that a physical state is expressed as cycles on the ladder lattice. For example, 
\begin{equation}\label{eq:state}
    \parbox{10.cm}{\includegraphics[scale=0.45]{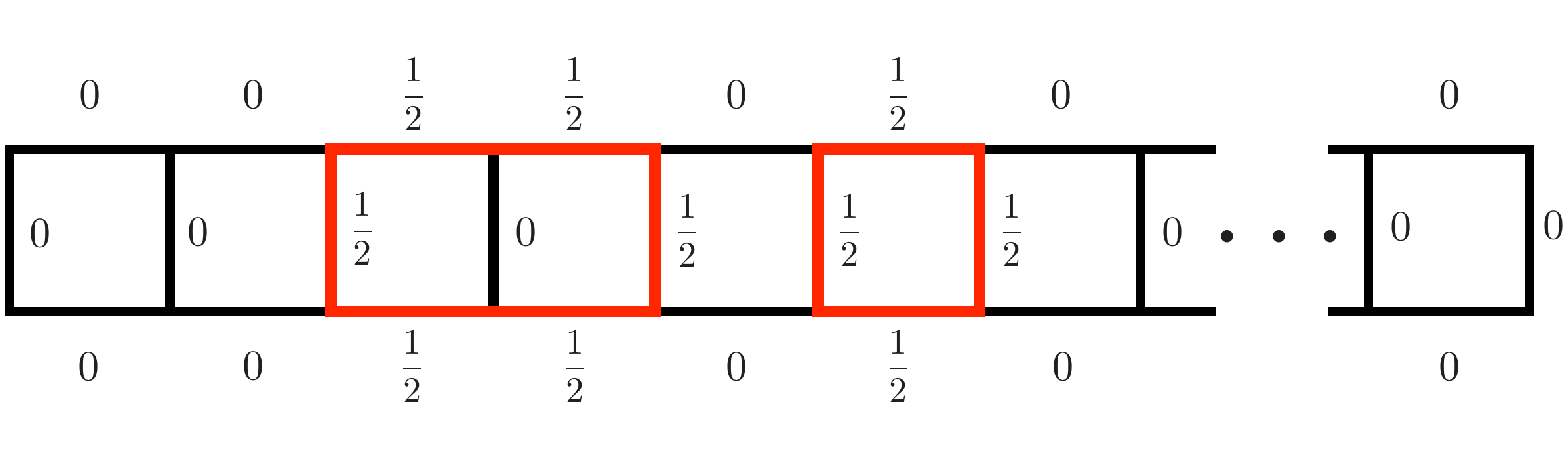}}
\end{equation}
represent a physical state.
As is in the case of the two-dimensional Ising model, a state represented by loops can be expressed by using a state of spins on the dual lattice,
and the basis of physical state is the same as Ising ones, $|\sigma_1,\sigma_2,\cdots \sigma_N,\rangle$, with $\sigma_i\in\{\up,\down\}$.
The above state in the $SU(2)$ basis \eqref{eq:state} can be expressed in the Ising basis as
\begin{equation}
    \parbox{9.5cm}{\includegraphics[scale=0.4]{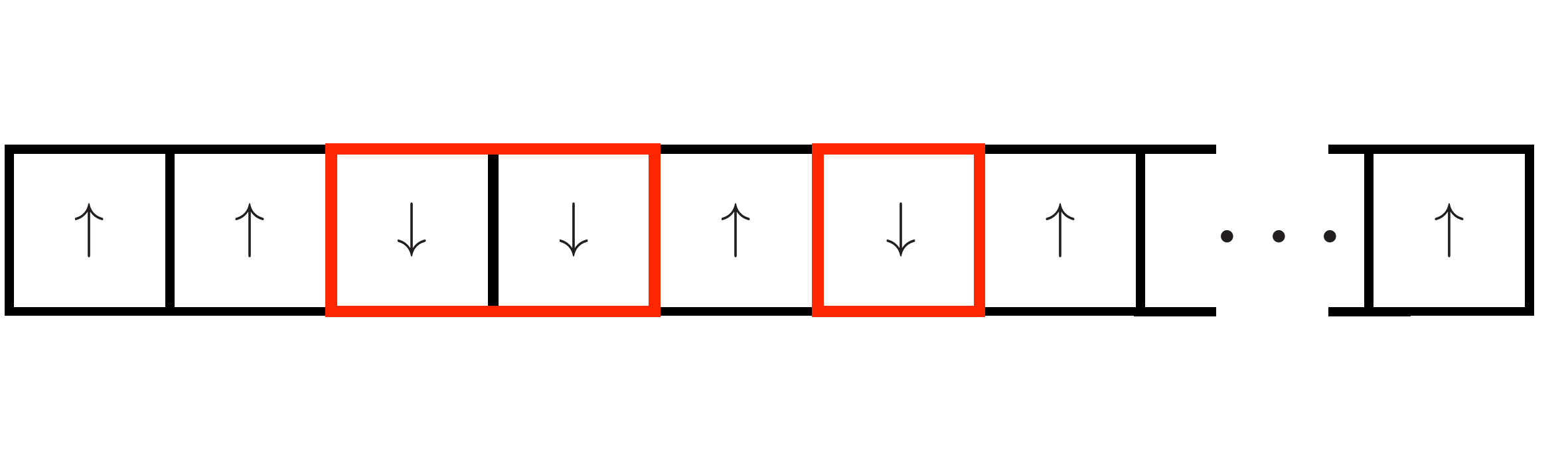}}
    =|\up\up\down\down\up\down\up\cdots\up\rangle.
\end{equation}
Let us now map the Hamiltonian of $j_{\rm max}=1/2$ to a spin chain on the dual lattice. 
The spin of the horizontal link can be replaced by
\begin{equation}
    j_i\to\frac{1}{4}(1-Z_i) ,
\end{equation}
and
\begin{equation}
    j'_i\to\frac{1}{4}(1-Z_i) .
\end{equation}
On the other hand, the vertical link is replaced by
\begin{equation}
    j''_{i}\to\frac{1}{4}(1-Z_{i-1}Z_{i}).
\end{equation}
The electric part of the Hamiltonian reduces to 
\begin{equation}
\begin{split}
     H_E &= \sum_{i=0}^{N}\frac{1}{2}\Bigl[2\frac{1}{4}(1-Z_i)\Bigl(1+\frac{1}{4}(1-Z_i)\Bigr)
    +\frac{1}{4}(1-Z_{i-1}Z_{i})\Bigl(1+\frac{1}{4}(1-Z_{i-1}Z_{i})\Bigr)
    \Bigr]\\
    &=\sum_{i=0}^{N}\frac{3}{16}( 3-2Z_{i}-Z_{i-1}Z_{i}),
\end{split}
\end{equation}
where $Z_{-1}=Z_{N}=1$ follow from the open boundary condition. This Hamiltonian is equivalent to that of the Ising model~\eqref{eq:Ising} with the vanishing transverse magnetic field.

Let us move on to the magnetic part. The matrix elements of the Hamiltonian are given in Eq.~\eqref{eq:magnetic part}.
For $j_{\rm max}=1/2$ model, there are four $\lambda_{ss'}$'s, which are explicitly given as
\begin{align}
  \lambda_{++}(0,0,0)&=1,\quad
  \lambda_{--}(1/2,1/2,0)=1,\\
 \lambda_{+-}(0,1/2,1/2)&=-1,\quad
 \lambda_{-+}(1/2,0,1/2)= \frac{1}{2}.
 \end{align}
Let us consider the action of $\mathrm{tr}U_{\square}(i)$ on $|\sigma_{i-1},\sigma_i,\sigma_{i+1}\rangle$. 
From Eq.~\eqref{eq:magnetic part}, we can explicitly obtain 
 \begin{align}
     \mathrm{tr}U_{\square}(i)|\up\up\up\rangle =&\lambda_{++}\lambda_{++}\lambda_{++}\lambda_{++}|\up\down\up\rangle=|\up\down\up\rangle,\\
     \mathrm{tr}U_{\square}(i)|\up\down\up\rangle =&\lambda_{--}\lambda_{--}\lambda_{--}\lambda_{--}|\up\up\up\rangle=|\up\up\up\rangle,\\
     \mathrm{tr}U_{\square}(i)|\up\up\down\rangle =&\lambda_{+-}\lambda_{-+}\lambda_{++}\lambda_{++}|\up\down\down\rangle=-\frac{1}{2}|\up\down\down\rangle,\\
     \mathrm{tr}U_{\square}(i)|\up\down\down\rangle =&\lambda_{-+}\lambda_{+-}\lambda_{--}\lambda_{--}|\up\up\down\rangle=-\frac{1}{2}|\up\up\down\rangle,\\
     \mathrm{tr}U_{\square}(i)|\down\down\up\rangle =&\lambda_{--}\lambda_{--}\lambda_{-+}\lambda_{+-}|\down\up\up\rangle=-\frac{1}{2}|\down\up\up\rangle,\\
     \mathrm{tr}U_{\square}(i)|\down\up\up\rangle =&\lambda_{++}\lambda_{++}\lambda_{+-}\lambda_{-+}|\down\down\up\rangle=-\frac{1}{2}|\down\down\up\rangle,\\
     \mathrm{tr}U_{\square}(i)|\down\up\down\rangle =&\lambda_{+-}\lambda_{-+}\lambda_{+-}\lambda_{-+}|\down\down\down\rangle=\frac{1}{4}|\down\down\down\rangle,\\
     \mathrm{tr}U_{\square}(i) |\down\down\down\rangle =&\lambda_{-+}\lambda_{+-}\lambda_{-+}\lambda_{+-}|\down\up\down\rangle=\frac{1}{4}|\down\up\down\rangle,
 \end{align}
 where we suppressed the indices of spins in $\lambda_{s,s'}$.
 From these expressions, we can express $\mathrm{tr}U_{\square}(i)$ as
 \begin{equation}
    \mathrm{tr}U_{\square}(i)= \frac{1}{16}X_i(1+3Z_{i-1})(1+3Z_{i+1}).
 \end{equation}
Eventually, the Hamiltonian of $j_{\rm max}=1/2$ model has the form,
\begin{align}
  H_{\rm YM} &=\sum_{i=0}^N\frac{3}{16}( 3-2Z_i-Z_{i-1}Z_{i})  -K\sum_{i=0}^{N-1} \frac{1}{16}X_i(1+3Z_{i-1})(1+3Z_{i+1}),
\end{align}
with $Z_{-1}=Z_{N}=1$.

\bibliographystyle{utphys}
\bibliography{HPbib}

\end{document}